\documentclass[5p]{elsarticle}

\usepackage{graphicx}
\usepackage{subfig}
\usepackage{epstopdf}
\usepackage{amsmath}
\usepackage{amssymb}
\usepackage{multirow}
\usepackage{hhline}
\usepackage{array}
\usepackage{subfloat}
\usepackage{float}
\usepackage{url}
\usepackage{balance}
\usepackage{longtable}
\usepackage{verbatim}

\begin{document}

\begin{frontmatter}

\title{Shedding Light on the Dark Corners of the Internet:\\
A Survey of Tor Research}
\author[seecs]{Saad Saleh\corref{cor1}}
\ead{saad.saleh@seecs.edu.pk}
\author[itu]{Junaid Qadir}
\ead{junaid.qadir@itu.edu.pk}
\author[seecs,uoj]{Muhammad U. Ilyas}
\ead{usman.ilyas@seecs.edu.pk, milyas@uj.edu.sa, usman@ieee.org}
\cortext[cor1]{Corresponding author}
\address[seecs]{School of Electrical Engineering and Computer Science (SEECS),\\ National University of Sciences and Technology (NUST), Islamabad-44000, Pakistan}
\address[itu]{Information Technology University, Lahore, Pakistan}
\address[uoj]{Department of Computer Science, Faculty of Computing and Information Technology,\\ University of Jeddah, Jeddah, Mecca Province 21589, Saudi Arabia}

\begin{abstract}
Anonymity services have seen high growth rates with increased usage in the past few years.
Among various services, Tor is one of the most popular peer-to-peer anonymizing service.
In this survey paper, we summarize, analyze, classify and quantify $26$ years of research on the Tor network.
Our research shows that `security' and `anonymity' are the most frequent keywords associated with  Tor research studies.
Quantitative analysis shows that the majority of research studies on Tor focus on `deanonymization' the design of a breaching strategy.
The second most frequent topic is analysis of path selection algorithms to select more resilient paths.
Analysis shows that the majority of experimental studies derived their results by deploying private testbeds while others performed simulations by developing custom simulators.
No consistent parameters have been used for Tor performance analysis.
The majority of authors performed throughput and latency analysis.

\end{abstract}


\end{frontmatter}

\section{Introduction}

The Internet has revolutionized the world by transforming it into a global entity.
Widespread advantages of Internet have spawned new industries and services.
However, this connectivity comes at the cost of privacy.
Every Internet client has a unique identity in the form of an Internet protocol (IP) address which can be translated to its location by the local Internet service provider (ISP).
This lack of privacy has serious implications, particularly for journalists, freedom fighters and ordinary citizens.



Lack of privacy has lead to the use of anonymous communication networks (ACN).
ACNs hide client IP addresses through various techniques.
There are a number of ACNs including Tor, Java anonymous proxy (JAP), Hotspot Shield and UltraSurf etc.
Among various ACNs, Tor is the one of the most popular network, owing to its distributed nature which makes it difficult to connect the two end points of a session.
Recently, Tor has been used for bomb hoax at Harvard \cite{Brandom2013}. Similarly, it has been used by the Russians to bypass online censorship \cite{Khrennikov2016}.
A number of attempts are being made by FBI and other organizations to breach Tor network \cite{Hern2016}\cite{Graham-Smith2016}\cite{Neal2016}.

In this paper, we survey various studies conducted on the Tor network covering the scope of these studies.
We quantify the studies into three broad but distinct groups, including (1) deanonymization, (2) path selection, (3) analysis and performance improvements, and several sub-categories.
To the authors' best knowledge, this is the most comprehensive attempt at analyzing Tor network research with a focus over its anonymity mechanism.
Table \ref{tab: Comparison of other Surveys with this survey}  presents a comparison of this survey with previous surveys covering the scope of researches and implementation (experiments), verification (simulations) and analysis of various research works. Categorization of first column is made by listing all Tor areas considered in our study. AlSabah and Goldberg \cite{alsabah2016performance} presented the most comprehensive study covering complete Tor network and our paper is complementary to their survey paper. However, our paper pays more focus to the anonymity and breaching aspects of Tor than their paper. Their research paper presents only twenty references related to anonymity while we present more than $120$\footnote{This paper has 146 references, some of the references are to tools rather than research works; in all, we are considering a research corpus of 120 references.} references.

\begin{table*}[t]
  \caption{Comparison of other surveys with this survey.}\label{tab: Comparison of other Surveys with this survey}
  \centering
  \small
  \begin{tabular}{|
  @{}>{\centering}p{2.3cm}@{}|
  @{}>{\centering\arraybackslash}p{2.5cm}@{\hspace{0.015in}}|
  @{}>{\centering\arraybackslash}p{2.2cm}@{\hspace{0.015in}}|
  @{}>{\centering\arraybackslash}p{1.2cm}@{\hspace{0.015in}}|
  @{}>{\centering\arraybackslash}p{1.4cm}@{\hspace{0.015in}}|
  @{}>{\centering\arraybackslash}p{0.9cm}@{\hspace{0.015in}}|
  @{}>{\centering\arraybackslash}p{1.5cm}@{\hspace{0.015in}}|
  @{}>{\centering\arraybackslash}p{1.2cm}@{\hspace{0.015in}}|
  @{}>{\centering\arraybackslash}p{1.4cm}@{\hspace{0.015in}}|
  @{}>{\centering\arraybackslash}p{1.3cm}@{\hspace{0.015in}}|
  @{}>{\centering\arraybackslash}p{1.15cm}@{\hspace{0.015in}}|
  }
     \hline
     \multirow{14}{*}{Scope} & \multicolumn{2}{|c|}{[Areas]$\downarrow$ / [Research, Year]$\rightarrow$} & $<$This Paper$>$ & AlSabah and Goldberg \cite{alsabah2016performance} & Koch \emph{et al.} \cite{koch2016anonymous} & AlSabah and Goldberg \cite{alsabah2015performance} & Conrad and Shirazi \cite{conrad2014survey} & Jagerman \emph{et al.} \cite{jagerman2014fifteen} & Ren and Wu \cite{ren2010survey}  & Johnson and Kapadia \cite{johnson2007chaum}   \\

     \cline{2-11}
     & \multicolumn{2}{|c|}{Year$\rightarrow$}                      & 2017    & 2016 & 2016   & 2015 & 2014 & 2014 & 2010     & 2007    \\
     \cline{2-11}

      & \multicolumn{2}{|c|}{Coverage (Studies)$\rightarrow$}       & 120 & 120 & 10 & 99 & 40 & 37 & 109 & 32    \\
     \cline{2-11}
           & \multirow{6}{*}{Deanonymization}   & Hidden Services   & \checkmark & \checkmark &  & \checkmark & &   & \checkmark & \checkmark  \\
           &                               & Finger printing        & \checkmark    & \checkmark &  & \checkmark &  &    &            &             \\
           &                               & Attacks                & \checkmark    & \checkmark &   & \checkmark & \checkmark & \checkmark  & \checkmark & \checkmark   \\
           &                               & Traffic Analysis       & \checkmark    & \checkmark & \checkmark  & \checkmark &  \checkmark & \checkmark &           & \checkmark   \\
           &                               & Improvements           & \checkmark    & \checkmark &   & \checkmark   &  &  &           &             \\
           &                               & Bypassing Tor          & \checkmark    &\checkmark  &  &  &  \checkmark & \checkmark & \checkmark & \checkmark    \\
     \cline{2-11}
           & \multirow{2}{*}{Path Selection}  & Algorithm design    & \checkmark & \checkmark &  & \checkmark   &  &   &         &            \\
           &                               & Analysis               & \checkmark & \checkmark &  & \checkmark  &  &   &              &             \\
     \cline{2-11}
           & \multirow{4}{*}{Analysis}              & General       & \checkmark &  & \checkmark & \checkmark & \checkmark &   & \checkmark & \checkmark   \\
           &                                      & Modelling       & \checkmark &  &  &           &  &  &            &    \\
           &                                      & Analysis        & \checkmark &  & \checkmark &       &  &       &            &   \\
           &                                      & Improvement     & \checkmark & \checkmark &  &         &  &    &            &    \\
           &                                      & Mobile Tor      & \checkmark &  &           &            &  &  &  &  \\
     \hline
     \multirow{5}{*}{Implementation} & \multirow{5}{*}{Experiments} & Private Setup     & \checkmark & &  &  &  &  &  &  \\
                    &                                               & PlanetLab         & \checkmark & &  &  &  &  &  &  \\
                    &                                               & Cloud Services    & \checkmark & &  &  &  &  &  &  \\
                    &                                               & OpenFlow          & \checkmark & &  &  &  &  &  &  \\
                    &                                               & UC Framework      & \checkmark & &  &  &  &  &  &  \\
     \cline{1-11}
     \multirow{4}{*}{Verification}   & \multirow{5}{*}{Simulations} & Cus. Simulator    & \checkmark & &  &  &  &  &  & \\
                    &  & ExperimenTor                                                   & \checkmark & &  &  &  &  &  &  \\
                    &  & Shadow Simulator                                               & \checkmark & &  &  &  &  &  &  \\
                    &  & ModelNet                                                       & \checkmark & &  &  &  &  &  &  \\
     \hline
     Analysis & Parameters  &                                                           & \checkmark & &  &  & &  &    & \checkmark   \\
     \hline

   \end{tabular}

\end{table*}

Analysis of keywords used in various studies shows that anonymity, security and privacy have been used the most.
Our study shows that majority of the research works have been made in the field of ``deanonymization'' track, followed by ``performance analysis and architectural improvements''.
In the deanonymization track, a major chunk of research is devoted to \emph{breaching attacks} followed by \emph{traffic analysis}.
In the path selection track, most research works focused on the development of new algorithms.
Relays, protocol messages and traffic interception have been the most frequently exploited factors in the Tor's deanonymization track.
In the path selection track, performance and anonymity have been the most commonly used factors.
Performance, relay selection and anonymity have been the most studied parameters in the performance analysis and improvement track.
Analysis over simulations and experiments shows that 60\% of studies used experiments and 86\% of those experiments were carried out on private testbed networks.
Among simulations, 75\% of the studies developed their own simulator to analyze Tor network.
Analysis of simulation parameters shows that there is no distinct pattern of parameters. However, majority of the studies used bandwidth and latency.

Table \ref{tab: Glossary of all abbreviations used in the text.} presents a glossary of the important abbreviations used in our survey paper. This paper is organized as follows:  Section \ref{sec: Tor Architecture} introduces the architecture of Tor network and its comparison with other anonymity services. Section \ref{sec:Tor Research Areas} presents the studies covering deanonymization, path selection, and performance analysis and architectural improvements. Section \ref{sec:Platforms for Tor Research} presents the simulations and experiments conducted in previous studies. Section \ref{sec: Discussion} presents the Tor performance metrics, our findings and open research areas in Tor. Finally, section \ref{sec: Conclusion} concludes the paper.

\begin{table*}[t]
    \centering
    \caption{Glossary of the important abbreviations used in the text.}\label{tab: Glossary of all abbreviations used in the text.}
    \small
    \begin{tabular}{|c|c|c|c|}
        \hline
        \multicolumn{4}{|c|}{Glossary}\\
        \hline \hline
        ACK & Acknowledgement &
        MRA & Multi-Resolution Analysis \\
        ACN & Anonymous Communication Network &
        NAT & Network address translation \\
        ADSL & Asymmetric digital subscriber line &
        NTP & Network Time Protocol \\
        AES & Advanced Encryption Standard &
        OP & Onion Proxy \\
        AS & Autonomous System &
        OR & Onion Router \\
        CGI & Computer-generated imagery &
        PGP & Pretty Good Privacy \\
        CSRF & Cross site request forgery &
        POP3 & Post Office Protocol 3 \\
        DHCP & Dynamic Host Configuration Protocol &
        PPTP & Point-to-Point Tunneling Protocol \\
        DNS & Domain Name System &
        P2P & Peer-to-Peer \\
        DoS & Denial of Service &
        QoE & Quality of Experience \\
        DS & Directory Server &
        QoS & Quality of Service \\
        DSL & Digital Subscriber Line &
        ROC & Region of Convergence \\
        EWMA & Exponentially weighted moving average &
        RRD & Round Robin Database \\
        FIFO & First In First Out &
        RTT & Round Trip Time \\
        FN & False Negative &
        SMTP & Simple Mail Transfer Protocol \\
        FP & False Positive &
        SSH & Secure Shell \\
        HTML & HyperText Markup Language&
        SVM & Support Vector Machine \\
        HTTP & Hypertext Transfer Protocol &
        TAP & Tor Authentication Protocol \\
        IMAP & Internet Message Access Protocol &
        TCP & Transmission Control Protocol \\
        ICMP & Internet Control Message Protocol &
        TMT & Tunable mechanism of Tor \\
        IP & Internet Protocol &
        Tor & The Onion Router \\
        ISP & Internet Service Provider &
        TP & True Positive \\
        I2P & Invisible Internet Project &
        TTL & Time To Live \\
        JAP & Java Anonymous Proxy &
        URL & Uniform Resource Locator \\
        JVM & Java virtual machine &
        VDE & Virtual Distributed Ethernet \\
        LAN & Local area network &
        VM & Virtual Machine \\
        L2TP & Layer 2 Tunneling Protocol &
        VPN & Virtual Private Network \\
        ML & Machine Learning &
        VPS & Virtual Private Server \\
        \hline
    \end{tabular}
\end{table*}

%

%
%
%
%
%
%
%
%


\section{Anonymity tools}
\label{sec: Tor Architecture}

In this section, we present and discuss Tor and other anonymity tools.
In the first part, we present the architecture of Tor network before presenting details of the research in Tor.
In the second part, we present the comparison and working mechanism of other anonymity tools which compete with Tor.

\subsection{Architecture of Tor network}

Tor network is composed of a decentralized distributed network of relays operated by volunteer users \cite{goldschlag1999onion}.
In July $2016$, nearly $10,000$ users (per day) participated in the Tor network (as Tor relays and Tor bridges) to provide anonymity services to nearly half a million users daily \cite{torMetricPortal2016}.
History of Tor dates back to late 1990's when Goldschlag, Reed and Syverson presented the architecture and implemented onion routing in several papers \cite{reed1996proxies, goldschlag1996hiding, reed1998anonymous, syverson2001towards} which laid the foundation of Tor network by providing proxy servers which are resilient to eavesdropping and effectively hide client's IP address.

The Tor network consists of routers which cooperate with each other to provide low latency anonymity services to users.
Central servers help Tor establish and update links between Tor routers.
User participation as Tor relays (router) is optional, but it is recommended because it improves the chances of staying anonymous, because it increases the traffic to the user.

Tor's architecture has three types of components, namely onion proxy (OP), onion router (OR) and directory server (DS).
OPs are used by Tor users to obtain up-to-date information of operating relays from DS.
OPs also creates connections using the information from a DS.
Users may configure OPs to select specific routers.

ORs are Tor relay routers, operated by volunteer users, to act as entry (guard), middle and exit relays.
Information of all online relays is available at DS.
To counter attacks on Tor that block Tor relays, a secret group of Tor relays exists with the DS, called \emph{bridges}.
A set of three bridge relays is available through unique \emph{Gmail} addresses.
Once a connection is established, every OR knows only immediate predecessor and successor node.

Nine authorities acting as Tor DSs keep an up-to-date record of all available ORs and broadcast the bandwidth, IP, public key, exit policies etc. to OPs.

\begin{figure}[t]
  \centering
  \includegraphics[width=0.9\columnwidth]{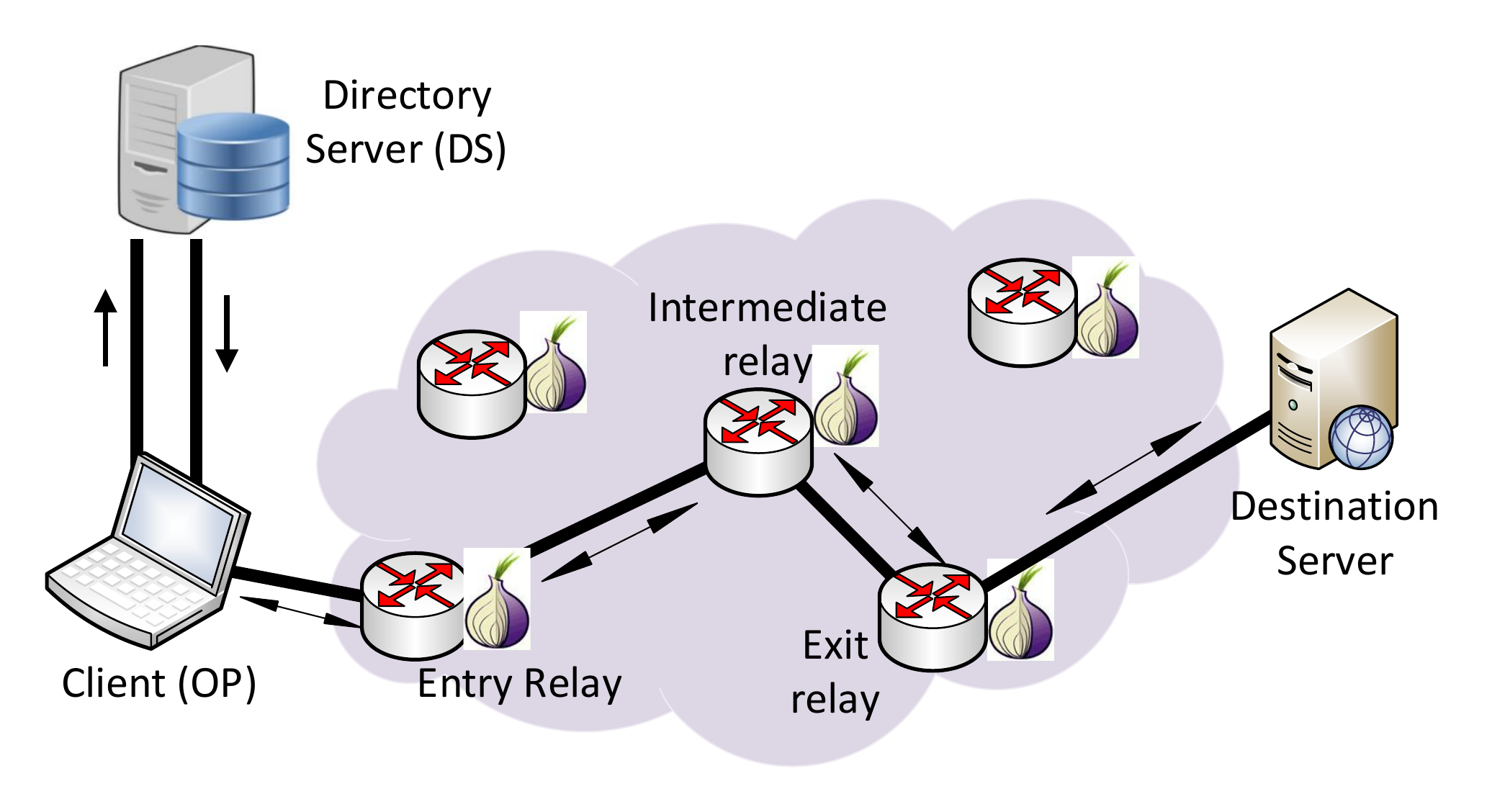}
  \caption{Architecture of Tor network.}
	\label{fig:tor_architecture}
\end{figure}

Figure \ref{fig:tor_architecture} shows a circuit established from a user to a server through three Tor relays.
The various steps involved in circuit establishment are listed below:
\begin{enumerate}
  \item OP sends HTTP requests to DS for information about Tor relays.
  \item OP selects three Tor routers (entry relay, intermediate relay and exit relay) using Tor's path selection algorithm considering maximum anonymity and performance.
  \item OP sends a \emph{Create Cell} request (containing half of Diffie-Hellman handshake \cite{bresson2001provably}) to the entry node. The entry node replies with the hash of the negotiated key.
  \item Next, OP sends a \emph{Extend Cell} request to the entry node containing the address of the intermediate relay and encryption key. Entry node forwards the cell to the intermediate node. Similar to previous case, the intermediate node replies with \emph{Created cell} response. Similar process continues till client negotiates the key with the exit relay.
  \item OP constructs IP packet $P1$ with source and destination IP addresses of exit relay and destination server, respectively, and packet size of $512$ bytes.
  \item OP encrypts the packet further with the key $E3$, negotiated between client and exit relay, and containing source and destination addresses of intermediate relay and exit relay.
  \item Next, OP encrypts with the key $E2$, negotiated between client and intermediate relay, and containing source and destination IPs of entry relay and intermediate relay.
  \item Finally, OP constructs packet $P2$ encrypted with key $E1$, negotiated between client and entry relay, and containing source and destination IPs of OP and entry relay.
  \item Packet $P2$ is transmitted from the entry relay, which decrypts the packet and forwards it to intermediate relay. All relays decrypt the packet using their specialized decryption keys and forward it towards the destination.
\end{enumerate}

\emph{Cell} refers to the \emph{Tor Packet}, comprising of payload data and headers, with an aggregate size of 512 bytes \cite{pries2008new, barbera2013cellflood, wang2013improved}.
%
%
Padding is used to fill cells with less data.

Tor relays communicate with each other by pairwise TCP connections.
Traffic multiplexing is used to transfer data between any pair of relays.
Tor employs token buckets to rate limit connections.
Buckets are filled and removed with tokens based on configured bandwidth limits and data read, respectively.
TCP buffers are read using a round-robin scheduling mechanism.
For flow control, edges (client and exit node) keep track of data flow by maintaining an active window about the packets in flight.
Data packets are processed in a first-in-first-out (FIFO) manner from the queues of Tor relays.
Multiplexing of packets, from Tor relays to relay links, is performed using exponentially-weighted moving average (EWMA) scheduler.

\subsection{Comparison with other anonymization services}

In this section, we present the features and working mechanisms of other deanonymization services which compete with Tor network. Table \ref{tab: Comparison of Tor   with other anonymization services} presents a comparison of Tor with other deanonymization services. Table \ref{tab: Comparison of Tor   with other anonymization services} shows that Tor is the only anonymity service which provides various services (http, https, visible TCP port, remote DNS, hides IP and user-to-proxy encryption) under all circumstances. On the contrary, JonDo, I2P, CGI and socks5 provide some services in limited circumstances only. A summary of various anonymity services is presented in following subsections.

\begin{table*}[t]
  \caption{Comparison of Tor with other anonymization services (`\checkmark $\star$' refers to `In limited circumstances').}\label{tab: Comparison of Tor with other anonymization services}
  \centering
  \begin{tabular}{|p{1.5cm}|p{1.5cm}|p{1.5cm}|p{1.5cm}|p{1.5cm}|p{1.5cm}|p{1.5cm}|p{1.5cm}|}
    \hline
    Proxy/ Anon. Service & HTTP & HTTPS & Visible TCP Port & UDP & Remote DNS & Hides IP & user-to-proxy encryption \\
    \hline
    \hline
    http & \checkmark &  &  &  & \checkmark & \checkmark $\star$ &  \\
    \hline
    https & \checkmark & \checkmark &  &  & \checkmark & \checkmark $\star$ &  \\
    \hline
    socks4 & \checkmark & \checkmark &  &  &  & \checkmark &  \\
    \hline
    socks4a & \checkmark & \checkmark &  &  & \checkmark & \checkmark &  \\
    \hline
    socks5 & \checkmark & \checkmark &  & \checkmark & \checkmark & \checkmark &  \\
    \hline
    CGI & \checkmark $\star$ & \checkmark $\star$ &  &  & \checkmark & \checkmark $\star$ & \checkmark $\star$ \\
    \hline
    I2P & \checkmark $\star$ & \checkmark $\star$ &  & \checkmark & \checkmark & \checkmark & \checkmark \\
    \hline
    JonDo & \checkmark & \checkmark &  & \checkmark $\star$ & \checkmark & \checkmark & \checkmark \\
    \hline
    \textbf{Tor} & \checkmark & \checkmark & \checkmark &  & \checkmark & \checkmark & \checkmark \\
    \hline
  \end{tabular}
\end{table*}

\subsubsection{Cross Platform Anonymity Tools}

A number of cross platforms anonymity tools are used now-a-days. In below lines, we summarize the basic working mechanism of prominent anonymity tools.

\begin{itemize}
  \item Java Anonymous Proxy (JAP or JonDonym) \cite{gordon2016official}: Users can select among several Mix Cascades, different from P2P.
  \item PacketiX.NET \cite{softether2016}: Virtual LAN card and Virtual HUB by Ethernet and can provide layer $2$ VPN virtualization.
  \item JanusVM \cite{janusvm2016}: Uses Virtual Private Network (VPN) connection - No update on project since $2010$.
  \item proXPN \cite{proxpn2016}: A personal VPN that provides you safety and privacy while using the Internet.
  \item USAIP \cite{USAIP2016}: A VPN service provider with servers in Switzerland, Luxembourg and Hungary etc.
  \item VPNReactor \cite{VPNreactor2016}: Uses a VPN connection with time limits for free and pro service and user logs are kept for $5$ days.
\end{itemize}

\subsubsection{Windows Based Anonymity Tools}

A number of windows based anonymity tools are large competitors of Tor network. Basic mechanisms of prominent anonymity tools are summarized in below lines:

\begin{itemize}
  \item xB Browser \cite{xbbrowser2009}: A browser designed to run over the Tor network and XeroBank anonymity network.
  \item Hotspot Shield \cite{hotspotshield2016}: Uses VPN. Hosts web servers accessible through proxy and has a central server that can be compromised.
  \item AdvTor \cite{AdvTor2016}: Acts as a portable client and server for the Tor network. Improvements include the UNICODE path, HTTP and HTTPS protocols, estimates AS paths etc.
  \item SecurityKISS \cite{securitykiss2016}: A VPN service based on OpenVPN, PPTP and L2TP.
  \item UltraSurf \cite{ultrsurf2016}: Uses HTTP proxy to bypass censorship and uses encryption protocols for privacy.
  \item CyberGhost VPN \cite{cyberghost2016}: OpenVPN based proprietary client, Centralized server with VPN using 1024-SSL encryption.
  \item Freegate \cite{freegate2016}: Uses range of proxy servers (called Dynaweb) along with encryption.
\end{itemize}

\subsubsection{Linux based solutions}

Linux, being the prominent platform, is used by various anonymity tools to guarentee anonymity to its users. Working mechanism of some tools is summarized in below lines:

\begin{itemize}
  \item Tails (Amnesic Incognito Live System) \cite{tails2016}: Has a Debian Linux distribution using Tor network.
  \item Privatix \cite{privatix2016}: Provides encryption with anonymous web browsing (using Tor, Torbutton and firefox).
\end{itemize}

\subsubsection{Anonymous Search Engines}

Several search engines also provided anonymity to their users by collecting no user information. Some of these anonymous search engines are as follows:

\begin{itemize}
  \item Ixquick \cite{ixquick2016}: Opens all search results through a proxy for anonymity.
  \item DuckDuckGo \cite{duckduckgo2016}: Provides searcher's privacy and avoids ``filter bubble''. Displays same information to all users for a given search.
\end{itemize}

\subsubsection{Anonymous Emails}

Anonymous emails are provided by many servers. In following lines, we have summarized the tools provided anonymous email capability to their users.

\begin{itemize}
  \item Anonymous E-mail \cite{anonymousemail2016}: Sends the emails with anonymous senders.
  \item Safe-mail \cite{safemail2016}: Provides a secure communication, storage, distribution and sharing system for the internet.
  \item HushMail \cite{hushmail2016}: Provides a PGP encrypted email service.
  \item 10 Minute Mail \cite{10minutemail2016}: Gives an email address which expires after 10 minutes to counter spam emails.
  \item Yopmail \cite{yopmail2016}: Provides a disposable email address with no registration and password.
\end{itemize}

\section{Tor Research Areas}
\label{sec:Tor Research Areas}

\begin{figure*}[t]
  \centering
  \includegraphics[width=1.5\columnwidth]{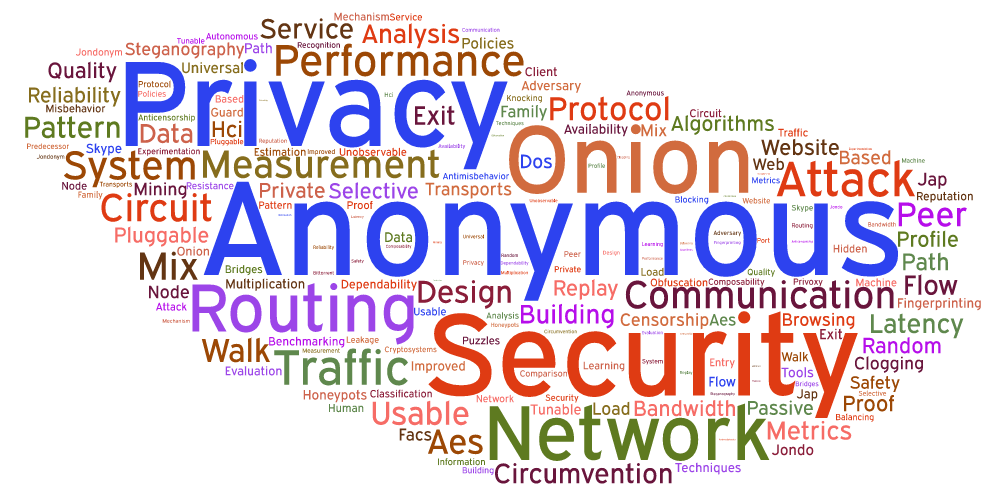}\\
  \caption{Word cloud of \emph{keywords} used in Tor research works.}\label{fig:Word cloud of keywords used in Tor researches}
\end{figure*}

A large number of networks have utilized Tor networks for various purposes.
To get an overview of the research areas dealt in various studies, we created a word cloud of the keywords of these studies.
Figure \ref{fig:Word cloud of keywords used in Tor researches} shows the word cloud of keywords (on log-scale) of all studies cited in this paper.
Data of keywords shows that anonymity, privacy and security are the most important terms dealt in various studies.

\begin{figure*}[t]
  \centering
  \includegraphics[width=2.0\columnwidth]{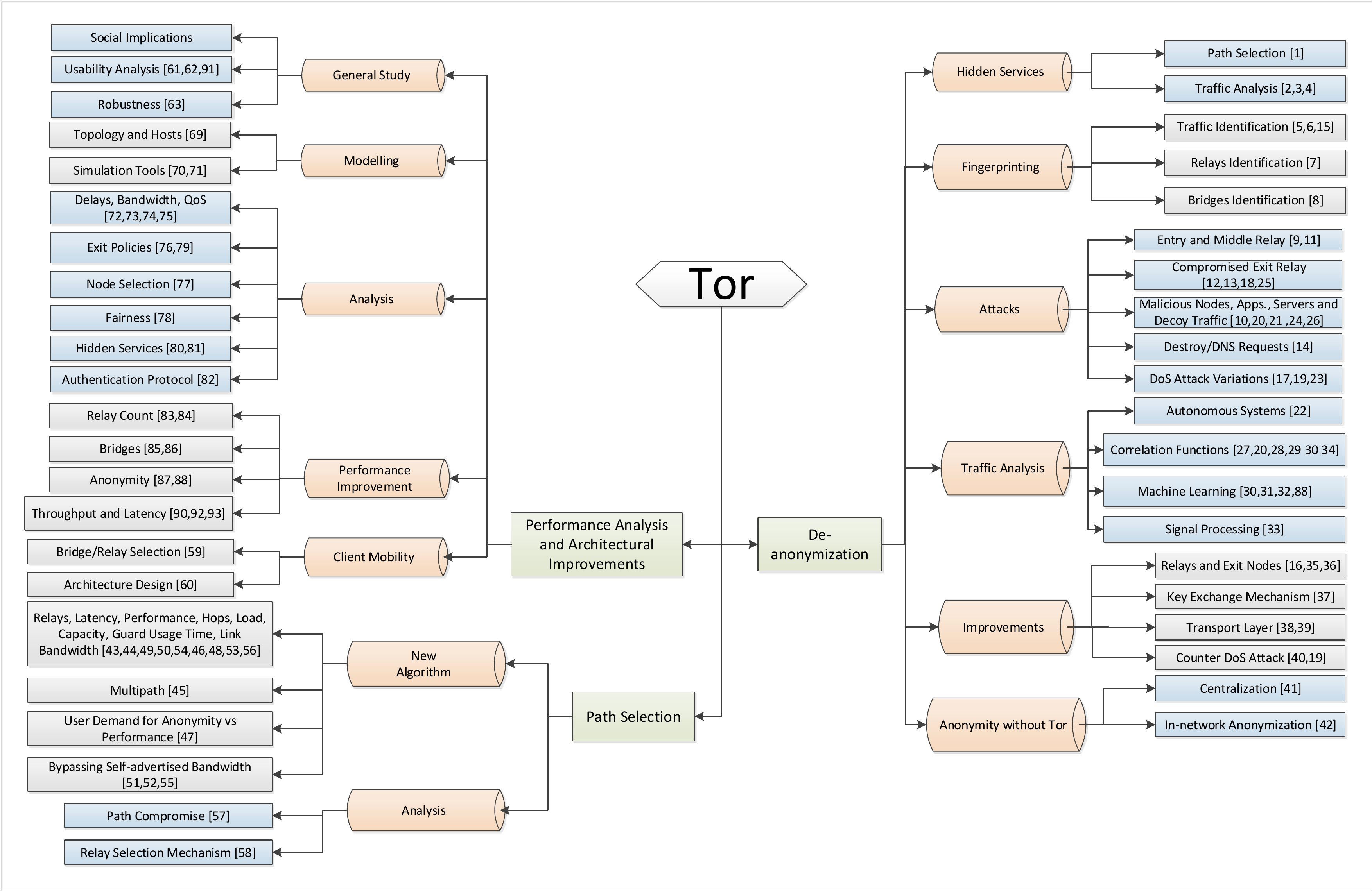}\\
  \caption{Taxonomy of Tor research. Tor literature can be broadly classified into three areas: deanonymization, path selection, and performance analysis and architectural improvements.}\label{fig: Taxonomy of Tor research}
\end{figure*}

In our review, we observed that research works on Tor could be broadly classified into three tracks/categories which include (1) deanonymization, (2) path selection, and (3) Analysis and performance improvement. Figure \ref{fig: Taxonomy of Tor research} shows the classification of our survey paper along with a list of all research works present in various subcategories.

In deanonymization track, research works were observed in six different categories covering (1) Hidden services which limit their scope to hidden servers identification, (2) Fingerprinting which are based on pinpointing Tor network, (3) Attacks which are focused over breaching Tor network, (4) Traffic analysis which analyze Tor traffic to pinpoint the weaknesses, (5) Studies studying improvements in Tor to avoid deanonymization, and  (6) Anonymity without Tor which suggest alternate methods to provide anonymity by pinpointing weaknesses in Tor.

In the path-selection track, all research works are either based upon (1) Development of new algorithms providing better efficiency and anonymity, and (2) Analysis of Tor's algorithm to study its strong and weak points in circuit establishment mechanism.

Lastly, analysis and performance improvement track focuses on four sub-areas which include (1) Generalized studies over Tor providing usability and social implications, (2) Modelling studies which focus on the development of model for analysis of Tor, (3) Analysis studies which cover QoS, relays, servers, etc., (4) Performance improvement studies provide modification in relays and architecture to provide better QoS, and (5) Development of efficient mechanisms for Tor clients with mobility.

\begin{figure}[b]
  \centering
  \includegraphics[width=0.9\columnwidth]{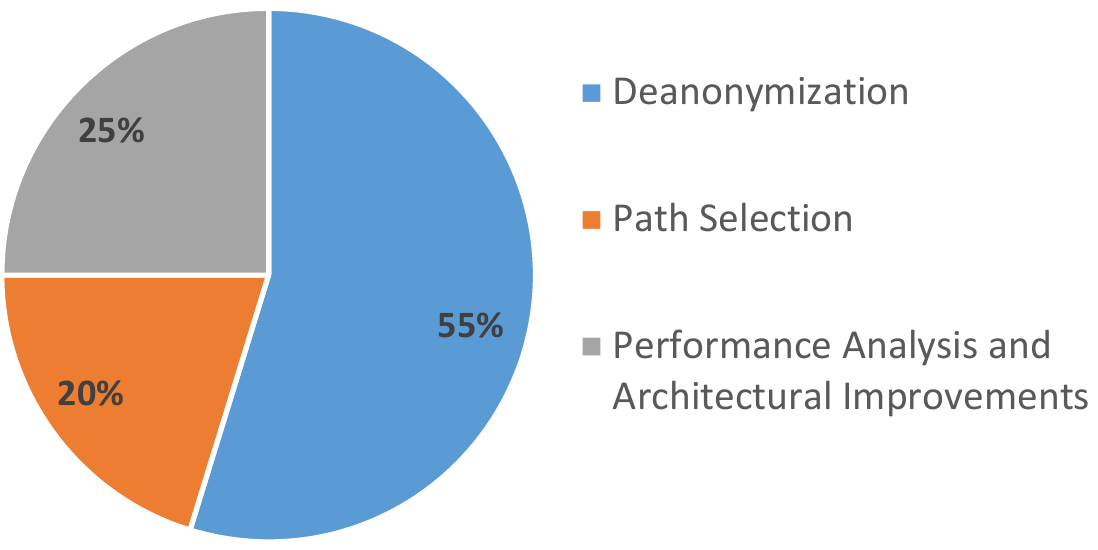}\\
  \caption{Classification of Tor research areas.}\label{fig: Classification of Tor research areas.}
\end{figure}

Figure \ref{fig: Classification of Tor research areas.} shows the classification of various research areas studied in the Tor network.
It is pertinent to mention that all numeric values used for all pie charts, figures and tables in this paper have been calculated by the authors.
Source of all numeric values is the `Reference' section at the end of the paper, which includes scholarly research articles. Moreover, references have been collected by the authors from Tor repository\footnote{https://www.freehaven.net/anonbib/} with a particular bias towards papers covering `Tor' network only. Span of collection varies from $2007$ to $2017$ in reputed international conferences and journals. We also included the important studies in this field before $2007$ which play helpful role in understanding of Tor network, such as \cite{reed1996proxies}\cite{reed1998anonymous}.
Many articles were also collected from `ACM digital library' and `IEEE Xplore digital library' with a particular focus towards anonymity and security in Tor.

%
%

\subsection{Tor Deanonymization}

\begin{figure}[b]
  \centering
  \includegraphics[width=0.8\columnwidth]{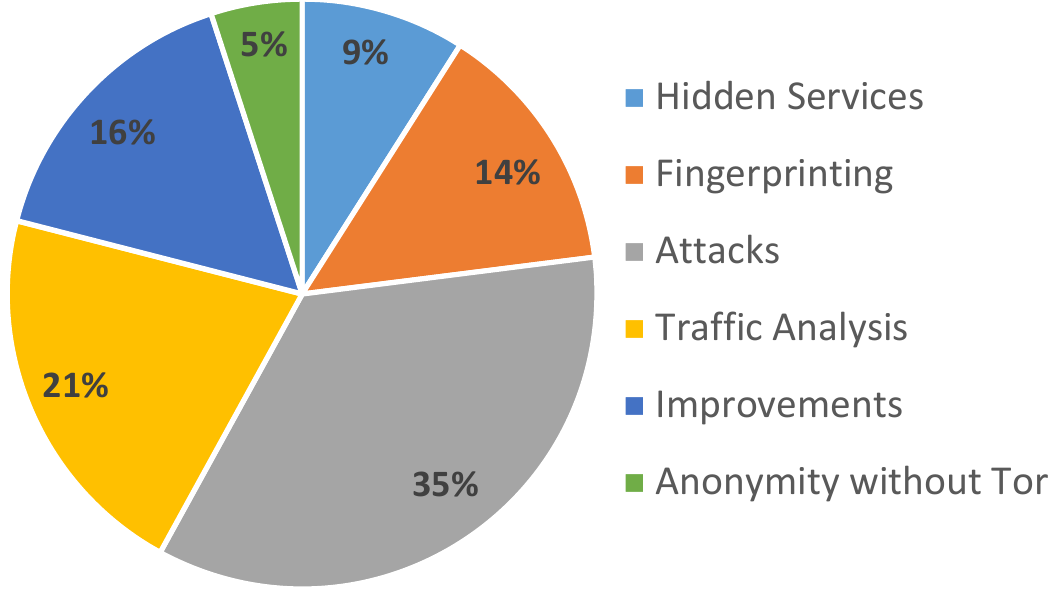}\\
  \caption{Classification of Tor's deanonymization approaches.}\label{fig: Classification of Tor's deanonymization approaches.}
\end{figure}

Breaching the Tor network is one of the most widely studied research problems.
In fact, the majority of the studies describe deanonymization attacks without identifying any counter-measures \cite{arp2014torben}.
%
%
Research works covering deanonymization can be subdivided into a number of sub-categories including (1) Hidden services identification, (2) Tor traffic identification, (3) Attacking Tor network, (3) Traffic fingerprinting, (4) Focusing over Tor improvements, and (6) Providing anonymity without Tor.
Classification of various research problems is shown in the pie chart in Figure \ref{fig: Classification of Tor's deanonymization approaches.}.

Table \ref{tab: Researches on Tor's Deanonymization.} presents a comparison of various research works in the Tor's deanonymization track.
Prominent patterns show that relay compromise and traffic interception are the most frequent factors in deanonymizing Tor.
This suggests that relays and traffic are more susceptible for exploitation than other factors.
Individual details of various research works in following subsections would explain this exploitation in much detail.

\begin{table*}
  \caption{Research works focused on Tor's deanonymization. Table entries symbolize attacks (Att.), counter-attacks (Cou. Att.), Analysis (Ana.), relays (Rel.), Autonomous systems (AS), browser (brows.), server (serv.), decoy traffic (Dec. Traf.), protocol messages (Prot. Mess.), traffic interception (Traf. Interc.), Flag cheating (Flag Cheat.).}
  \label{tab: Researches on Tor's Deanonymization.}
  \footnotesize
  \centering

  \begin{tabular}{|
  @{}>{\centering}p{3.5cm}@{}|
  @{}>{\centering\arraybackslash}p{0.4cm}@{\hspace{0.03in}}|
  @{}>{\centering\arraybackslash}p{0.5cm}@{\hspace{0.03in}}|
  @{}>{\centering\arraybackslash}p{0.5cm}@{\hspace{0.03in}}|
  @{}>{\centering\arraybackslash}p{0.4cm}@{\hspace{0.03in}}|
  @{}>{\centering\arraybackslash}p{0.4cm}@{\hspace{0.03in}}|
  @{}>{\centering\arraybackslash}p{0.7cm}@{\hspace{0.03in}}|
  @{}>{\centering\arraybackslash}p{0.5cm}@{\hspace{0.03in}}|
  @{}>{\centering\arraybackslash}p{0.5cm}@{\hspace{0.03in}}|
  @{}>{\centering\arraybackslash}p{0.6cm}@{\hspace{0.03in}}|
  @{}>{\centering\arraybackslash}p{0.7cm}@{\hspace{0.03in}}|
  @{}>{\centering\arraybackslash}p{0.7cm}@{\hspace{0.03in}}|
  @{}>{\centering\arraybackslash}p{8.5cm}@{\hspace{0.03in}}|
  }
  \hline
  Research  & \multicolumn{3}{|c|}{Focus}  & \multicolumn{8}{|c|}{Exploited Tor's weakness} & Idea\\
  \cline{2-12}
            & Att. & Cou. & Ana. & \multicolumn{4}{|c|}{Compromised}                        & Dec.    & Prot. & Traf.  & Flag & \\
  \cline{5-8}
            &        & Att. &          & Rel.       & AS          & brows.     & serv.     & traf.   & mess. & interc. & Cheat. &\\

  \hline \hline
  Overlier and Syverson \cite{overlier2006locating}  & \checkmark & & \checkmark & & & & & & & \checkmark & &Timing signature analysis attack, service location attack, predecessor attack and distance attack \\
  Elices \emph{et al.} \cite{elices2011fingerprinting}   & \checkmark & & & & & & & & & \checkmark & &Fingerprint analysis attack \\
  Zhang \emph{et al.} \cite{zhang2011application}    & \checkmark & & & & & & & & & \checkmark & & Application level time correlation attack \\
  Biryukov \emph{et al.} \cite{biryukov2013trawling} & \checkmark & \checkmark & & & & & & & & & \checkmark & Using corrupted relay node and cheating Tor's flag marking mechanism \\

  \hline
  \multicolumn{13}{|c|}{Tor's Traffic identification} \\
  \hline
  Bai \emph{et al.} \cite{bai2008traffic}         & \checkmark & & & & & & & & & \checkmark & &Packet examination, context checking and matching \\
  Barker \emph{et al.} \cite{barker2011using}      & \checkmark & & & & & & & & & \checkmark & &Using unsupervised macine learning tachniques over packet sizes \\
  AlSabah \emph{et al.} \cite{alsabah2012enhancing}     & \checkmark & & & & & & & & & \checkmark & &Application level time correlation attack \\
  Houmansadr \emph{et al.} \cite{houmansadr2013parrot}  & \checkmark & & & & & & &\checkmark & & & &Passive and active attacs to bypass traffic imitation technique \\
  Chakravarty \emph{et al.} \cite{chakravarty2008identifying} & \checkmark & \checkmark & & & & & & & & & & Observing bandwidth fluctuations through compromised node\\
  Winter and Lindskog \cite{winter2012great}  & \checkmark & & & & & & & & &\checkmark & & Using port tuples\\

  \hline
  \multicolumn{13}{|c|}{Tor attacks} \\
  \hline

  Sulaiman and Zhioua \cite{sulaiman2013attacking} & \checkmark & & & \checkmark & & & & & & & &Using unpopular ports over compromised relays \\
  Chan-Tin \emph{et al.} \cite{chan2013revisiting}    & \checkmark & & & \checkmark & & & \checkmark & & & & &Using malicious servers to observe traffic fluctuations over relays \\
  Pries \emph{et al.} \cite{pries2008new} & \checkmark & & & \checkmark & & & & \checkmark & & & &Passing duplicate cells through compromised entry relay \\
  Wang \emph{et al.} \cite{wang2009novel}       & \checkmark & & & \checkmark & & & & & & & &Returning malicious page through compromised exit relay \\
  Wagner \emph{et al.} \cite{wagner2012breaking}  & \checkmark & & & \checkmark & & \checkmark & & & & & &Compromised exit node to send images which is used by semi-supervised learning algorithm \\
  Benmeziane and Badache \cite{benmeziane2010tor}    & \checkmark & & & & & & & & \checkmark & & &Using destroy and DSN requests for man-in-the-middle attack \\
  Jansen \emph{et al.} \cite{jansen2014sniper}  & \checkmark & & & & & & & & \checkmark & & &Using valid protocol messages over relays to perform denial of service attack \\
  Abbot \emph{et al.} \cite{abbott2007browser} & \checkmark & & & \checkmark & & \checkmark & & & & & &Use compromised relay and user's browser for man-in-the-middle attack \\
  Evans \emph{et al.} \cite{evans2009practical}  & \checkmark & & & \checkmark & & & & & & & &Use exit relay to inject javascript for DoS attack \\
  Bauer \emph{et al.} \cite{bauer2007low}       & \checkmark & & & \checkmark & & & & & & & &Advertise low bandwidth to divert traffic towards malicous nodes \\
  Edman and Syverson \cite{edman2009awareness}      & \checkmark & & & & \checkmark & & & & & & &Using autonomous systems to breach Tor traffic \\
  Barbera \emph{et al.} \cite{barbera2013cellflood}     & \checkmark & & & & & & \checkmark & & & & &Perform DoS attack by placing large load on Tor routers \\
  Le Blond \emph{et al.} \cite{blond2011one}    & \checkmark & & & \checkmark & & & & & & & &Using peer-to-peer applications with compromised exit relays to deanonymize users \\
  Geddes \emph{et al.} \cite{geddes2013low}      & \checkmark & & & \checkmark & & & & & & & & Exploiting compromised exit node to advertise high stats to attract large traffic \\
  Chakravarty \emph{et al.} \cite{chakravarty2011detecting} &  & \checkmark & & & & & & &\checkmark & & & Using decoy traffic to detect traffic interception \\

  \hline
  \multicolumn{13}{|c|}{Tor traffic analysis attacks} \\
  \hline
  Johnson \emph{et al.} \cite{johnson2013users}     & \checkmark & & & & & & & & &\checkmark & & Correlation based attacks using a single compromised relay \\
  Murdoch and Danezis \cite{murdoch2005low}       & \checkmark & & & \checkmark & & & & & & \checkmark & & Use timing signature attack by passing large traffic from corrupted node to Tor relays \\
  Chakravarty \emph{et al.} \cite{chakravarty2014effectiveness} & \checkmark & & & & & & \checkmark & & & \checkmark & & Generate traffic between two servers and map relays through correlation \\
  Zhang \emph{et al.} \cite{zhang2008novel}       & & \checkmark  & & & & & & & & \checkmark & & Suggested priority queue algorithm to bypass correlation between load and latency \\
  Song \emph{et al.} \cite{song2013anonymize}        & \checkmark  & & & & & & & & &\checkmark & & Use time and stream size with k-means algorithm to deanonymize users \\
  Panchenko \emph{et al.} \cite{panchenko2011website}   & \checkmark  & & & & & & & & &\checkmark & & Use volume, time and direction for classification \\
  Wang and Goldberg \cite{wang2013improved}         & \checkmark  & & & & & & & &\checkmark & & & Use Tor cells for website fingerprinting \\
  Jin and Wang \cite{jin2009effectiveness}              & \checkmark  & & & & & & & & & \checkmark & & use wavelet based decomposition to estimate timing distortion \\
  Gilad and Herzberg \cite{gilad2012spying}        & \checkmark  & & & & & & & & \checkmark & & & Breach by off-path TCP connection or eavesdrop on clients \\
  \hline
  \multicolumn{13}{|c|}{Tor Improvements} \\
  \hline

  Gros \emph{et al.} \cite{gros2010protecting}     & & \checkmark  & \checkmark  & \checkmark & & & & & & & & Proposed Honeywall to rank node's reliability \\
  Winter and Lindskog \cite{winter2014spoiled}    & & \checkmark  & & \checkmark & & & & & & & & Proposed exit relay scanner to avoid misuse of exit node \\
  Xin and Neng \cite{xin2009design}           & & \checkmark  & & \checkmark & & & & & & & & Proposed a tuning mechanism to keep a track of reliable nodes \\
  Backes \emph{et al.} \cite{backes2012provably}   & & \checkmark  &\checkmark  & & & & & & \checkmark & & & Identified flaws in current key exchange mechanisms \\
  Marks \emph{et al.} \cite{marks2010unleashing}    & & \checkmark  & & & & & & & \checkmark & & & Suggested separate bi-directional TCP links to increase anonymity \\
  Nowlan \emph{et al.} \cite{nowlan2013reducing}   & & \checkmark  &\checkmark  & & & & & & \checkmark & & & Suggested use of uTCP and uTLS to avoid head of line blocking problem \\
  Danner \emph{et al.} \cite{danner2012effectiveness}   & & \checkmark  &\checkmark  & \checkmark & & & & & & & & Investigated DoS attack and proposed improvements to avoid it \\

  \hline
  \multicolumn{13}{|c|}{Anonymity without Tor} \\
  \hline
  Herzberg \emph{et al.} \cite{herzberg2011camouflaged} & &\checkmark  &\checkmark  & & & & & \checkmark & & & & Suggested camouflaged web server by mimicking GMAIL traffic \\
  Mendonca \emph{et al.} \cite{mendonca2012flexible} & &\checkmark  & & & \checkmark & & & & & & & Proposed concealed source identifier through network service provider \\
  \hline
  \end{tabular}
\end{table*}

\subsubsection{Tor Hidden Services}

An important feature of the Tor network is provisioning of Tor service through a hidden server.
A series of protocols used by hidden server and Tor users can make location of hidden server invisible to client \cite{murdoch2006hot}.
However, several studies listed below address the deanonymization of hidden servers.

\emph{Locating Hidden Server:} Overlier and Syverson \cite{overlier2006locating} presented new attack strategies to detect the location of hidden servers using only one Tor node. They proposed changes in route selection and relay selection to increase anonymity. The average duration of the attack varied from minutes to a few hours. The various attacks they considered included the timing signature analysis attack, service location attack, predecessor attack and distance attack. Their proposed solution included introducing middleman nodes to connect to rendezvous points, introducing dummy traffic, extending hidden server path to rendezvous point and using guard entry nodes.

\emph{Timing Signature Attack:} Elices \textit{et al.} \cite{elices2011fingerprinting} presented a fingerprint analysis attack for Tor's hidden services. They used timestamps from logs of machines hosting hidden services on the Tor network to generate detectable fingerprints. The authors studied delay properties of the Tor network and other users' log entries to make the fingerprint attack feasible.

\emph{Application Layer Correlation Attack:} Zhang \textit{et al.} \cite{zhang2011application} described an application level HTTP-based attack for Tor's hidden services. Time correlation was used to assess the resemblance between web accesses and the traffic generated in a compromised Tor router. This attack assumes that the compromised onion router can operate as an entry relay.

\emph{Detection, Measurement and Deanonymization of hidden services:} Biryukov \textit{et al.} \cite{biryukov2013trawling} analyzed weaknesses in hidden services which can be exploited by attackers to detect, measure and deanonymize hidden services running over the Tor network. Services of three different applications were analyzed, (1) Botnet for command and control, (2) Silk Road\footnote{Silk road was an online market place which provided anonymity to its customers by way of the Tor network. It was used in great part for the sale of drugs and illegal materials and was shutdown by the FBI. Defunct Website: \url{http://silkroad6ownowfk.onion}} and (3) DuckDuckGo\footnote{DuckDuckGo is a search engine that does not track its users and provides anonymity to users by giving same search results for any query to all users. Website: \url{https://duckduckgo.com}}. The study identified major flaws of Tor that included the inflation and cheating of bandwidth by a corrupted relay node, and cheating marking mechanism of flags in Tor network from attacker relay node.

\subsubsection{Tor Traffic Detection}

A number of studies focus their research on the identification of Tor traffic form other network traffic. These studies suggest that differentiation of traffic can ultimately be used to block Tor traffic, as done by China a number of times in the recent past \cite{winter2012great}. Various approaches for traffic identification are summarized as follows.

\emph{Tor Traffic Identification from Network Traffic:} Bai \textit{et al.} \cite{bai2008traffic} studied the traffic identification mechanisms of popular anonymity tools, i.e., Tor and Web-Mix. Authors used fingerprint identification (packet examination and packet context checking) followed by matching to identify the traffic. Key attributes used for traffic identification from other network traffic include specific strings, packet length and packet transmission frequency in the network.

\emph{Differentiate Tor Traffic from Encrypted Traffic:} Barker \textit{et al.} \cite{barker2011using} showed that traffic from the Tor network can be differentiated form encrypted traffic in the network. They captured regular HTTPS, Tor HTTPS and HTTP traffic routed through Tor and analyzed their packet sizes and developed an unsupervised machine learning (ML) classifier that operates only on packet size attribute with $97.54\%$ true positive (TP) and $1.06\%$ false positive (FP) rates.

\emph{Differentiating Tor Traffic:} AlSabah \textit{et al.} \cite{alsabah2012enhancing} developed an ML classifier to differentiate web traffic from bulk download traffic. AlSabah \textit{et al.} used the following four features to classify Tor traffic: (1) Circuit lifetime, (2) data transferred, (3) cell inter-arrival times, and (4) recently sent cells. They tested na\"ive Bayes, Bayesian networks and decision tree classifiers. Using the proposed classification method, they reported $75\%$ improvement in responsiveness and $86\%$ decrease in download rates.

\emph{Fingerprinting Tor traffic:} Houmansadr \textit{et al.} \cite{houmansadr2013parrot} aimed to differentiate the traffic of anonymous networks from other network traffic. They claimed that mimicking other traffic is an obsolete way for anonymity. They devised a number of passive and active attack strategies to breach anonymous networks. Their study suggested the use of partial imitation and use of new strategies by incorporating popular protocols like HTTPS email etc.

\emph{Tor Proxy Node Identification:} Chakravarty \textit{et al.} \cite{chakravarty2008identifying} described a novel attack that identifies all Tor relays participating in a given circuit. The attack modulates the bandwidth of an anonymous connection through a compromised server, router or an entry point and observes the resultant fluctuations in the Tor network using \textit{LinkWidth} \cite{chakravarty2008linkwidth}. LinkWidth sends a train of pulses comprising of alternate TCP-SYN and TCP-RST packets and capacity is computed at the receiver end by estimating packet dispersion. Authors reported a $59.46\%$ TP rate and $10\%$ true negatives rate for compromised Tor relays using the proposed strategy.

\emph{Identification of Tor Bridges:} Winter and Lindskog \cite{winter2012great} conducted an extensive investigation into the  blocking of Tor relays and bridges by China. Their investigation showed that Tor bridges were blocked by port tuples, rather than IP addresses and that bridges were blocked only when they were active. Their investigation also showed that adversaries did not conduct traffic fingerprinting for domestic traffic and that packet fragmentation could be used to circumvent China's firewall.

\emph{Fingerprinting Keywords in Search Queries:} Oh \emph{et al.} \cite{oh2017fingerprinting} investigated the viability of keyword fingerprinting attacks in the Tor network. Study showed that effective feature selection can help any passive adversary in figuring out the identity of the user. Time and volume of traffic play the most crucial role in determining the identity of the user. Among other features keyword sets, incremental search and high security search are other features used for classification. Experimental results demonstrated recall, precision and accuracy of 80\%, 91\% and 48\%, respectively for one of 300 targeted keywords of Google.

\subsubsection{Tor Attacks}

Attacking the Tor network is an interesting research dimension which ultimately aims to block access to it. Several attempts by China and other countries have failed in the recent past because Tor is being improved continuously \cite{winter2012great}. In this subsection, we summarize various studies covering Tor attacks.

\emph{Unpopular Ports Attack:} Sulaiman and Zhioua \cite{sulaiman2013attacking} described an attack they developed which can compromise circuits in the Tor network. Their attack takes advantage of unpopular ports in the Tor network. Sulaiman and Zhioua added a small number of compromised entry /exit relays to the Tor network ($\thicksim30$ relays) which permit the use of unpopular ports. By doing so, $50\%$ of developed circuits can be compromised, which significantly decreases the anonymity of the Tor network.

\emph{Circuit Clogging Attack:} Chan-Tin \textit{et al.} \cite{chan2013revisiting} proposed an attack that can identify the Tor routers used in any circuit. For the proposed attack a client connects to a malicious server which sends data to the client in large bursts and in small amounts. During large bursts, Tor routers take long times to process the extra amount of data. Authors showed that continuous monitoring of all Tor relays can identify the Tor relays used in the particular circuit. A mechanism to detect the behavior of malicious routers by the client was also evaluated, which measured network latency of the client.

\emph{Replay Attack:} Pries \textit{et al.} \cite{pries2008new} suggested a replay attack to detect the exit routers in the Tor network. The replay attack assumes that the entry onion router is compromised. The replay attack duplicates packets coming from a sender. Tor uses counter-mode Advanced Encryption Standard (AES-CTR)\cite{daemen2013design}\footnote{Advanced Encryption Standard (AES) is an encryption standard which is based upon substitution-permutation technique. It has three members Rijndael family each with block size of 128 bits and key lengths of 128, 192 and 256 bits.\cite{standard2001announcing}} for encryption and decryption, any duplicate cells will give a cell recognition error at the exit routers. This behavior leaks exit router information to the entry router by simple correlation.

\emph{Flow Multiplication Attack:} Wang \textit{et al.} \cite{wang2009novel} designed a flow multiplication attack  similar to a man-in-the-middle attack. The attack assumes that the exit router is compromised. Whenever a client sends a request to target server, the exit router returns a malicious page which triggers certain fetch requests in the client browser over the same circuit. An accomplice at the entry router can see the requests, and together with knowledge of the exit relay, identify the complete Tor circuit.

\emph{Attack Using Game Theory and Data Mining:} Wagner \textit{et al.} \cite{wagner2012breaking} proposed an attack which exploits the exit malicious exit node to cluster observed traffic flows using an active tag injection scheme. The proposed method has two steps, (1) image tags are injected into HTML replies from the exit node to the user, and (2) a semi-supervised learning algorithm based upon deep data mining is used to reconstruct the entire browsing session of the user. The authors model the Tor network in form of a game theoretical concept where all Tor users and rogue nodes play a game for identification of malicious node. Once a rogue node has been identified, it's game is over because no other user uses it due to presence of special flag in it. Authors main aim is to work over over the equilibrium between rogue nodes and Tor users.

\emph{Attack using Destroy and DNS Requests:} Benmeziane and Badache \cite{benmeziane2010tor} investigated possible breaches of Tor targeting its network requests. They exploited \textit{destroy requests} (Tor's circuit destruction requests) and DNS requests to break anonymity. Destroy requests are not encrypted, which poses a serious threat to Tor. Moreover, a local eavesdropper can use the man-in-the-middle attack strategy against DNS requests, which are unprotected.

\emph{The Sniper Attack:} Jansen \textit{et al.} \cite{jansen2014sniper} presented the Sniper attack, a low-resource denial-of-service (DoS) attack against the Tor network which can disable arbitrary relays. The adversary builds a Tor circuit through the target relay and starts obtaining a large file by continuously sending the SENDME cells (protocol messages for continuously receiving the file), which increases the congestion window size. By repeating over multiple circuits, memory of host of target relay would exhaust which can disrupt the functioning of Tor relay. Experiments showed that an adversary can consume $2,187$ KB/s memory of a victim relay at the cost of very little bandwidth and decrease Tor network bandwidth by as much as $35\%$.

\emph{Browser-based Attacks:} Abbot \textit{et al.} \cite{abbott2007browser} proposed a novel attack that tricks a user's web browser into sending a distinctive signal over the Tor network (by installing a Java or HTML script). An attacker that controls an exit relay can use it in a man-in-the-middle attack to mirror and forward duplicated traffic to a malicious server. By analyzing the data, the malicious server can deanonymize the Tor user. However, this study makes two significant assumptions: the ability to control the exit relay and the ability to configure / compromise a targeted user's web browser.

\emph{Congestion Attack using Long Paths:} Evans \textit{et al.} \cite{evans2009practical} proposed an extension to the congestion attack proposed by Murdoch and Danezis \cite{murdoch2005low} owing to the enormity of the current Tor network. Evans \textit{et al.} proposed the combination of Javascript injection and DoS attack. A Tor exit relay is used by the attacker to inject Javascript code into a user's browser, which makes the browser send a response every second. They suggested modifications such as disabling JavaScript, thwarting DoS attack by disabling ability to control latency of routers. In the modified design, routers keep a track of all paths with flags and disable any request for latency by using flags.

\emph{Exploiting Routing Algorithm:} Bauer \textit{et al.} \cite{bauer2007low} exploited Tor's routing algorithm to steer a disproportionate number of users towards selecting their entry and exit relays from a set of malicious Tor routers. Bauer \textit{at al.} suggested that low-latency constraints force Tor's routing algorithm to prefer nodes advertising high bandwidths. Instead of performing complex traffic analysis techniques, the authors suggested to collect detailed flow logs from malicious nodes (both entry and exit nodes) and use the information of node selection to deanonymize flows.

\emph{Analyzing Autonomous Systems for Tor Path Selection:} Edman and Syverson \cite{edman2009awareness} analyzed the effect of autonomous systems (AS) for path selection in Tor network. They studied the selection of AS residing in different countries and found it quite effective. Traffic analysis of the Tor network showed that majority of traffic passes through a few ASs because all established paths focus over latency and anonymity which occurs better in some ASs. Analysis shows that increase in relays has not increased the diversity to a large extent.

\emph{DoS Attack using Cell Flooding}: Barbera \textit{et al.} \cite{barbera2013cellflood} presented a novel attack which generates a few circuits requiring large computing and networking resources. Their study showed that this attack requires only $0.2\%$ resources for old routers and $1-16\%$ router resources for new attacks, which makes it an inexpensive attack to execute. Barbera \textit{et al.} proposed a mitigation scheme by placing an upper cap on the utilization of resources at routers.

\emph{Exploiting peer-to-peer application:} Le Blond \textit{et al.} \cite{blond2011one} suggested that peer-to-peer applications can be exploited to trace IP addresses of users running Tor. Moreover, scan of malicious Tor exit relays should be used to correlate various user streams for deanonymization. Experiments showed that their `bad apple' attack was able to identify $193\%$ more streams, including $27\%$ HTTP streams, and reveal IP addresses of $10,000$ Tor users. This constituted $9\%$ of all the flows passing through the exit relays under their control.

\emph{Induced Throttling Attacks:} Geddes \textit{et al.} \cite{geddes2013low} proposed a new attack which breaches the Tor network by exploiting its selection bias in favor of high capacity relay nodes. Authors showed that induced throttling at the corrupt exit node by exploiting congestion or traffic shaping algorithms can induce similar traffic patterns at other relays associated with the corrupted exit relay.

\emph{Using Decoy Traffic:} Chakravarty \textit{et al.} \cite{chakravarty2011detecting} used decoy traffic on anonymous networks to detect traffic interception. The proposed strategy is based on the idea of injecting traffic containing bait credentials for decoy services requiring user authentication. Chakravarty \textit{et al.} set up decoy IMAP and SMTP servers and identified ten instances of traffic interception over ten months.

\subsubsection{Tor Traffic Analysis Attacks}

A few studies have focused on the analysis of Tor traffic for breaching this network. Analysis shows that traffic analysis can provide an efficient mechanism for deanonymization. A few of these studies are summarized as follows.

\emph{Traffic Correlation Attacks:} Johnson \textit{et al.} \cite{johnson2013users} conducted a thorough analysis of the Tor network with a deep focus on the development of a threat model. They built the Tor path simulator (TorPS) to assess Tor's vulnerability to correlation based attacks. Their study suggested that a single Tor relay adversary can deanonymize $80\%$ of users within six months. This research showed that set of relays is dependent upon the user's application which reduces security of the Tor network.

\emph{Using Timing Signature:} Murdoch and Danezis \cite{murdoch2005low} presented a simple mechanism to evaluate the Tor nodes being used in a circuit. In the proposed scheme, a malicious node sends probe data to the Tor relays. All Tor relays used in the circuit will experience a delay and client-server communication will be modulated. Hence, correlation between delay and modulation gives insight about the relays being used in a circuit.

\emph{Traffic Analysis Attack:} Chakravarty \textit{et al.} \cite{chakravarty2014effectiveness} used NetFlow data to analyze the effectiveness of traffic analysis attacks against Tor network. Their proposed attack creates variations in traffic at the server end and observes the effects at a colluding server at the other end. They reported $81.4\%$ accuracy in real-world experiments with $6.4\%$ FP rates.

\emph{Queue Scheduling and Resource Allocation:} Zhang \textit{et al.} \cite{zhang2008novel} proposed a priority queue scheduling mechanism to reduce the correlation between high load and high latency which would ultimately increase the level of anonymity. However, increase in anonymity comes at the cost of latency which degrades quality of service at the user end. Extensive experiments using the proposed mechanism showed an increase in anonymity due to decrease in correlation between load and latency.

\emph{Correlation Using K-means Algorithm:} Song \textit{et al.} \cite{song2013anonymize} applied machine learning techniques to deanonymize Tor flows at the first hop and last hop in the network. They used the time / stream size tuple of attributes together with the \emph{k-means} algorithm to deanonymize by matching first hop traffic with last hop traffic. Their results showed that as little as $8$ packets are enough to deanonymize a Tor stream with greater than $99\%$ accuracy.

\emph{Website Fingerprinting Using Machine Learning:} Panchenko \emph{et al.} \cite{panchenko2011website} suggested the use of machine learning approaches requiring feature selection and classification for website fingerprinting. Authors used Support Vector Machine (SVM) classifier with various features including packet sizes (except 52 size packets because of excess use in acknowledgements), packet size markers to express direction of flow, HTML markers, total transmitted bytes, number markers, occurring packet sizes, percentage incoming packets and number of packets. Extensive research showed that volume, time and direction of the traffic were the most promising features and classification of close-world and open-world dataset gave $55\%$ detection rate. However, camouflaging the traffic decreased the detection rate to $3\%$.

\emph{Website Fingerprinting Using New Metrics:} Wang and Goldberg \cite{wang2013improved} proposed the use of Tor cells as a unit of data transfer rather TCP/IP packets for website fingerprinting. Authors collected data using realistic assumptions on adversaries from client to entry guard node. The study suggested the removal of SENDME cells as they do not play any significant role in improving performance. Proposed metrics use the observation that dynamic content is present in only incoming packets and it is present at the end of the packets. Upto $95\%$ recall rate and $0.2\%$ FP rate is observed using SVM classifier.

\emph{Wavelet Decomposition Attack:} Jin and Wang \cite{jin2009effectiveness} suggested a wavelet based decomposition mechanism to estimate the distortion in timing at the receiver end of Tor network. Authors showed that wavelet based multi-resolution analysis (MRA) captures the variability of the timing distortion at all levels, with better granularity than traditional estimation of timing distortion. Deanonymization rate of 96\% was obtained for Tor at a packet rate of 4 pkts/sec in 3 minutes without changing established paths (circuits) of Tor. Analysis showed that Tor circuit rotation could decrease the accuracy of deanonymization to 72\% after 5 Tor circuit rotations in $3$ minutes.

\emph{Exploiting Side-channels to Identify Clients:} Gilad and Herzberg \cite{gilad2012spying} exploited three kinds of side-channels including (1) globally incrementing IP identifiers, (2) packet processing delays, and (3) bogus-congestion events. Sequential port allocation is also used to identify the clients. Two scenarios for breaching have been presented including (1) fully off-path attack to detect TCP connections, and (2) detecting Tor connections by eavesdropping on clients.

\subsubsection{Tor Improvements}

In this section, we present some miscellaneous studies about improvement in Tor network by focusing on Tor relays, path selection mechanisms, transport layer protocols and application layer improvements.

\emph{Misusing Tor Exit Node:} Gros \emph{et al.} \cite{gros2010protecting} studied the abuse and misuse of Tor exit nodes to compromise the anonymity of the Tor network. Authors proposed a mechanism, called \emph{Honeywall}, to avoid misuse of any Tor exit node. According to Honeywall, whenever any exit node detects a malicious behavior, it lowers the reputation of the immediate predecessor router and also sends an alert to it. Similarly, the intermediate router lowers the reputation of its predecessor router. Through this strategy, all ``bad'' nodes eventually end up with lower reputations and all ``good'' nodes have higher reputation.

\emph{Exposing Exit Relays:} Winter and Lindskog \cite{winter2014spoiled} detected malicious exit Tor relays and profiled their behavior. An exit relay scanner was built to identify all outgoing Tor traffic and identify the malicious nodes and avoid man-in-the-middle attacks. Patches were built for the Tor browser bundle to collect certificates through multiple paths to check authenticity of the destination server.

\emph{Tuning mechanism for Tor:} Xin and Neng \cite{xin2009design} showed that Tor lacks the evaluation system for the node store. Authors presented and theoretically analyzed a tuning mechanism for Tor. The proposed tuning system included the establishment of an evaluation system and optimization of Tor node store and output mode. Through the evaluation system, all nodes are ranked based on their anonymity, uptime, bandwidth and latency. In the optimization stage authors suggested to use a fixed number of circuits such that traffic load has least effect on latency.

\emph{Increasing Security of Tor Network:} Backes \emph{et al.} \cite{backes2012provably} conducted research on the security of Tor network for anonymous browsing and presented a novel security protocol. Authors elaborated the concept of security in anonymity softwares. Their study showed that current key exchange algorithms are inefficient and a number of security enhancements were suggested including cryptographic requirements for secure browsing.

\emph{Transport Layer Improvements:} Marks \emph{et al.} \cite{marks2010unleashing} studied TCP based deficiencies in the Tor network. By studying the transmission mechanism of Tor, authors proposed to split bidirectional links into two separate TCP links. Experiments with separate TCP links showed 100\% increase in throughput with a decrease in throughput variance from 43000 KB/s to 10000 KB/s \cite{marks_unleashingtor_thesis}.

\emph{Switching to uTCP and uTLS:} Nowlan \emph{et al.} \cite{nowlan2013reducing} probed into the cross-stream head of line blocking problem of TCP in the Tor network. Their study suggested the use of unordered TCP (uTCP) and unordered TCL (uTLS) for reducing inter-dependence in inter-leaving streams, due to the requirement of low latency in the Tor network.

\emph{Feasibility of DOS attack over Tor:} Danner \emph{et al.} \cite{danner2012effectiveness} conducted a deep investigation on the feasibility of Denial-of-Service (DoS) attack (proposed in a previous study by Borisov \emph{et al.} \cite{borisov2007denial}) over Tor network. Authors showed through simulations and analytical evaluations that corrupted relay nodes can be used to exploit Tor network and perform DoS attack. Authors suggested the use of reliable guard nodes (entry and exit) which can decrease the probability of selection of a corrupt Tor relay.

\subsubsection{Anonymity without Tor}

To present a glimpse of studies providing anonymity without Tor, we present a few studies focusing on packet encapsulation and central server based anonymity mechanism.

\emph{Anonymity Using a Central Server:} Herzberg \emph{et al.} \cite{herzberg2011camouflaged} proposed a camouflaged browsing design using a camouflaged server. The basic idea is to communicate with the camouflaged server using a manner similar to popular web services. Encrypted communication, URLs of GMAIL with packet frequency and sizing similar to GMAIL can easily pass unnoticed through any adversary. Although this design provides better anonymity, it suffers from a single point of failure.

\emph{In-Network IP Anonymization Service:} Mendonca \emph{et al.} \cite{mendonca2012flexible} presented a novel idea of user anonymity by working with a network service provider. Proposed service \emph{AnonyFlow} used an in-network IP anonymization service. The fundamental idea was to conceal the source identifier from the other side of the network. An OpenFlow based implementation was used for performance evaluation. However, anonymity could be breached by compromising the network service provider.

\subsection{Tor Path Selection}

\begin{figure}[b]
  \centering
  \includegraphics[width=0.8\columnwidth]{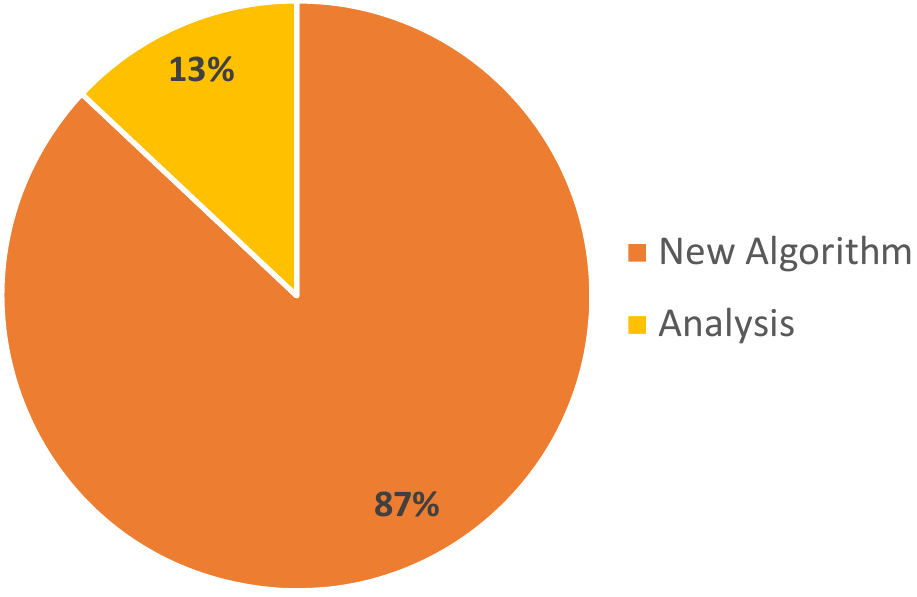}\\
  \caption{Focus of published research on Tor's Path Selection approaches.}\label{fig: Classification for Tor's Path Selection approaches.}
\end{figure}

Tor selects three relays based upon its path selection algorithm which incorporates anonymity and reliability characteristics of relays and users \cite{dingledine2004tor}. By compromising the path selection mechanism, the complete anonymity mechanism of Tor can be breached. In this subsection, we present studies covering (1) new algorithms for path selection, and (2) analysis of path selection algorithms. An overview showing the classification of major studies is shown in Figure \ref{fig: Classification for Tor's Path Selection approaches.}.

\begin{table*}[t]
  \caption{Research works on Tor's path selection. Table entries symbolize New algorithms (New Algo), Analysis (Anal.), Autonomous Systems (AS), Relay Locations (Relay loc.), Hops, Performance-Latency-Bandwidth (Perf., Lat, BW), Multi-path, Load, Relay Capacity (Rel. Cap.) and Anonymity (Anon).}
  \label{tab: Researches on Tor's Path Selection.}
  \centering
  \small
    \begin{tabular}{|
  @{}>{\centering}p{3cm}@{}|
  @{}>{\centering\arraybackslash}p{0.55cm}@{\hspace{0.035in}}|
  @{}>{\centering\arraybackslash}p{0.6cm}@{\hspace{0.035in}}|
  @{}>{\centering\arraybackslash}p{0.4cm}@{\hspace{0.035in}}|
  @{}>{\centering\arraybackslash}p{0.65cm}@{\hspace{0.035in}}|
  @{}>{\centering\arraybackslash}p{0.6cm}@{\hspace{0.035in}}|
  @{}>{\centering\arraybackslash}p{0.6cm}@{\hspace{0.035in}}|
  @{}>{\centering\arraybackslash}p{0.7cm}@{\hspace{0.035in}}|
  @{}>{\centering\arraybackslash}p{0.6cm}@{\hspace{0.035in}}|
  @{}>{\centering\arraybackslash}p{0.55cm}@{\hspace{0.035in}}|
  @{}>{\centering\arraybackslash}p{0.65cm}@{\hspace{0.035in}}|
  @{}>{\centering\arraybackslash}p{8.0cm}@{\hspace{0.035in}}|
  }
  \hline
            & \multicolumn{2}{|c|}{Focus} & \multicolumn{8}{|c|}{Path Selection Parameters} & Idea\\
  \cline{2-11}
  Research  & New  & Anal. & AS    & Relay & Hops & Perf.   & Multi- & Load & Rel. & Anon.      & \\
            & Algo &       &       & loc.  &      & Lat, BW & path   &      & Cap. &                &      \\
  \hline
  \multicolumn{12}{|c|}{New Path Selection Algorithms} \\
  \hline
  Akhoondi \emph{et al.}  \cite{akhoondi2012lastor}        & \checkmark & & \checkmark &\checkmark & & & & & & & Included relay locations and autonomous system reliability\\
  Chen \emph{et al.} \cite{chen2010toward}             & \checkmark & & &\checkmark &\checkmark &\checkmark & & & & \checkmark & Included hops and geographic distance in path selection\\
  Karaoglu \emph{et al.} \cite{karaoglu2012multi}         &\checkmark & & & & & \checkmark & \checkmark &  & & & Studied multipath design\\
  Panchenko \emph{et al.} \cite{panchenko2012improving}        &\checkmark & & & & & \checkmark & & \checkmark & \checkmark & & Studied Load and Capacity at nodes\\
  Li \emph{et al.} \cite{li2012tmt}              &\checkmark & & & & & \checkmark & & & & \checkmark & Proposed tunable mechanism varying between anonymity and performance\\
  Panchenko \emph{et al.} \cite{panchenko2008performance}        &\checkmark &\checkmark & & \checkmark & & \checkmark & & &\checkmark & & Studied latency, link capacity and load at nodes\\
  Liu and Wang \cite{liu2009random}                   &\checkmark & & & & &\checkmark & &\checkmark & &\checkmark & Proposed random walk based circuit building protocol\\
  Liu and Wang \cite{liu2009improved}                   &\checkmark & & & & &\checkmark & & & &\checkmark & Proposed new relay selection mechanism with backup circuit algorithm\\
  Snader and Borisov \cite{snader2011improving}             &\checkmark & & & & &\checkmark &\checkmark & & &\checkmark & Studied malicious nodes, proposed balance between anonymity and performance\\
  Li \emph{et al.} \cite{li2012relay}               &\checkmark & & & & &\checkmark & & & &\checkmark & Proposed relay recommendation system\\
  Tang and Goldberg \cite{tang2010improved}              &\checkmark & & & & &\checkmark & & & & & Suggested the use of bursty circuits instead of busy paths\\
  Wang \emph{et al.} \cite{wang2012congestion}            &\checkmark & & & & &\checkmark & & & & & Included latency as a measure of congestion in path selection\\
  Snader and Borisov \cite{snader2008tune}   &\checkmark & & & & &\checkmark & & & &\checkmark & Suggested opportunistic bandwidth measurement with priority based traffic handling\\
  Elahi \emph{et al.} \cite{elahi2012changing} &\checkmark &\checkmark & &\checkmark & & & & & & & Discouraged short term entry guard churn and time-based entry guard rotation\\

  \hline
  \multicolumn{12}{|c|}{Analysis of Path Selection} \\
  \hline

  Bauer \emph{et al.} \cite{bauer2009predicting}            & &\checkmark & &\checkmark & & & & &\checkmark & & Suggested random or Snader-Borisov approach for router selection\\
  Wacek \emph{et al.} \cite{wacek2013empirical}            & &\checkmark & & & &\checkmark & & & &\checkmark & Suggested bandwidth weighted relay selection and avoidance of congested circuits\\

  \hline
  \end{tabular}
\end{table*}

Table \ref{tab: Researches on Tor's Path Selection.} presents a comparison of various research works in Tor's path selection track. Comparison shows that performance and anonymity were the most frequently studied parameters for path selection. However, majority studies neglected autonomous systems, relay locations, hop counts, multi-path mechanisms, load and relay capacity. Moreover, majority research works focus on the development of new algorithms while few studies analysed the current path selection algorithms.

\subsubsection{New Path Selection Algorithms}

\emph{LASTor - Low Latency with Better Anonymity Algorithm:} Akhoondi \emph{et al.} \cite{akhoondi2012lastor} proposed a new path selection algorithm namely \emph{LASTor}. LASTor incorporates the locations of relays before choosing paths and does not always select the shortest path as it reduces the entropy of path selection. Moreover, LASTor avoids paths passing through ASs which can compromise anonymity of the system by traffic correlation.

\emph{Optimizing Hops, Performance flags and Geographic Distance:} Chen and Pasquale \emph{et al.} \cite{chen2010toward} studied the path selection mechanism by varying the number of hops, performance ratings and changing the geographic distance between routers. Trade-offs between anonymity and other parameters (latency etc.) were extensively evaluated. The authors concluded that reduction in hops and geographic distance can increase throughput and decrease anonymity.

\emph{Using MultiPath Routing} Karaoglu \emph{et al.} \cite{karaoglu2012multi} evaluated the multipath design for Tor network to avoid congestion and overcome the limitations in Tor's circuit construction. Evaluations revealed a four-fold increase in throughput with better load balancing and traffic mixing. However, high buffer costs at the Tor proxies were the major limitations of multipath design.

\emph{Path Selection Using \emph{Load} and \emph{Capacity} of Nodes:} Panchenko \emph{et al.} \cite{panchenko2012improving} studied the delays in the Tor network and provided new measures in path selection to improve user experience. Two factors used for path selection design are ``load'' at the nodes and maximum ``capacity'' at the nodes. Authors showed that these factors can increase the performance by $70\%$. Their study concludes that nodes, not edges, are the deciding factors for performance.

\emph{Tunable Mechanism of Tor:} Li \emph{et al.} \cite{li2012tmt} emphasized the development of a tunable mechanism for Tor users depending on \emph{anonymity} and \emph{performance} required by users. Authors used ``path length'' as a metric to tune user requirements based upon anonymity and performance followed by client side modifications of Tor protocol. Results showed that browsing time deteriorates quickly from $14.4$ to $140.1$secs with a $37.3\%$ increase in failure rate by increasing the path length from $2$ to $6$. The proposed mechanism requires only client side modification.

\emph{Using Latency and Link Capacity:} Panchenko \emph{et al.} \cite{panchenko2008performance} evaluated the impact of different factors on the performance of the Tor network. Factors considered included overloaded nodes and links and geographical diversity of nodes. Authors presented a novel path selection algorithm based on latency experienced by the nodes and link capacity. Metrics used for evaluations included circuit setup duration, round trip time (RTT), stream throughput and influence of penetration.

\emph{Random Walk Based Circuit Building Protocol:} Liu and Wang \cite{liu2009random} presented a random walk based circuit building protocol (RWCBP) which is a two-step method: circuit construction, followed by application message transmission. Network latency, computational latency and transmission loads were used to analyze the performance of the proposed protocol. Using indexes of performance and anonymity, resilience of the proposed protocol was analyzed.

\emph{New Circuit Building Protocol:} Liu and Wang \cite{liu2009improved} studied the current protocol of the Tor network and proposed a novel circuit building design with two phases: selection of user selectable relay nodes and circuit construction. Authors presented enhancements in the selection of relay nodes, fast circuit construction and backup circuit algorithm. Better performance and user experience are obtained with the new protocol while achieving the same level of anonymity.

\emph{Tunable Path Selection for Better Security and Performance:} Snader and Borisov \cite{snader2011improving} addressed the issue of the selection of malicious nodes in the path selection due to self-advertised bandwidth. Authors proposed an algorithm which is based upon the anonymity and performance in the Tor network. Significant performance gains were observed using the proposed strategy with single and multipath route selection.

\emph{Relay Recommendation System:} Li \emph{et al.} \cite{li2012relay} proposed a relay recommendation system to provide reliable information about all relays for building circuits (paths). Its main goals include the mitigation of low-resource attacks, better performance and tradeoffs between anonymity and performance. Authors proposed path selection algorithms for increased anonymity. Significant performance gains with increase in anonymity were observed in the simulations of the proposed scheme.

\emph{Preferring Bursty Circuits over Busy Circuits:} Tang and Goldberg \cite{tang2010improved} proposed a new algorithm which suggests the use of bursty circuits instead of busy circuits. Authors suggest that bursty circuits (such as web browsing) can provide less latency than the busy circuits (used for bulk data transfer). Proposed circuit selection algorithm uses exponentially weighted moving average (EWMA) of cells sent on any path and uses the path with lowest EWMA (because new and bursty paths have high EWMA).

\emph{Incorporating Congestion in Path Selection:} Wang \emph{et al.} \cite{wang2012congestion} proposed a novel path selection algorithm which incorporates the latency of nodes as a measure for congestion. The proposed algorithm favors nodes which provide lower latency. Study suggests that node latency is greater than the link latency in majority of the cases. Authors conclude that the proposed algorithm can reduce latency by upto $40\%$.

\emph{Opportunistic Bandwidth Measurement Algorithm:} Snader and Borisov \cite{snader2008tune} addressed Tor's shortcoming of favoring high bandwidth nodes based on advertised bandwidth. Their study showed that an opportunistic measurement of bandwidth for all routers by other connected routers can reduce the vulnerability risk by any adversary in Tor. Moreover, priority based traffic handling, i.e., high performance or high anonymity can reduce the risk of partitioning attacks.

\emph{Analyzing and Improving Entry Guard Selection:} Elahi \emph{et al.} \cite{elahi2012changing} conducted an in depth investigation on the selection of entry guards in Tor network. The study showed that short-term entry guard churn and explicit time-based entry guard rotation result in an increased usage of entry guards in clients, which results in a greater number of profiling attacks.

\emph{Trust-Aware Path Selection Algorithm:} Johnson \emph{et al.} \cite{johnson2017avoiding} proposed a path selection algorithm which uses the probability based distribution to keep itself aware of the location of adversaries in the Tor network. In developing trust based model, authors take the relays uptime as the most trustworthy factor in determining the selection of the path. Bypassing the paths containing adversaries can mitigate the traffic analysis attacks conducted by the adversaries.

\subsubsection{Analysis of Path Selection}

\emph{Predicting Path Compromise:} Bauer \emph{et al.} \cite{bauer2009predicting} showed that the current mechanism of Tor is vulnerable to path compromise because Tor selects paths based on bandwidth capabilities of routers. Study shows that the application level protocol is a significant factor to predict path compromise. Research suggests that router selection should be random or through Snader-Borisov approach to avoid any bias in router selection. Study showed that most robust applications for path compromise are HTTP and HTTPs applications while the most vulnerable are peer-to-peer applications.

\emph{Empirical Evaluation of Relay Selection:} Wacek \emph{et al.} \cite{wacek2013empirical} evaluated the relay selection mechanism of Tor to estimate latency. Performance and anonymity were analyzed for a number of relay selection techniques under varying load conditions. The authors suggest that a combination of bandwidth-weighted relay selection and avoidance of congested circuits can provide better throughput and less latency.

\subsection{Tor Analysis and Performance Improvements}

\begin{figure}[b]
  \centering
  \includegraphics[width=0.8\columnwidth]{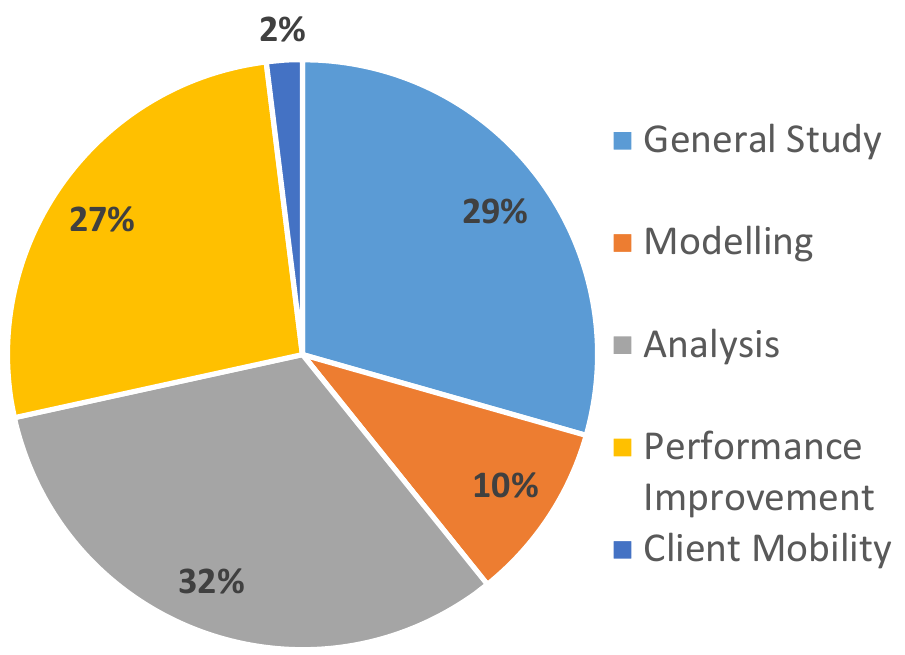}\\
  \caption{Focus of various research works on the analysis of Tor.}\label{fig: Classification of approaches on Tor's Analysis.}
\end{figure}

In this section, we cover studies on Tor dealing with its analysis and performance improvement mechanisms. Classification of various studies is shown in Figure \ref{fig: Classification of approaches on Tor's Analysis.}.

Table \ref{tab: Researches on general study and modelling of Tor} presents a comparison of various generalized studies covering the modelling of Tor network. Comparison shows that majority studies focused over analysis of Tor network. Moreover, usability analysis and anonymity analysis were the most frequently studied topics followed by performance analysis. Very few research works focused over the sociability issues of Tor network.

\begin{table*}
  \caption{Research works on general study and modelling of Tor. Table entries symbolize Discussion (Dis.), Analysis (Anal.), Propose (Propos.), Sociability Issues (Soci. Issu.), Usability Issues (Usab.), Performance Latency (Perf. Lat.), Performance - bandwidth (Perf. BW), Anonyimty (Anon.).}\label{tab: Researches on general study and modelling of Tor}
  \centering
  \small
  \begin{tabular}{|
  @{}>{\centering}p{3cm}@{}|
  @{}>{\centering\arraybackslash}p{0.6cm}@{\hspace{0.035in}}|
  @{}>{\centering\arraybackslash}p{0.6cm}@{\hspace{0.035in}}|
  @{}>{\centering\arraybackslash}p{0.9cm}@{\hspace{0.035in}}|
  @{}>{\centering\arraybackslash}p{0.6cm}@{\hspace{0.035in}}|
  @{}>{\centering\arraybackslash}p{1.1cm}@{\hspace{0.035in}}|
  @{}>{\centering\arraybackslash}p{0.6cm}@{\hspace{0.035in}}|
  @{}>{\centering\arraybackslash}p{0.7cm}@{\hspace{0.035in}}|
  @{}>{\centering\arraybackslash}p{0.68cm}@{\hspace{0.035in}}|
  @{}>{\centering\arraybackslash}p{8.5cm}@{\hspace{0.035in}}|
  }
  \hline
            & \multicolumn{3}{|c|}{Study focus} & \multicolumn{5}{|c|}{Research Parameters} & Idea\\
  \cline{2-10}
  Research                          & Dis.  & Anal. & Propos.    & Soci.  & Usab.         & Perf.    &  Perf.     & Anon. &       \\
                                    &       &       &            & issu.  & [St./eas]     & lat      &   BW       &       &       \\

  \hline
  \multicolumn{10}{|c|}{General Study over Tor}\\
  \hline

  Dingledine \emph{et al.}  \cite{dingledine2007deploying}   &\checkmark &           &           & \checkmark & & & & & Discussed challenges and social issues, and studied Tor network\\

  Abou-Tair \emph{et al.}  \cite{abou2009usability}      &\checkmark &\checkmark &           &&\checkmark & &\checkmark &\checkmark & Studied usability, bandwidth and anonymity over anonymous networks\\

  Clark \emph{et al.}  \cite{clark2007usability}           &           &\checkmark &           &&\checkmark & & &\checkmark & Performed usability analysis of Tor with other anonymity tools\\

  Edmundson \emph{et al.}  \cite{edmundson14security}      &           &\checkmark &           && &\checkmark & &\checkmark & Compared anonymity and performance of \emph{Safeplug} with Tor\\

  Barthe \emph{et al.}  \cite{barthe2010robustness}          &           &\checkmark &           && &\checkmark &\checkmark & & Studied robustness in Tor network\\

  Mulazzani \emph{et al.}  \cite{mulazzani2010anonymity}       &           &\checkmark &           &&\checkmark & & &\checkmark & Analysed monitoring and anonymity issues in Tor\\

  Huber \emph{et al.}  \cite{huber2010tor}          &           &\checkmark &           &&\checkmark & & &\checkmark & Studied anonymity using HTTP usage statistics\\

  McCoy \emph{et al.}  \cite{mccoy2008shining}           &           &\checkmark &           &&\checkmark & & &\checkmark & Studied applications, usage statistic and misusage of Tor\\

  Loesing \emph{et al.} \cite{loesing2010case}  &           &\checkmark &           &&\checkmark & & &\checkmark & Studied country and port usage of Tor\\

  Chen \emph{et al.}  \cite{chen2009xpay}            &           &           &\checkmark &\checkmark & & & &\checkmark & Proposed anonymous payments over anonymous network\\

  \hline
  \multicolumn{10}{|c|}{Modelling Tor network}\\
  \hline

  Jansen \emph{et al.}  \cite{jansen2012methodically}       & &\checkmark &\checkmark && &\checkmark &\checkmark & & Proposed graph based Tor topology\\

  Jansen and Hopper  \cite{jansen2011shadow}          & & &\checkmark && &\checkmark &\checkmark & & Developed discrete event Tor simulatork\\

  Bauer \emph{et al.}  \cite{bauer2011experimentor}        & & &\checkmark && &\checkmark &\checkmark & & Developed emulation toolkit \emph{ExperimenTor} for Tor\\

  \hline
  \end{tabular}
\end{table*}

\subsubsection{General Studies of Tor}

Several studies covering pros and cons of Tor and analyzing statistics of Tor referring to users' quality of experience are summarized in the paragraphs below.

\emph{Understanding Challenges and Social Factors:} In \cite{dingledine2007deploying}, Dingledine \emph{et al.} described the challenges in implementation of Tor and discussed social issues. Tor network design and its details were also discussed with reference to the previous state-of-the-art. Possible avenues for improvements in the Tor network and flaws in the current system were presented including abuse, security implications and perceived social value.

\emph{Who is More User Friendly ?} Abou-Tair \emph{et al.} \cite{abou2009usability} focused on the usability of different anonymizing solutions including Tor, I2P\footnote{I2P is an anonymous overlay network which supports both TCP and UDP traffic. Web: \url{https://geti2p.net/en/}}, JAP/JonDo (Java Anonymous Proxy)\footnote{Java Anon Proxy allows web browsing with pseudonymity using its proxy based system. Web: \url{https://anonymous-proxy-servers.net/}} and Mixmaster\footnote{Mixmaster is a Chaumian mix network which is an anonymous remailer providing security against traffic analysis and sender deanonymization. Web: \url{http://mixmaster.sourceforge.net/}}. The installation of all softwares was analyzed with regard to ease-of-use. They measured the bandwidth consumption of all softwares. The authors concluded that I2P and Mixmaster provide better anonymity but are more complex. On the contrary, Tor and JAP are easy to use but comprise somewhat on the degree of anonymity they provide.

\emph{Usability Analysis of Tor:} Clark \emph{et al.} \cite{clark2007usability} conducted usability analysis for deployment of Tor and software tools associated with Tor including Vidalia, Privoxy, Torbutton and FoxyProxy. Research showed that all implementations have associated pros and cons. The study presented guidelines for future implementations for maximum usability of anonymity tools. Research spanned over the installation, configuration, usage menu, verification and switch-off features of various anonymity tools.

\emph{Safeplug vs. Tor:} Edmundson \emph{et al.} \cite{edmundson14security} analyzed the security provided by Safeplug in comparison to the Tor network. Safeplug\footnote{\url{https://pogoplug.com/safeplug}} is a plug-and-play network device which is plugged into the router and it acts as an HTTP proxy by directing all web traffic through the Tor network. Safeplug was launched to provide ease in access for Tor users. On the contrary, Tor network can be accessed through Tor browser bundle provided by Tor. Study showed that Safeplug was vulnerable to first and third-party trackers, through which users can be deanonymized. Attacker can modify the settings of Safeplug externally through cross-site request forgery (CSRF). Safeplug provided more latency and less protection than Tor.

\emph{Robustness of Tor:} Barthe \emph{et al.} \cite{barthe2010robustness} argued that \emph{robustness} has always been neglected while \emph{privacy} is the issue that receives most attention. Authors defined general and flexible definitions for robustness and studied the Golle and Juels protocol. By identifying the weaknesses in the current protocol, novel enhancements were also proposed for robustness.

\emph{Anonymity and Monitoring on Tor:} Mulazzani \emph{et al.} \cite{mulazzani2010anonymity} addressed the \emph{Monitoring} and \emph{Anonymity} issues in the current Tor network. A dataset was collected over a period of six months. Analysis showed that a sinusoidal pattern in users is observed with half of servers located in Germany and United States. A proposed implementation has been added into \emph{TorStatus}\footnote{\url{https://torstatus.blutmagie.de}}, which is the project displaying Tor network status, available routers, bandwidths, hosts and availability history.

\emph{Tor Traffic Statistics:} Huber \emph{et al.} \cite{huber2010tor} analyzed the HTTP usage of the Tor network. Research showed that $78\%$ of Tor users do not use Tor using TorButton, which can be used for deanonymization. $1\%$ of Tor requests are vulnerable to piggybacking attacks. $7\%$ requests, pertaining to social networks, contain identifiable information. The authors suggested the use of HTTPS instead of HTTP for secure communication.

\emph{Tor Usage Statistics:} McCoy \emph{et al.} \cite{mccoy2008shining} studied the applications, user countries and usage of Tor network. Statistics collected from Tor showed that non-interactive protocols (BitTorrent traffic), comprising of a minority of connections, consumed majority of resources. Non-secure protocols like HTTP can be exploited by the exit router to log sensitive information. The study suggested a protocol for identification of all exit routers capturing POP3 traffic. Usage statistics revealed that USA, Germany and China are major users of Tor.

\emph{Statistical Data of Tor:} Loesing \emph{et al.} \cite{loesing2010case} collected the statistics from the live Tor network to measure two aspects of communication, i.e., (1) country wise usage, and (2) traffic port numbers for exiting traffic. Both these statistics can be used for future improvements in the Tor network for better anonymity services. The study also revealed that port $80$ receives most traffic.

\emph{Micropayments Using Tor:} Chen \emph{et al.} \cite{chen2009xpay} proposed a novel mechanism of anonymous payments for network services. The proposed mechanism allows users to make untraceable micro-payments to each other. Authors included features of offline verification, overspending prevention, aggregation and low overheads. Experiments showed only $4\%$ overhead for the proposed strategy.

\subsubsection{Modeling Tor Network}

In this section, we present the modeling techniques used for analyzing Tor.

\emph{Modeling Topology and Hosts of Tor:} Jansen \emph{et al.} \cite{jansen2012methodically} developed a model of Tor which closely resembled the Tor network. Authors developed a graph for Tor topology where vertexes related to downstream bandwidth, upstream bandwidth and packet loss, and edges related to latency, jitter and packet loss. All hosts including relays, authorities, clients and Internet servers were mapped to the developed graph based on characteristics obtained from Tor.

\emph{Shadow: Simulating Tor Network} Jansen and Hopper \cite{jansen2011shadow} developed an open source discrete event simulator for simulating the network layer of Tor on a single machine. Authors compared the performance of \emph{Shadow} with real-world simulation results from the PlanetLab testbed.

\emph{Emulation Toolkit for Tor Experimentation:} Bauer \emph{et al.} \cite{bauer2011experimentor} developed ExperimenTor, an emulation toolkit for Tor network. Their research was focused on the toolkit rather than the analysis of the Tor network.

\subsubsection{Analysis of Tor}

Analysis of Tor network has been a part of many studies covering delays, bandwidth, quality of service, relay selection and authentication protocols.
Several studies covering these areas are summarized in following sections.

Table \ref{tab: Researches on analysis and performance improvements of Tor} presents a comparison of various research works in analysis and performance improvement track. Comparison shows that relay selection and latency analysis are the most frequently studied topics followed by anonymity, bandwidth and quality of service analysis. Very few studies focused over queues, traffic shaping techniques and protocol messages.

\begin{table*}
  \caption{Research studies on analysis and performance improvements of Tor. Table entries symbolize New algorithms (New Algo), Analysis (Anal.), Relay Selection (Relay Sel.), Performance Latency (Perf. Lat.), Performance Bandwidth (Perf. BW.), Quality of Service (QoS), Queues, Protocol Messages (Prot. Msgs.), Traffic Shaping (Traff. Shap.), Anonymity (Anon.).}\label{tab: Researches on analysis and performance improvements of Tor}
  \centering
  \small
  \begin{tabular}{|
  @{}>{\centering}p{2.8cm}@{}|
  @{}>{\centering\arraybackslash}p{0.65cm}@{\hspace{0.035in}}|
  @{}>{\centering\arraybackslash}p{0.65cm}@{\hspace{0.035in}}|
  @{}>{\centering\arraybackslash}p{0.75cm}@{\hspace{0.035in}}|
  @{}>{\centering\arraybackslash}p{0.55cm}@{\hspace{0.035in}}|
  @{}>{\centering\arraybackslash}p{0.65cm}@{\hspace{0.035in}}|
  @{}>{\centering\arraybackslash}p{0.65cm}@{\hspace{0.035in}}|
  @{}>{\centering\arraybackslash}p{0.95cm}@{\hspace{0.035in}}|
  @{}>{\centering\arraybackslash}p{0.65cm}@{\hspace{0.035in}}|
  @{}>{\centering\arraybackslash}p{0.75cm}@{\hspace{0.035in}}|
  @{}>{\centering\arraybackslash}p{0.73cm}@{\hspace{0.035in}}|
  @{}>{\centering\arraybackslash}p{7.35cm}@{\hspace{0.035in}}|
  }

  \hline
  \multirow{3}{*}{Research} & \multicolumn{2}{|c|}{Focus} & \multicolumn{8}{|c|}{Path Selection Parameters} & \multirow{3}{*}{Idea}\\
  \cline{2-11}
    & New  & Anal.                            &  Relay & Perf.  & Perf. &  QoS & Queues & Prot. & Traff. & Anon. &       \\
            & Algo &                                  &  Sel.  & Lat.    & BW.  &      &        & Msgs. & Shap.  &       &       \\

  \hline
  \multicolumn{12}{|c|}{Analysis of Tor network}\\
  \hline

  Dhungel \emph{et al.}  \cite{dhungel2010waiting}       & &\checkmark      & \checkmark & \checkmark & & & & & & & Analysed latency for Tor relays\\
  Loesing \emph{et al.}  \cite{loesing2008performance}       & &\checkmark      & & \checkmark & & \checkmark & & & & & Analysed latency, QoS and performance of Tor\\
  Ehlert \emph{et al.}  \cite{ehlert2011i2p}       & &\checkmark      & &\checkmark &\checkmark & & & & & & Studied bandwidth and latency in Tor\\
  Pries \emph{et al.}  \cite{pries2008performance}        & &\checkmark      & & & \checkmark & \checkmark & & & & & Investigated bandwidth for various path selection algorithms\\
  Liu and Wang  \cite{liu2009anti}                &\checkmark &      &\checkmark &\checkmark & &\checkmark & & & &\checkmark & Proposed relay reliability mechanism considering performance anonymity and QoS\\
  Wang \emph{et al.}  \cite{wang2013empirical}          & & \checkmark     &\checkmark & & & & & & & & Performed an empirical analysis over family nodes\\
  Tschorsch and Scheuermann \cite{tschorsch2011tor}   & \checkmark &      & &\checkmark & & & \checkmark & & & & Proposed fairness model for efficient and fair resource allocation\\
  Chaabane \emph{et al.}  \cite{chaabane2010digging}       & & \checkmark    &\checkmark & & && & & & & Studied misuse of Tor's exit nodes as proxies\\
  Hopper \emph{et al.}  \cite{hoppershort}      & & \checkmark      & & & && &\checkmark & & & Analysed Tor's performance considering key exchange mechanisms\\
  Lenhard \emph{et al.}  \cite{lenhard2009performance}       & & \checkmark     & &\checkmark & && & & & & Studied communication overhead in low bandwidth networks for Tor's hidden services\\
  Goldberg \cite{goldberg2006security}                     & & \checkmark     &\checkmark & & && &\checkmark & & & Analysed anonymity with Tor;s authentication protocol\\

  \hline
  \multicolumn{12}{|c|}{Tor performance improvement}\\
  \hline

  Jansen \emph{et al.}  \cite{jansen2010recruiting}       & \checkmark &      &\checkmark & & && & & & & Proposed token based performance mechanisms for recruiting more relays\\
  Dingledine \emph{et al.} \cite{dingledine2010building}    & \checkmark &      &\checkmark & & && & & & & Proposed priority based traffic handling for relays\\
  Wang \emph{et al.}  \cite{wang2013rbridge}       & \checkmark &        &\checkmark & & && & & &\checkmark & Proposed node reliability mechanism to avoid blockage of bridges\\
  Smits \emph{et al.} \cite{smits2011bridgespa}       & \checkmark &        &\checkmark & & && &\checkmark & &\checkmark & Proposed packet authorization based mechanism to protect bridges from eavesdroppers\\
  Moghaddam \emph{et al.}  \cite{mohajeri2012skypemorph}       & \checkmark &   & & & && & &\checkmark & \checkmark & Proposed traffic morphing (using Skype traffic) to avoid censorship\\
  Weinberg \emph{et al.}  \cite{weinberg2012stegotorus}       & \checkmark &    & & & && & &\checkmark & \checkmark & Proposed traffic shaping (by assembling regular HTTP traffic) to avoid deanonymization\\
  Gopal and Heninger \cite{gopal2012torchestra}       & \checkmark &         & &\checkmark & && & & & & Suggested latency reduction by separate TCP connections for interactive and bulk traffic\\
  AlSabah \emph{et al.}  \cite{alsabah2011defenestrator}       & \checkmark &     & &\checkmark & &&\checkmark & & & & Proposed traffic morphing (using Skype traffic) to avoid censorship\\
  Jansen \emph{et al.}  \cite{jansen2012throttling}       & \checkmark &      & &\checkmark & && & & &\checkmark & Proposed throttling mechanisms for reducing latency by avoiding bulk traffic\\

  \hline
  \end{tabular}
\end{table*}

\emph{Understanding Delays in Tor:} Dhungel \emph{et al.} \cite{dhungel2010waiting} analyzed the delays in the entire Tor network. Authors suggested that overlay network plays the most significant role in Tor. The study revealed that $11\%$ of Tor routers are overloaded with traffic which resulted in very high delays. In $7.5\%$ of circuits, overall latency introduced a $450$ms delays. \emph{Guard} routers incorporate more delay than \emph{non-guard} routers. There is high fluctuation in delay for all routers except for those having high bandwidths.

\emph{Measurement and Statistics:} Loesing \emph{et al.} \cite{loesing2008performance} studied the latencies inside the Tor network. A deep investigation was conducted to evaluate the individual delays and QoS properties. The authors showed that circuit building time (Introduction and Rendezvous) is the most crucial delay period in Tor. Fr\'{e}chet and exponential distributions were combined to analyze the response times.

\emph{Comparing Bandwidth with Latency:} Ehlert \cite{ehlert2011i2p} compared the bandwidth and latency performance of Tor network with the popular I2P network. Authors measured the core latency (HTTP GET requests durations), average latency (webpage download times including external threads and pictures) and bandwidth (download speeds). This research showed that I2P network provides lower core latency and Tor network excels in average latency and bandwidth, owing to the nodes distribution and penetration of the Tor network.

\emph{Tor QoS with Path Selection Strategy:} Pries \emph{et al.} \cite{pries2008performance} suggested that TCP suffers severe performance degradation from the random path selection of Tor. Slight QoS improvement is achieved with Tor's bandwidth weighted path selection algorithm. The main reason attributed for small improvements is low bandwidth of Tor routers.

\emph{Investigating Tor's Exit Policies:} Liu and Wang \cite{liu2009anti} studied the exit policies of the exit nodes and addressed the short-comings in the current Tor architecture. A new protocol was proposed which comprised of three parts: (1) reporting misbehavior protocol, (2) building global blacklist protocol, and (3) blocking misbehavior protocol for users. User experience, performance and anonymity were the key indexes used for evaluation.

\emph{Behavior of Family Nodes:} Wang \emph{et al.} \cite{wang2013empirical} presented an empirical analysis of Tor family nodes. A rich dataset of live Tor network comprising of three years was used to study the impact of family nodes. The study suggested that family nodes provide stable and better service than other nodes. Moreover, attacks on family nodes can disrupt the Tor network more severely than random Tor nodes.

\emph{Fairness in Tor:} Tschorsch and Scheuermann \cite{tschorsch2011tor} analyzed the fairness issues in the current Tor network. Large unfairness was observed in the current resource allocation mechanism of the Tor network. Authors proposed a max-min fairness based model for efficient and fair scheduling of resources. The proposed design was analyzed with Tor's $N23$ congestion feedback mechanism.

\emph{Misuse of Tor:} Chaabane \emph{et al.} \cite{chaabane2010digging} showed that Tor network was being used for transmitting P2P traffic (Bit torrent etc.) over the Tor network. HTTP and Bit torrent were analyzed on the Tor network. The study showed that Tor exit nodes are being used as one hop SOCKS proxies through tunneling. New techniques were devised to detect such abnormalities in exit nodes' behavior. Research showed that simple crawling over exit nodes can be used to collect as many bridge identities as needed.

\emph{Challenges for Hidden Services of Tor:} Hopper \cite{hoppershort} conducted research on the poor performance of Tor based on the fact that users of Tor increased from $1$ million to nearly $6$ million but no dramatic change was observed in the network. Hopper attributed the poor performance to the key exchange mechanism of Tor, which was later updated. Study showed the possible research dimensions of limiting request rates from botnets, throttling entry guard, reusing failed partial circuits and isolating hidden services circuits.

\emph{Tor Hidden Services in Low Bandwidth Access Networks:} Lenhard \emph{et al.} \cite{lenhard2009performance} conducted a measurement and statistical analysis for estimating the communication overhead of Tor hidden services in low bandwidth access networks. Research showed that boot strapping time, RTT and circuit building time were the major bottlenecks to performance. Due to numerous delays, an increase in timeout value was suggested to avoid repeated retransmissions.

\emph{Analysis of Tor Authentication Protocol:} Goldberg \cite{goldberg2006security} analyzed the security of Tor's authentication protocol (TAP). The authors argued that any security breach by a single malicious Tor relay can deanonymize users' sessions. Through empirical evaluations, research showed that TAP is secure in random oracle model.

\emph{Statistics Collection Mechanism of Tor:} Mani and Sherr \cite{mani2017historvarepsilon} analysed the data collection mechanism of Tor through `PrivEx'. They showed that statistics of PrivEx can be easily compromised by the present of adversary nodes in the Tor network. As a result of shortcomings of PrivEx, authors proposed `HisTor', a privacy preserving statistics collection mechanism of Tor which is much more diverse than PrivEx. HisTor uses the count of queries by exit nodes and relays in form of a histogram where individual nodes have little control over the aggregate statistics.

\subsubsection{Tor Performance Improvement}

Owing to the increasing demand for Tor, various studies have proposed performance improvements to cope with future demands. In this section, we present these studies covering Tor node selection, traffic distribution and latency management etc.

\emph{Node Recruitment for Tor:} Jansen \emph{et al.} \cite{jansen2010recruiting} focused their research on recruitment of new Tor relays, motivated by the fact that only $1.5\%$ nodes participate as relays. Authors proposed BRAIDS which is a token based mechanism providing high bandwidth to those users who employ BRAIDS. Proposed scheme characterizes traffic into high throughput, low latency and normal traffic. Based upon usage of BRAIDS and node networking stats, tickets are generated which can be used to increase bandwidth.

\emph{Encouraging Tor nodes for traffic relaying:} Dingledine \emph{et al.} \cite{dingledine2010building} proposed a mechanism to encourage Tor nodes for traffic relaying. Study suggested a priority based traffic handling, which gives more weight (in form of bandwidth and delays) to those nodes contributing resources to Tor. However, all Tor relays carry an additional load of priority based traffic handling. Directory authorities need to assign priority levels to all Tor users participating in Tor relays.

\emph{Improving Distribution Mechanism of Tor Bridges:} Wang \emph{et al.} \cite{wang2013rbridge} improved the distribution mechanism of Tor bridges by implementing node reliability statistics to avoid the blockage of bridges by corrupt nodes. The uptime of assigned bridges is used to give reputation points to users. In case of any blockage of a bridge, a new bridge address is given on payment of earned credit. To ensure anonymity, reputation information is stored on users' systems by using a privacy-preserving technique which cannot be circumvented by malicious users.

\emph{Packet Authorization for Tor Bridges:} Smits \emph{et al.} \cite{smits2011bridgespa} proposed BridgeSPA, a packet authorization based mechanism, to protect users of Tor hosting Bridges. All Tor user hosting bridges are susceptible to traffic analysis attacks. To counter this attack, the authors suggest the transmission of a bridge key by bridge distribution authorities which is valid for a limited time, as determined by the bridge. For any communication with the bridge, Tor users should use that key within the assigned time period.

\emph{SkypeMorph - Tor traffic Shaping:} Moghaddam \emph{et al.} \cite{mohajeri2012skypemorph} proposed a new mechanism namely SkypeMorph to avoid the censorship of Tor bridges. The fundamental idea was to hide Tor traffic as Skype video traffic (a widely used protocol). SkypeMorph, which runs side by side with Tor, makes it hard to distinguish Tor traffic from Skype traffic. Two schemes were suggested for traffic morphing, (1) using the target stream attributes, (2) incorporating both source and destination streams to incorporate packet timings. Both streams provided nearly identical performance, but the former had lower complexity.

\emph{StegoTorus - Steganographing Tor Traffic:} Weinberg \emph{et al.} \cite{weinberg2012stegotorus} proposed a novel technique to bypass censorship on Tor. Their scheme is based upon the idea of chopping Tor traffic into multiple streams, resembling HTTP traffic, before passing through the censor. StegoTorus acted as a proxy on Tor clients.

\emph{Torchestra - Separate connections for Interactive and Bulk Traffic:} Gopal and Heninger \cite{gopal2012torchestra} proposed the transmission of interactive and bulk traffic over two separate TCP connections among all nodes in the Tor network. Exponentially weighted moving average (EWMA) algorithm was used to distinguish between interactive and bulk traffic on all circuits. Upto $40\%$ reduction in delays was observed as compared to standard Tor for the proposed strategy.

\emph{Reducing Latency in Tor:} AlSabah \emph{et al.} \cite{alsabah2011defenestrator} proposed a mechanism for congestion control and flow control in order to reduce latency in the Tor network. The study suggested the use of small fixed size windows and small dynamic windows which can reduce the packets in flight. For flow control, the study proposed an N23 algorithm which caps the queue lengths of Tor routers and provided a $65\%$ increase in webpage responses and $32\%$ decrease in page loading time.

\emph{Throttling Tor bulk users:} Jansen \emph{et al.} \cite{jansen2012throttling} addressed the poor performance of Tor network using bulk data transfers. Three dynamic throttling algorithms were proposed for reducing network congestion and latency. The guard relay capped the bandwidth capacity of nodes, so, only local relay information was used. Simulations showed that throttling reduces the web page latency and increases the anonymity of Tor network.

\subsubsection{Tor Client Mobility}

In this section, we study the research works focused on the mobility of Tor network with a particular emphasis on anonymity.
Table \ref{tab:   Researches on Tor's mobile devices} shows the research works in path selection track and shows that performance and anonymity have been the most frequently studied parameters. Details are presented in below paragraphs.

\begin{table*}
  \caption{Research works on Tor's client mobility. Table entries symbolize New algorithms (New Algo), Analysis (Anal.), Autonomous Systems (AS), Relay Locations (Relay loc.), Hops, Performance-Latency-Bandwidth (Perf., Lat, BW), Multi-path, Load, Relay Capacity (Rel. Cap.) and Anonymity (Anon).}\label{tab: Researches on Tor's mobile devices}
  \centering
  \small
  \begin{tabular}{|
  @{}>{\centering}p{3cm}@{}|
  @{}>{\centering\arraybackslash}p{0.6cm}@{\hspace{0.035in}}|
  @{}>{\centering\arraybackslash}p{0.65cm}@{\hspace{0.035in}}|
  @{}>{\centering\arraybackslash}p{0.45cm}@{\hspace{0.035in}}|
  @{}>{\centering\arraybackslash}p{0.7cm}@{\hspace{0.035in}}|
  @{}>{\centering\arraybackslash}p{0.65cm}@{\hspace{0.035in}}|
  @{}>{\centering\arraybackslash}p{0.65cm}@{\hspace{0.035in}}|
  @{}>{\centering\arraybackslash}p{0.75cm}@{\hspace{0.035in}}|
  @{}>{\centering\arraybackslash}p{0.65cm}@{\hspace{0.035in}}|
  @{}>{\centering\arraybackslash}p{0.6cm}@{\hspace{0.035in}}|
  @{}>{\centering\arraybackslash}p{0.7cm}@{\hspace{0.035in}}|
  @{}>{\centering\arraybackslash}p{8.0cm}@{\hspace{0.035in}}|
  }
  \hline
  \multirow{3}{*}{Research}          & \multicolumn{2}{|c|}{Focus} & \multicolumn{8}{|c|}{Path Selection Parameters} & \multirow{3}{*}{Idea}\\
  \cline{2-11}
            & New  & Anal. & AS    & Relay & Hops & Perf.   & Multi- & load & Rel. & Anon.      &      \\
            & Algo &       &       & Loc.  &      & Lat, BW & path   &      & Cap. &            &      \\
  \hline
  Doswell \emph{et al.}  \cite{doswell2013novel}       & &\checkmark & & & &\checkmark & & & & & Suggested bridge relays to avoid bandwidth issues while roaming\\
  Andersson \emph{et al.} \cite{andersson2007practical}      &\checkmark & & & & & \checkmark & & & & \checkmark &  Proposed trade-of between anonymity and performance\\

  \hline
  \end{tabular}
\end{table*}

\emph{Using Bridge Relays:} Doswell \emph{et al.} \cite{doswell2013novel} analyzed the performance of Tor for wireless devices roaming across multiple networks. Analysis showed that the \emph{speed} of mobile wireless devices significantly affects the circuit building time and Tor's performance. Authors studied the use of bridge relays to provide persistent Tor connections for mobile devices.

\emph{New Architectural Designs:} Andersson \emph{et al.} \cite{andersson2007practical} proposed several new architectural designs for a mobile Tor network. A trade-off between anonymity and performance was evaluated. Several criteria used in performance estimation included usability, availability, trust and practicality. The study concluded that the single Tor client option offers lowest degree of anonymity.

\begin{figure}[b]
  \centering
  \includegraphics[width=0.7\columnwidth]{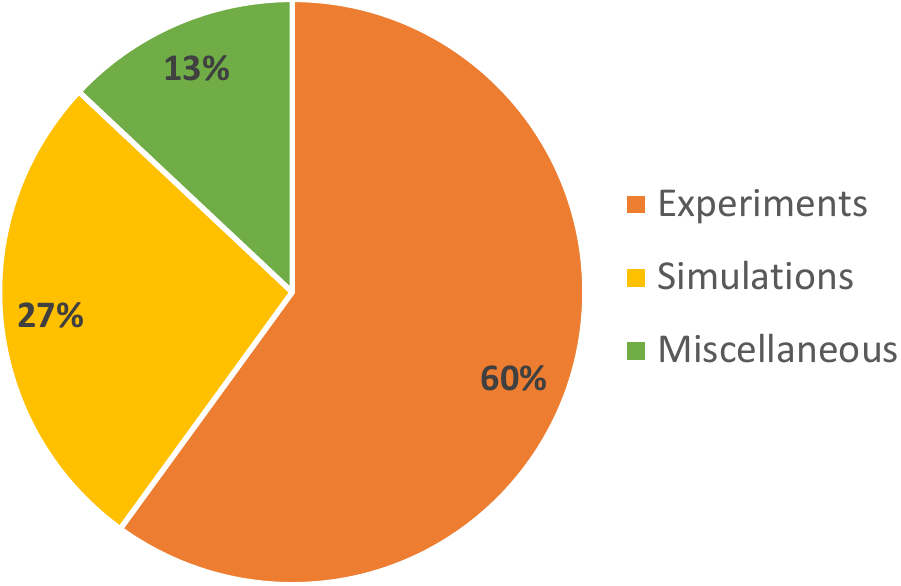}\\
  \caption{Classification of platforms for Tor's research.}\label{fig: Classification of platforms for Tor's research.}
\end{figure}

\begin{figure*}[t]
  \centering
  \includegraphics[width=1.2\columnwidth]{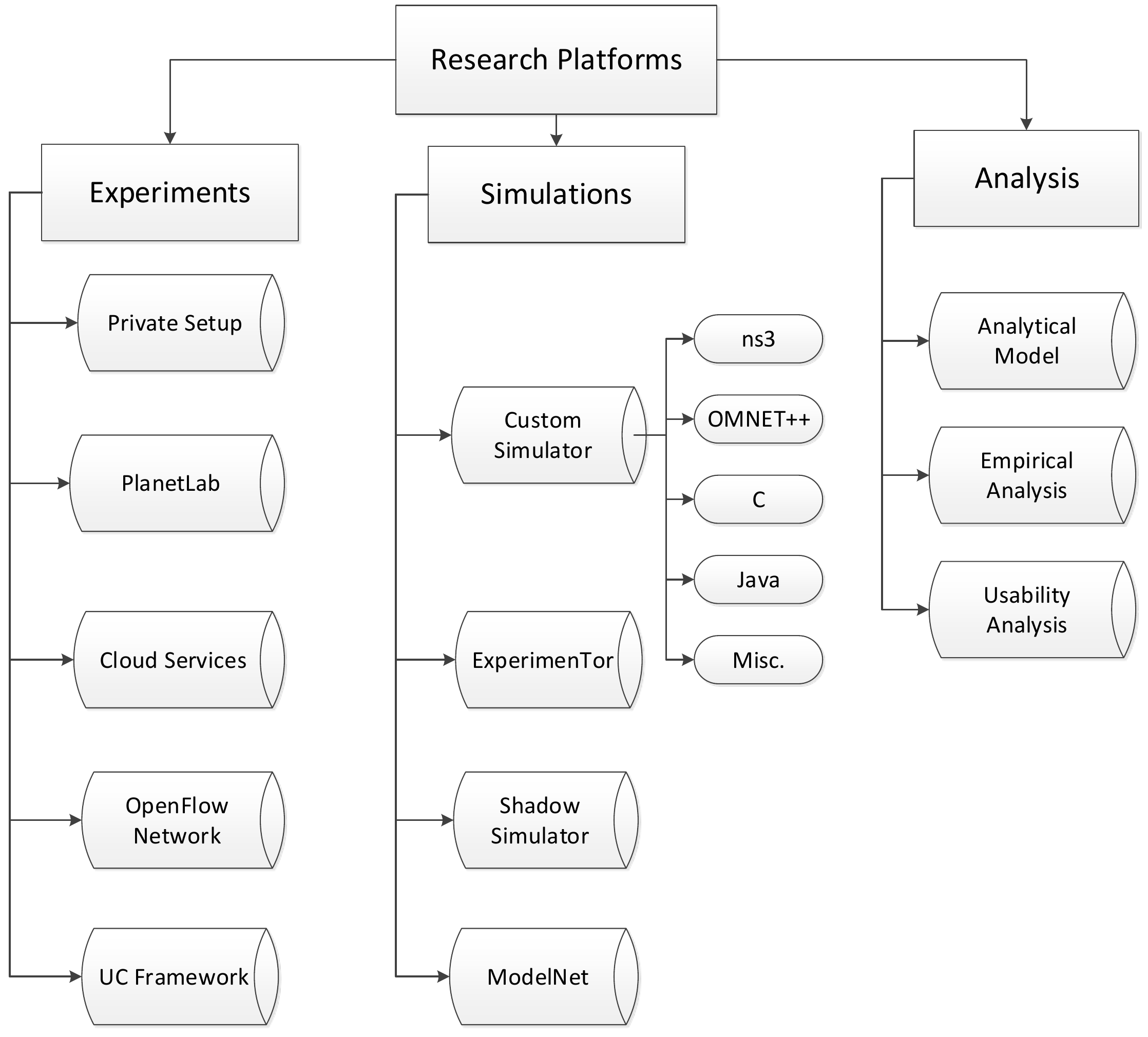}\\
  \caption{Taxonomy of platforms employed in Tor research.}\label{fig: Taxonomy of platform for Tor researches}
\end{figure*}

\section{Platforms for Tor Research}
\label{sec:Platforms for Tor Research}

In this section, we study the platforms used to study Tor network. Our observations spanning over decades of anonymity research shows that all research works have studied the Tor network using three different techniques, (1) Experiment, (2), Simulations, and (3) Analysis. Figure \ref{fig: Classification of platforms for Tor's research.} shows that 60\% of the studies used in this paper conducted experiments. Only 27\% of the studies conducted experiments. In the experiment section, majority studies developed their own testbed followed by experiments on cloud services and PlanetLab testbeds. In the simulations section, majority research works used extensive simulations to study Tor network. Finally, some studies analyze Tor network by collecting statistics and discussing the sociability and usability issues of Tor network. These three classification categories are elaborated in Figure \ref{fig: Taxonomy of platform for Tor researches} which shows the platforms used to study Tor network.

\subsection{Tor Experiments}

Studies covering Tor experiments have focused over several areas including (1) private setup establishment, (2) PlanetLab experiments, (3) cloud services (4) OpenFlow networks and (5) universal composability framework.

\begin{figure}[b]
  \centering
  \includegraphics[width=0.8\columnwidth]{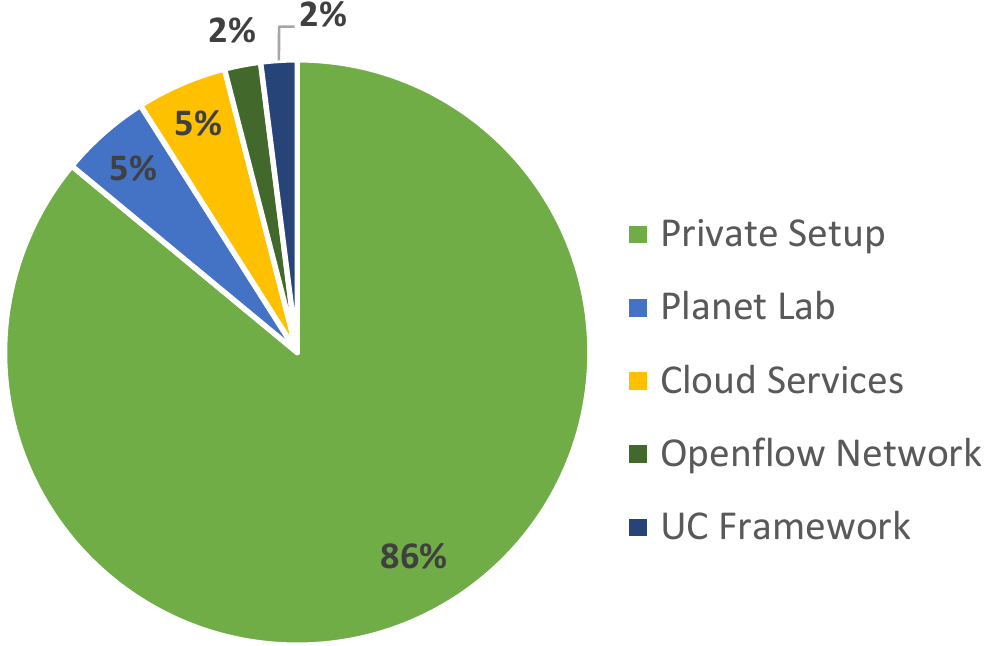}\\
  \caption{Classification of platforms used in experimental Tor research.}\label{fig: Classification of experiments on Tor.}
\end{figure}

Table \ref{tab: Experimental setups used in different researches} presents the clients, relays, servers, Tor services and Tor implementation used by various research works. Comparison shows that majority studies deployed their private testbeds with 1-2 clients and 1-2 servers. Several studies deployed limited number of relays for experiments. Number of clients were increased drastically in the PlanetLab and cloud setup for Tor experiments. Moreover, traffic analysis was the most frequently studied topic. Majority research works used the default Tor setup without any modifications. Figure \ref{fig: Classification of experiments on Tor.} shows the classifications of experiments on Tor. Analysis of figure shows that majority of studies deployed their own private testbeds.

\begin{table*}
  \centering
  \caption{Experimental setups used in different research works.}
  \label{tab: Experimental setups used in different researches}
  \footnotesize
  \begin{tabular}{| p{4cm} | p{1.3cm} | p{1.3cm}| p{2cm}| p{2.7cm}| p{3.3cm} |}
    \hline
    Research & Servers  & Relays    & Clients & Service         & Tor implementation\\
    \hline
    \hline
    \multicolumn{6}{|c|}{\emph{Private Setup}}\\
    \hline
    Overlier and Syverson \cite{overlier2006locating}      & 2        &           & 1       & Hidden service  & default Tor\\
    \hline
    Andersson and Panchenko \cite{andersson2007practical}      & 1       &           &   1       & Mobile Tor    &   Onion Coffea\\
    \hline
    Panchenko \emph{et al.} \cite{panchenko2008performance}    &       &               &   1       &   Download Service    & Def. Tor + Onion Coffea\\
    \hline
    Pries \emph{et al.} \cite{pries2008performance}    &   1   &               &   1       &  Download Service     &   Privoxy\\
    \hline
    Wagner \emph{et al.} \cite{wagner2012breaking}    &   2   &   1           &   1       &  Log processing       &   WebProxy\\
    \hline
    Chan-Tin \emph{et al.} \cite{chan2013revisiting}    &   1   &               &   2       &  Traffic Analysis     & Def. Tor\\
    \hline
    Pries \emph{et al.} \cite{pries2008new}    &   1   &   2           &   1       &   TCP data coll.      &   Tor mod.\\
    \hline
    Herzberg \emph{et al.} \cite{herzberg2011camouflaged}    &       &               &   1       &   Web page download   &   Def. Tor\\
    \hline
    Bauer \emph{et al.} \cite{bauer2009predicting}    &       &   6+       &   1+      &   Path compromise     &   Def. Tor\\
    \hline
    Song \emph{et al.} \cite{song2013anonymize}    &       &   6           &   1       & Traffic Analysis      &   Tor mod.\\
    \hline
    Dhungel and Steiner \cite{dhungel2010waiting}    &   1   &   2           &   1       & Traffic Analysis      &   Def. Tor\\
    \hline
    Gros \emph{et al.} \cite{gros2010protecting}    &       &               &   2+      & Traffic Analysis      &   Honeywall\\
    \hline
    Wang \emph{et al.} \cite{wang2009novel}    &       &   2           &           & Traffic Analysis      &   Privoxy\\
    \hline
    Zhang \emph{et al.} \cite{zhang2011application}     &   1   &   3           &   1       & Hidden Service        &   Polipo\\
    \hline
    Loesing \emph{et al.} \cite{loesing2008performance}    &       &   1           &   1+      & Access Attempt        &   Def. Tor\\
    \hline
    Chen and Pasquale \cite{chen2010toward} &   1   &               &   10      & Download              &   Def. Tor\\
    \hline
    Panchenko \emph{et al.} \cite{panchenko2012improving}    &   $1+$ &               &   1+      & Traffic Analysis      &   Def. Tor\\
    \hline
    Houmansadr \emph{et al.} \cite{houmansadr2013parrot}    &   $4$ &               &   $3$      & Traffic Analysis   &   $-$\\
    \hline
    Li \emph{et al.} \cite{li2012tmt}    &   1   &               &   1       & Download Analysis     &   Def. Tor\\
    \hline
    Chakravarti \emph{et al.} \cite{chakravarty2008identifying}     &   1   &               &   2       & Download Analysis     &   Def. Tor\\
    \hline
    Mulazzani \emph{et al.} \cite{mulazzani2010anonymity}    &       &               &   1+      & Traffic Analysis      &   Tor Status\\
    \hline
    Chaabane \emph{et al.} \cite{chaabane2010digging}    &  $1+$     &       6       &   $1+$     & Traffic Analysis      &   Def. Tor\\
    \hline
    Bai \emph{et al.} \cite{bai2008traffic}     &   2   &               &   6       & Traffic Analysis      &   Def. Tor\\
    \hline
    Barker \emph{et al.} \cite{barker2011using}     &   3   &       15      &   1+      & Traffic Analysis      &   Def. Tor\\
    \hline
    Marks \emph{et al.} \cite{marks2010unleashing}    &   3   &               &   3       & Download Analysis     &   \\
    \hline
    Jin and Wang \cite{jin2009effectiveness}    &   1   &               &   1       & Traffic Analysis      &   Tor mod.\\
    \hline
    Tang and Goldberg \cite{tang2010improved}    &   1   &   1           &   1       & Download Analysis     &   Def. Tor\\
    \hline
    Alsabah \emph{et al.} \cite{alsabah2012enhancing}     &       &               &   1 (3 Apps)& Traffic Analysis    &   Def. Tor\\
    \hline
    Moghaddam \emph{et al.} \cite{mohajeri2012skypemorph}    &       &               &   2+      & Traffic Analysis      &   SkypeMorph\\
    \hline
    Weinberg \emph{et al.} \cite{weinberg2012stegotorus}    &   1   &               &   1       & Download Analysis     &   StegoTorus\\
    \hline
    Evans \emph{et al.} \cite{evans2009practical}    &       &               &   1       & Traffic Analysis      &   Def. Tor\\
    \hline
    Wang \emph{et al.} \cite{wang2012congestion}    &       &               &   1       & Traffic Analysis      &   Def./Mod. Tor\\
    \hline
    Ehlert \cite{ehlert2011i2p}    &       &               &   1       & Traffic Analysis      &   Def. Tor\\
    \hline
    Barbera \emph{at al.} \cite{barbera2013cellflood}    &       &   2           &   4       & Traffic Analysis      &   Def. Tor\\
    \hline
    Winter and Lindskog \cite{winter2012great}    &       &   2           &   2+      & Traffic Analysis      &   Tor mod.\\
    \hline
    Edmundson \emph{et al.} \cite{edmundson14security}    &       &               &   1       & Download Analysis     &   Def. Tor\\
    \hline
    Huber \emph{et al.} \cite{huber2010tor}    &       &   1           &           & Traffic Analysis      &   Def. Tor\\
    \hline
    Blond \emph{et al.} \cite{blond2011one}    &       &   6           &   1+      & Traffic Analysis      &   Def. Tor\\
    \hline
    Lenhard \emph{et al.} \cite{lenhard2009performance}    &       &               &   1+      & Hidden Service        &   Def. Tor\\
    \hline
    McCoy \emph{et al.} \cite{mccoy2008shining}    &       &   3           &           & Traffic Analysis      &   Tor mod.\\
    \hline
    Chakravarty \emph{et al.} \cite{chakravarty2011detecting}    &       &               &   1       & Traffic Analysis      &  Def. Tor\\
    \hline
    Snader and Borisov \cite{snader2008tune}    &   1   &               &   1       & Traffic Analysis      & Tunable Tor + Vanilla\\
    \hline
    Gilad and Herzberg \cite{gilad2012spying}    &   1   &               &   1       & Traffic Analysis      & Def. Tor\\
    \hline
    Loesing \emph{et al.} \cite{loesing2010case}    &      &              &   1+       & Traffic Analysis      & Def. Tor\\
    \hline
    Chen \emph{et al.} \cite{chen2009xpay}    &       &   3+ (VMs)    &   2+ (VMs)& Traffic Analysis      & Def. Tor\\
    \hline
    Panchenko \emph{et al.} \cite{panchenko2011website}    &       &               &   1+      & Traffic Analysis      & Def. Tor\\
    \hline
    Wang and Goldberg \cite{wang2013improved}    &       &               & 200 cores & Traffic Analysis  & Def. Tor\\
    \hline \hline
    \multicolumn{6}{|c|}{\emph{PlanetLab Setup}}\\
    \hline
    Akhoondi \emph{et al.} \cite{akhoondi2012lastor}        &           &           & 50        & Traffic Analysis  & LASTor \\
    \hline
    Murdoch and Danezis \cite{murdoch2005low}        & 1         &           & 2         & Traffic Analysis  & Tor Mod. \\
    \hline
    Bauer \emph{et al.} \cite{bauer2007low}        & 6         & 2-6       & 40-90     & 40 node network   &  \\
    \cline{2-6}
                & 6         & 3-6       & 60-90     & 60 node network   &  \\
    \hline \hline
    \multicolumn{6}{|c|}{\emph{Cloud Setup (Amazon EC2)}}\\
    \hline
    Sulaiman and Zhioua \cite{sulaiman2013attacking}        &   1       &           &           & Traffic Analysis  & Def./Mod Tor     \\
    \hline
    Karaoglu \emph{et al.} \cite{karaoglu2012multi}        & 1         &           & 4         & Traffic Analysis  & Def. Tor           \\
    \hline
    Biryukov \emph{et al.} \cite{biryukov2013trawling}         &          &            & 50        & Hidden Services   & Def. Tor        \\
    \hline
  \end{tabular}
\end{table*}

\subsubsection{Private Setup connected with Tor}

Overlier and Syverson \cite{overlier2006locating} performed experiments by setting up two nodes (one in Europe and other in US) running hidden services at two ends of the Tor network. Access to webpages and images was provided using these services. The client PC was setup both as a client and a middleman node, and all sampling takes place at this client node.

Andersson and Panchenko \cite{andersson2007practical} performed experiments to verify the performance of their proposed mobile protocol. Mobile Tor was setup on a laptop connected to the Tor network. The content server hosting the files was placed at Karlstadt University. Experiments used OnionCoffee, which is a Java project developed under the PRIME project.

Panchenko \emph{et al.} \cite{panchenko2008performance} performed experiments using a Pentium Dual Core $1$GHz CPU with $2$GB RAM as a client nodes. Two existing Tor implementations (default Tor and OnionCoffee) were used on the client nodes. The Internet connection had a $10$Gbps bandwidth while the local backbone was $100$Gbps. Actual Tor relays were used to analyze the performance.

Pries \emph{et al.} \cite{pries2008performance} performed experiments by downloading a $458$kB file from a school web server. Command line utility \emph{wget} was used as the downloading tool. \emph{wget's} http-proxy and ftp-proxy were configured to download all files through \emph{Privoxy} from the server. Tor release \texttt{0.1.1.26} was configured on the exit and entry nodes.

Wagner \emph{et al.} \cite{wagner2012breaking} implemented a novel architecture using Tor. Three machines were setup running Tor exit node, BIND (DNS server with tcpdump), and Apache webserver, respectively. All machines were synchronised by NTP. Connected to Tor network, WebProxy was implemented in Perl. \texttt{iptables} was used to re-route traffic from Tor exit node to Perl proxy server. All processing of web server logs and proxy logs was performed using Perl, sqlite and modified tcpick.

Chan-Tin \emph{et al.} \cite{chan2013revisiting} setup a limited network for probing Tor network using client, burst server and probe machines. Entry and middle routers were chosen randomly while exit node was forced by choice. Four Tor relays were probed for the experiment and data of probes was collected after every $5$secs. Five connections were setup by the client using multi-threading.

Pries \emph{et al.} \cite{pries2008new} setup client, server, entry malicious router and exit malicious router by setting up four devices. A TCP client application was built which sent and received TCP data. Test server used port $41$ and received and displayed data on the screen. The client used \texttt{tsocks} to transport its TCP stream through onion proxy. The Tor configuration file was configured to select designated Tor entry and exit routers.

Herzberg \emph{et al.} \cite{herzberg2011camouflaged} implemented their proposed camouflaged browsing design over a test machine with an ADSL connection to the Internet with $1,269$kBps downlink and $103$ kBps uplink bandwidth. Four different URLs were tested with $100$ measurements and access time for browsers was recorded. \texttt{wget} was used to download web pages.

Bauer \emph{et al.} \cite{bauer2009predicting} built an extensive experimental setup by establishing circuits for different kinds of applications with a number of malicious routers. The simulator generated $10,000$ circuits with $6$ to $106$ malicious routers. The path compromise rate for different applications was estimated by the selection of malicious routers.

Song \emph{et al.} \cite{song2013anonymize} used an \emph{Au3} script to capture Tor traffic. Six nodes located at distinct places (India, Romania, Luxemburg, New Zealand, Chile, and Russia) were deployed as exit nodes. Onion proxy running on a local PC was configured to use the deployed exit nodes. Traffic of all routers was captured to analysis.

Dhungel and Steiner \cite{dhungel2010waiting} measured delay of Tor network by setting up two relays instead of three. Client, exit router and destination server were fixed while the entry router was selected from the list of available routers. To cope with varying network characteristics, the experiment was repeated for eight months with each duration of $40$ minutes. All $1,426$ available routers were pinged five times for measurements.

Gros \emph{et al.} \cite{gros2010protecting} performed experiments by using the proposed Honeywall mechanism. All vulnerable clients using Tor were placed on one side of the Honeywall and Internet cloud was present on the other side of the Honeywall. All Tor clients had distinct private addresses while Honeywall had a single public IP address.

Wang \emph{et al.} \cite{wang2009novel} conducted experiments in a partially controlled environment. The OP code was modified to use the designated entry and exit Tor routers. Entry and exit Tor nodes were configured to record the data relayed through them. Internet Explorer was used at the Tor client through Privoxy. The middle Tor router was selected through Tor router selection algorithms.

Zhang \emph{et al.} \cite{zhang2011application} used Mozilla Firefox on Fedora $11$ to access the hidden service using \emph{Polipo}. The hidden server was configured to use the bridge whose traffic was being logged continuously. Clients and bridges were configured to record the circuit ID, command, stream ID and arrival time.

Loesing \emph{et al.} \cite{loesing2008performance} conducted experiments on Tor by configuring the Tor client to use fixed first / entry relay, which was being monitored continuously. Second and third Tor relays were chosen randomly by Tor's router selection algorithm. A single access attempt was performed by creating new Tor clients after every five minutes over a $72$ hour duration.

Chen and Pasquale \cite{chen2010toward} analyzed the throughput by downloading a $100$kB file through nearly $100$ unique paths with $10$ times repeated downloads over each path. $10$ Tor clients were configured over PlanetLab testbed distributed around the globe. A file server containing $100$kB file was hosted in the US using \texttt{thttpd}. \texttt{cURL} was used to conduct downloads. Python was used to write measurement scripts using the \texttt{TorCtl} library for Tor control port. Tor circuits were configured to be replaced after every $30$ minutes instead of $10$ minutes so that no middle replacement takes place.

Panchenko \emph{et al.} \cite{panchenko2012improving} performed experiments in the current Tor network with the estimation of delay and throughput. In first experiment, onion routers are fixed but links connecting the circuit are variable. $2,000$ sets were built with random onion routers. In the second experiment, ICMP Ping was used to measure the delay between sending a \emph{SYN} and receiving a \emph{SYN-ACK} packet. Each ping was iterated $20$ times to calculate the mean value.

Houmansadr \emph{et al.} \cite{houmansadr2013parrot} conducted a deep experimental investigation on the Tor network. Application layer softwares (Skype, CensorSpoofer) were executed in VirtualBox virtual machines (VMs) on a Funtoo Linux machine. Various VMs were connected through virtual distributed Ethernet (VDE). Authors built their own plugin for VDE which could drop packets at variable rates and also modify packet contents. Various VDE switches were connected to the central switch which provides DHCP connectivity to the Internet.

Li \emph{et al.} \cite{li2012tmt} tested their proposed tunable mechanism of Tor (TMT) over the real Tor network. Two virtual private servers (VPSes), acting as client and server, were configured on \emph{Linode}. The client was configured with the TMT enhancement while  the server hosted a web page. Time to load the file, number of attempts and number of failure attempts were measured to estimate the performance of TMT.

Chakravarti \emph{et al.} \cite{chakravarty2008identifying} setup their own client, server and probing host machine at three distinct locations inside US. $100$MB file was placed at the web server which gave sufficient downloading time to the client. Linux traffic controller was used to shape the client-server bandwidth. $26$ distinct Tor circuits were created and probed through different locations and compromised links were detected.

Mulazzani \emph{et al.} \cite{mulazzani2010anonymity} collect data by using \emph{TorStatus} and updating its script \emph{tns-update.pl} and \emph{network-history.php}. \emph{RRDtool} was used to store the values in a round robin database (RRD). The collected dataset was used for basic network monitoring.

Chaabane \emph{et al.} \cite{chaabane2010digging} conduct a deep traffic analysis of Tor using HTTP and Bit Torrent protocols. The authors created and monitored six Tor relay nodes (placed in US, Germany, France, Japan, Taiwan) advertising $100$kB available bandwidth for $23$ days. On average $20$GB of data is provided by each server on every day. Data was collected at entry and exit relays.

Bai \emph{et al.} \cite{bai2008traffic} setup eight PCs with one PC running Tor and one PC running java anonymous proxy (JAP). Dummy traffic was generated from the other six PCs. Traffic was captured through \texttt{ethereal}. \emph{Winsock Packet Editor} was used to record packets generated by a specific application. Duration of the test was about $120$mins with five repetitions.

Barker \emph{et al.} \cite{barker2011using} collected Tor network traces by developing a complete Tor setup. Firefox running on Ubuntu was used on all machines. Using the Selenium browser testing framework, $170$ simulations were executed by accessing $30$ websites. Three directory servers with $15$ relays were configured to be used for experiments. Regular HTTPs traffic and HTTP and HTTPs traffic through private Tor network were collected.

Marks \emph{et al.} \cite{marks2010unleashing} conducted a simple experiment using three PCs running Linux kernel (\texttt{2.6.26} and CUBIC TCP). All three machines were connected via an Ethernet switch. All Ethernet interfaces were configured to be $10$Mbps full-duplex links. The first and last two devices setup TCP connections. First device sent data to the second for $250$secs while the second retransmitted after $50$secs delay for a duration of $250$secs.

Jin and Wang \cite{jin2009effectiveness} conducted extensive experiments by monitoring both anonymous traffic and Tor traffic during two experiments. In the first experiment, an Apache webserver on a Dell PC using Redhat Enterprise 4 Linux was configured. A watermark encoder was installed on the Apache proxy. A Dell Precision $390$ was configured as a NAT router to route traffic between client and the anonymous server. In the second experiment, an SSH server and watermark encoder were installed on one machine acting as server. SSH client and watermark decoder were installed on another machine. Three random characters were sent every second from one machine to the other through Tor. Entry and exit relays were fixed.

Tang and Goldberg \cite{tang2010improved} setup their own node (acting as the middle node). Entry and exit nodes were selected from the Tor nodes of the directory server. Authors avoided the use of PlanetLab testbed because majority nodes were providing only $100$KB/s. \emph{Webfetch} was used to download the target file ($87$KB) from author's web server. Connecting circuits and load was varied to verify the proposed path selection strategy.

Alsabah \emph{et al.} \cite{alsabah2012enhancing} performed real world experiments by collecting offline data of $200$ circuits from three distinct application traces. All three applications (BitTorrent client, web browsing client and stream client) were setup on the same machine which was configured to use a specified Tor node as the entry node. All $200$ circuits included browsing ($122$), BitTorrent ($49$) and streaming circuits ($28$). All applications collected $24$ hours of data over a $6$ week period with periodic intervals.

Moghaddam \emph{et al.} \cite{mohajeri2012skypemorph} implemented their proposed SkypeMorph technique on Linux using C and C++ with boost libraries. Authors collected traces of Skype data set for modeling using multiple machines. The proposed SkypeMorph scheme was tested by downloading multiple files with and without it.

Weinberg \emph{et al.} \cite{weinberg2012stegotorus} implemented the proposed scheme \emph{StegoTorus} by deploying an experimental setup. The client was a desktop PC in California with DSL link to the Internet (downstream $5.6$Mb/s, upstream $0.7$Mb/s) and the virtual host was situated in New Jersey inside a commercial data center. $1$MB files were downloaded over several trials to test the performance.

Evans \emph{et al.} \cite{evans2009practical} performed experiments on the real Tor network for their proposed congestion attack. The victim user (to be breached) was using Javascript on her browser. Entry node was fixed but the other two Tor relay nodes were selected at random (by Tor's router selection algorithm).

Wang \emph{et al.} \cite{wang2012congestion} measured the network delays during congestion by collecting delay readings of all Tor routers for $72$ hours in August $2011$. At the next stage, authors collected RTT measurements of the modified and unmodified Tor client to setup $255$ circuits. In all experiments, client machines were modified to incorporate the proposed algorithm and measure the delay.

Ehlert \cite{ehlert2011i2p} measured the performance of I2P and Tor network. For I2P network, experimental setup consisted of two machines, acting as dedicated outproxy and client. $500$ most visited websites were used for downloading webpages. For Tor, a client machine was connected to the Tor network and performance parameters were measured similar to I2P proxy.

Barbera \emph{at al.} \cite{barbera2013cellflood} conducted controlled experiments by setting up $100$Mb/s network connected to four hosts (possessing $2.66$GHz Core 2 Duo CPUs). For real time network experiments, the authors used their two OR nodes, acting as Tor relays. CellFlood attacks were performed on these routers and performance of attack and mitigation scheme was analyzed.

Winter and Lindskog \cite{winter2012great} deployed one relay in Russia and two bridges in Singapore and Sweden. Multiple clients were present in China for connection setup to Tor through designated bridges and relays. In Singapore, a Tor relay was hosted in an Amazon EC2 cloud. Bridge and relay in Sweden and Russia were hosted by an institution and data center, respectively. For vantage points in China, $32$ SOCKS proxies and a VPS running Linux was used.

Edmundson \emph{et al.} \cite{edmundson14security} analyzed the security of Safeplug and Tor by conducting separate experiments for both applications. Authors measured the latency of the system, and investigated the effect of cookies and third party trackers over both applications.

Huber \emph{et al.} \cite{huber2010tor} deployed a Tor exit node which logs the HTTP requests. Nine million HTTP requests were recorded in several weeks. All requests were analyzed for available patterns and statistics were presented in the research.

Blond \emph{et al.} \cite{blond2011one} conducted experiments by deploying Tor exit nodes. Authors instrumented and monitored six Tor nodes for a period of three weeks. One exit node was configured to accept TCP connection for Bit torrent, in order to perform the hijacking attack.

Lenhard \emph{et al.} \cite{lenhard2009performance} ran Tor processes on their devices connected to the Tor network. The hidden services were accessed through low bandwidth access network edge. A modem provided a data rate of $56$kb/s downstream and $44$kb/s upstream. For EDGE, data rate was around $230$kb/s. The broadband network provided $100$Mb/s.

McCoy \emph{et al.} \cite{mccoy2008shining} setup their router connected to $1$Gb/s network link with a rank of top $5\%$ Tor routers and flagged as \emph{Running}. At most $20$bytes were logged to avoid information breaching laws. Setup was configured for both experiments separately covering (1) exit router and (2) non-exit router. Entrance and middle router traffic was logged for $15$ days comprising of time stamp, previous hop's IP, TCP port, next hop's IP and circuit identifier. For exit traffic logging, \texttt{tcpdump} was used over the router which relayed $709$GB of traffic and only the first $150$bytes of packet were logged. \texttt{Ethereal} was used for protocol analysis.

Chakravarty \emph{et al.} \cite{chakravarty2011detecting} transmitted decoy traffic over a  custom client supporting IMAP and SMTP protocols. The client was implemented using Perl and service protocol emulation was provided by \emph{Net::IMAPClient} and \emph{Net::SMTP}. The client hosted on Intel Xeon CPU running Ubuntu Server Linux \texttt{v8.04}.

Snader and Borisov \cite{snader2008tune} performed experiments on Tor by downloading $1$MB files over HTTP connections through various exit routers. All other entities including guard routers, client and web server remain fixed for the entire duration of the experiment. $20,000$ and $40,000$ trials were performed for tunable Tor and standard Tor respectively spanning a duration of two months.

Gilad and Herzberg \cite{gilad2012spying} conducted an empirical investigation for the performance of proposed attacks in the Tor network. Indirect rate reduction attack was evaluated by experiments in the live network. For experiments, a Linux machine ran an Apache web server. Data at the rate of $0.5$KBps was transmitted.

Loesing \emph{et al.} \cite{loesing2010case} collected Tor statistics by following the legal requirements, user privacy, ethical approvals, informed consent and community acceptance. Authors collected data from the Tor network and evaluated the port numbers and country of origin of the obtained IP addresses.

Chen \emph{et al.} \cite{chen2009xpay} developed ORPay which uses out-of-band communication for payment primitives and control messages. The ``bank'' was built using C language and OpenSSL for encryption. Authors performed controlled experiments consisting of a set of interconnected PCs running directory servers and Tor routers on VMs. Inter-client bandwidth was $500-600$KB/s with $1-2$ms average latency and $0.5$ms for inter-VMs on the same machine. One micropayment was made for every $20$ packets.

Panchenko \emph{et al.} \cite{panchenko2011website} using standard PCs for fetching websites using Firefox with disabled active components (Java, Flash etc.) and Chickenfoot used as the default plugin. The closed-world dataset was collected from previous studies, to obtain labeled ground truth dataset.

Wang and Goldberg \cite{wang2013improved} performed experiments on SHARCNET, a parallel computing cluster. Upto $200$ cores were used for computation of SVM kernel matrix. \texttt{torrc} was configured to close the circuits manually instead of fixed $10$mins duration and fixed entry guard selection was disabled. iMacros and Tor controller was used to automate site accesses. For closed world circuits, fingerprinting was performed on $100$ sites with $40$ instances each and using $10$-fold cross validation. For open-world experiments, Alexa's top $1,000$ sites list was used.

\subsubsection{PlanetLab Experiments}

Akhoondi \emph{et al.} \cite{akhoondi2012lastor} performed experiments in the real Tor network by modifying the Tor Client with their proposed \emph{LASTor} protocol. \texttt{LASTor} is a Java application controlling the Tor client through \emph{Control Port}. $50$ \emph{PlanetLab} nodes running \texttt{LASTor} were used as Tor clients to access top $200$ websites. Both latency and anonymization were tested by collecting the traces of data set at the client nodes.

Murdoch and Danezis \cite{murdoch2005low} performed experiments on the Tor network by setting up a probe PC. A modified version of Tor was used in the probe PC to choose routes of length one. A TCP client was also established at the node which connects to the SOCKS interface of Tor using \emph{socat}. Original Tor relays were used with a corrupt destination Tor server recording the traffic traces. The probe server ran at the University of Cambridge Computer Laboratory while victim and corrupt server were run on PlanetLab nodes. Data from $13$ probing Tor nodes was collected and analyzed in \emph{GNU R}.

Bauer \emph{et al.} \cite{bauer2007low} performed experiments over PlanetLab testbed by setting up two independent node networks comprising of $40$ and $60$ nodes, respectively. Two and six malicious nodes were added in the $40$ node network while three and six malicious nodes were added in the $60$ node network. Traffic was generated by six machines running $60$ and $90$ clients (requesting files of less than $10$MB size using HTTP protocol) in the $40$ and $60$ node network, respectively. To avoid flooding of network requests, clients sleep in the $0-60$sec interval for random periods after every random number of web requests.

\subsubsection{Cloud Services}

Sulaiman and Zhioua \cite{sulaiman2013attacking} performed extensive experiments using Amazon \emph{EC2} cloud services. An Apache web server was used to host a simple web page. \texttt{socket.io} with \texttt{node.js.Socket.io} was installed which supported WebSocket to help users' browsers in using OP and using unpopular ports. For path selection, simulations were also performed for entrance router selection algorithm and non-entrance router selection algorithm. Several experiments were conducted on a number of unpopular ports with $1,500$ circuits established per experiment and compromised links were detected.

Karaoglu \emph{et al.} \cite{karaoglu2012multi} implemented a unidirectional scenario of client uploading a file to a web server. A client established multiple socket connections for multipath transmissions. A $1.5$MB file was uploaded through clients. To incorporate geo-diversity, client softwares were installed in the US and at Amazon EC2 sites in Singapore, Ireland and North Virginia. A web server, placed at the Emulab Utah facility, listened on multiple ports.

Biryukov \emph{et al.} \cite{biryukov2013trawling} performed deanonymization by spending less than $100$ USD on Amazon EC2 cloud. $50$ Amazon EC2 instances were generated which captured $59,130$ publication requests. Data from $120$ running hidden services from the Tor network was collected. Collected data was used to identify the vulnerability of Tor hidden services.

\subsubsection{OpenFlow Enabled Network}

Mendonca \emph{et al.} \cite{mendonca2012flexible} used OpenFlow implementation for their proposed AnonyFlow scheme. An experimental testbed used Linux to connect the two subnetworks. Each subnetwork was connected to two OpenFlow enabled switches and two Net FPGA based switches. All these OpenFlow switches were governed by a NOX controller. \texttt{iperf} was used at the two client hosts with each running for nearly $25$secs.

\subsubsection{Universal Composability framework}

Backes \emph{et al.} \cite{backes2012provably} provided security enhancements to the currently used Tor network. New algorithms were provided and setup in the universal composability (UC) framework.

\subsection{Tor Simulations}

\begin{figure}[b]
  \centering
  \includegraphics[width=0.8\columnwidth]{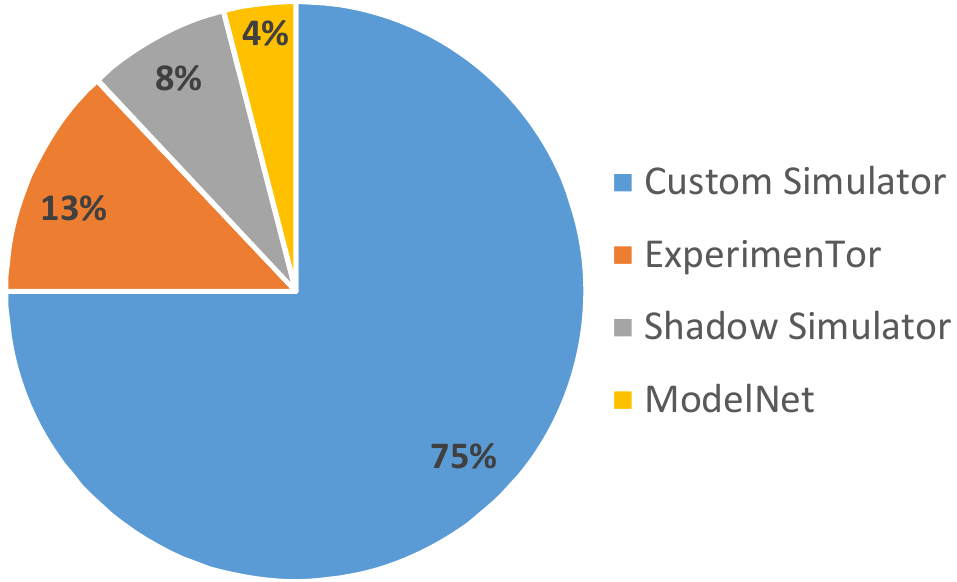}\\
  \caption{Classification of simulations on Tor.}\label{fig: Classification of Simulations on Tor.}
\end{figure}

Tor simulations have been performed by (1) developing custom simulator (2) using ExperimenTor, (3) employing Shadow simulator, and (4) using ModelNet, as shown in Figure \ref{fig: Classification of Simulations on Tor.}.
Figure \ref{fig: Classification of Simulations on Tor.} shows that 75\% of the researches (comprising of simulations) used in this study developed their custom simulator. Only $13\%$ used ExperimenTor. $8\%$ and $4\%$ of the studies used Shadow simulator and Modelnet, respectively.

\subsubsection{Custom Simulator}


Tschorsch and Scheuermann \cite{tschorsch2011tor} conducted simulations on \emph{ns-3} to implement Tor network with and without \emph{N23} modifications. To replicate the onion routers environment, all onion routers were connected to a central node. Access links of all onion routers had an $80$ms delay and $100$Mbps bandwidth. Sending hosts generate data at a rate of $400$kbps and Tor nodes had a maximum bandwidth limit of $600$kbps.


Doswell \emph{et al.} \cite{doswell2013novel} used the generic network simulator OMNET++ to simulate mobile Tor. Wireless access points were placed $75$m apart and results were estimated using linear mobility. Average throughput (kbps) was selected as the performance metric. A $300$kB webpage was downloaded after every $2$secs over the time-frame of $600$secs. An artificial latency was also introduced to incorporate congestion.


Edman and Syverson \cite{edman2009awareness} implemented the multi-thread path selection algorithm in C. Relationships between different ASs were borrowed from predecessor studies. RIBs collected by University of Oregon's RouteViews project were used.


Ngan \emph{et al.} \cite{dingledine2010building} built a discrete event simulator, in Java, for the Tor network. $64$-bit AMD Opteron $252$ dual core processors were used with $4$GB RAM and operating on Sun's JVM and RedHat Enterprise Linux. Tor network with $150$ relays was simulated and all cells from every client were simulated at every hop. Link latency was $100$ms and link capacity was $500$KB/s. All scenarios were tested comprising of Tor's original design, proposed design and a hybrid mechanism.


Benmeziane and Badache \cite{benmeziane2010tor} built their own simulator which incorporates public communications, DNS requests, and anonymous communications by Tor. The authors used $500$ senders using $100$ Tor relay nodes with $10$ executions per sender. Authors increased the number of recipients to $200$. Much of the simulator details were skipped.

Li \emph{et al.} \cite{li2012relay} developed their own discrete event simulator for Tor network. Key data structures and algorithms of Tor were used to simulate several thousand nodes. However, authors did not perform encryption, decryption and data transmission to avoid complexity. Moreover, simulations were driven by initialization and termination events. For a closer look, realistic values of bandwidth and uptime were obtained from the Tor metrics portal. Effective bandwidth of relays was set to $155$kBps with a $750$ standard deviation. $3000$ relays with millions of clients were used for simulations.

Snader and Borisov \cite{snader2011improving} developed a custom flow-level simulator for the Tor network. Using the Tor metric portal, bandwidth of actual Tor relays was used to simulate a $1,000$ node network. $10,000$ flows were simulated for each time unit of the simulator. Fair queueing was used for flow scheduling.

Jansen \emph{et al.} \cite{jansen2010recruiting} built a discrete event simulator for the Tor network comprising of $19,400$ web clients, $300$ Tor relays, $2,000$ servers and $600$ file sharing nodes. For file sharing, web traffic comprised of $12$Mbps downstream with $1.3$Mbps upstream bandwidth.

Johnson \emph{et al.} \cite{johnson2013users} built the \emph{TorPS} simulator for selection of Tor paths. Simulations were carried out for six months with an adversary model containing one guard relay and one exit relay having $83$MBps and $16.7$MBps. For analysis of client behavior, $50,000$ Monte Carlo simulations were carried out spanning a period of three months.

Nowlan \emph{et al.} \cite{nowlan2013reducing} developed a setup for a small virtual Tor network to estimate the performance of the proposed modification. Tor network comprises of three directory authorities, three relay servers and single onion proxy. The link delay had a mean of $50$ms with a $5\%$ path loss for the second onion router.

Jansen \emph{et al.} \cite{jansen2012methodically} performed extensive simulations for their proposed model on both small and large-scale networks. Loss rate and latency have been borrowed from Ookla and iPlane estimation services. For small scale network, $50$ relays and $500$ clients have been configured using $50$ HTTP file servers. For large scale networks, $100$ relays and $1,000$ clients are linked with $100$ HTTP servers. Files of $320$KB and $5$MB are downloaded for performance analysis.

Danner \emph{et al.} \cite{danner2012effectiveness} carried out extensive simulations of their proposed analytical model. However, authors do not focus on experiments or discrete event simulations.

Wang \emph{et al.} \cite{wang2013rbridge} analyzed the performance of their proposed bridge distribution mechanism on an event-based simulator. Aggressive blocking, conservative blocking and event-driven based blocking of bridges were tested. The authors also developed an analytical model for performance prediction.

Jansen and Hopper \cite{jansen2011shadow} developed a discrete event simulator to replicate the real-world Tor network in software running on a single machine. Performance was validated against $402$ node PlanetLab network. Through HTTP client and server plugins, data was transferred through Shadow for verifications of simulations.

Smits \emph{et al.} \cite{smits2011bridgespa} developed an open source implementation of the proposed mechanism. Implementation is based on Linux version \emph{2.6.4}. Bridge distribution authorities needed to be reconfigured for distribution of keys.

Elahi \emph{et al.} \cite{elahi2012changing} simulated Tor entry guard selection and rotation mechanism on multicore servers with each simulation run comprising of $80,000$ users. The entry guard data was collected from real Tor network spanning a duration of eight months.

Zhang \emph{et al.} \cite{zhang2008novel} developed a complete Tor setup containing client, server and three onion routers. A probe server and user nodes were deployed in different network segments. Tor code in the nodes was configured to use the designated three relay nodes. Data from the probe server was sent in bursts after every $0.2$secs while corrupt server sends data after every $10-15$secs.

\subsubsection{ExperimenTor}

Bauer \emph{et al.} \cite{bauer2011experimentor} built a toolkit for emulation of Tor network named by \emph{ExperimenTor}. The \emph{ModelNet} network emulation platform has been used as the baseline approach. Scalability is one of the issues in ExperimenTor, owing to high resource consumption for large number of nodes.

Wacek \emph{et al.} \cite{wacek2013empirical} performed network experiments over ExperimenTor for a variety of network topologies. Authors also performed simulations on their simulator which modeled a $1,524$ relay network.

Gopal and Heninger \cite{gopal2012torchestra} used ExperimenTor framework for simulations of their proposed Torchestra approach. ExperimenTor was setup on two physical machines working as edge node and emulator. For performance analysis, small and large files were downloaded starting from $300$KB. In the following stage, web and SSH traffic were simulated.

\subsubsection{Shadow Simulator}

Geddes \emph{et al.} \cite{geddes2013low} used the Shadow simulator with real Tor code on a simulated network. The simulated network consisted of $160$ exit relays, $240$ non-exit relays, $2375$ web clients, $125$ bulk clients, $150$ small and medium \texttt{Torperf} clients and $400$ HTTP servers. Experiments consisted of downloads of a $320$KB file from random servers after random delays ($1-60$secs). Bulk clients downloaded $5$MB file without any wait time. For \texttt{TorPerf} clients, $50$KB, $1$MB and $5$MB files were downloaded after every ten minutes.

Jansen \emph{et al.} \cite{jansen2011shadow} performed simulations over Shadow with a setup of $200$ HTTP servers, $950$ Tor web clients, $50$ Tor bulk clients and $50$ Tor relays. Bulk clients downloaded $5$MB file while web clients downloaded $5$KB page. Latency of network was borrowed from the latency of PlanetLab nodes. Performance was estimated by varying the load from $25$ (light) to $50$ (medium) to high ($100$) bulk users.

\subsubsection{ModelNet}

AlSabah \emph{et al.} \cite{alsabah2011defenestrator} used the \texttt{ModelNet} network emulation platform along with practical traffic models for performance evaluations. For small-scale experiments, $200$ downloads are made of the $300$KB and $5$MB files by two clients in two separate experiments. For large scale experiment, $20$ Tor routers are deployed with real Tor networks' bandwidth. Each link has $80$ms RTT delay. Ten clients download $1$--$5$MB file and $190$ clients download $100$--$500$KB file.

\subsection{Tor's Analysis}

A number of studies limited their research works to the analysis of current Tor network instead of simulations and experiments of Tor.
Analysis occurs in the subfields of usability of Tor, path selection mechanism, empirical analysis and development of theoretical model.
In below lines, we present the individual studies comprising of Tor's analysis.

\subsubsection{Analytical Model} Several studies develop analytical model for analysis of Tor network. These studies are presented in following lines.

\emph{Anti-misbehaviour Policy Analysis:} Liu and Wang \cite{liu2009anti} proposed anti-misbehavior policies and analyzed it with the original Tor architecture. No simulations or experiments were conducted.

\emph{Security Analysis:} Goldberg \cite{goldberg2006security} built an analytical model for analyzing the security of Tor's authentication protocol. Authors focused on analytical evaluations rather than simulations or experiments.

\emph{Botnet Abuse Analysis:} Hopper \cite{hoppershort} analyzed the various possibilities for avoiding botnet abuse in the Tor network. Majority schemes were discussed only, and a few schemes were tested to verify the performance. Several schemes were analyzed analytically.

\emph{Anonymity Model:} Xin \emph{et al.} \cite{xin2009design} developed a theoretical model to increase the anonymity of the Tor network. They aimed to implement the proposed system on PlanetLab testbed, in future.

\subsubsection{Empirical Analysis} A number of studies performed empirical analysis of Tor network without performing simulations or experiments. In following lines, we present the findings of these studies.

\emph{Statistical Analysis:} Wang \emph{et al.} \cite{wang2013empirical} performed empirical analysis and used data available from ``Tor Metric Portal'' for analysis. There were no simulations or experiments performed in the research. In another study, Elices \emph{et al.} \cite{elices2011fingerprinting} analyzed their attack on Tor using empirical analysis. Access logs from seven web servers were obtained to analyze user request pattern. Moreover, Abbott \emph{et al.} \cite{abbott2007browser} conducted a statistical evaluation by measuring the probabilities of breaching Tor using the proposed scheme.

\emph{Robustness Analysis:} Barthe \emph{et al.} \cite{barthe2010robustness} analyzed the robustness of the Tor network and proposed enhancements in the current network. Cryptographic enhancements were evaluated without any simulation or experimental validations.

\emph{Path Selection Protocol:} Liu and Wang \cite{liu2009improved} presented an improved circuit building protocol with no simulations or experiments. Proposed algorithm was analyzed considering various aspects. In another study, Liu and Wang \cite{liu2009random} presented random walk based algorithm for Tor circuit construction. Anonymity and performance were the key metrics evaluated in their study. However, the scope of this study did not cover simulations or experimental evaluations.

\subsubsection{Usability Analysis} Usability analysis of Tor has not been carried out by a lot of studies. However, some studies referring to usability analysis are summarized in below lines.

Clark \emph{et al.} \cite{clark2007usability} conducted a usability analysis by installing various components of Tor including Vidalia, Privoxy, Torbutton and Foxyproxy on a standard machine. 
In another study, Abou-Tair \emph{et al.} \cite{abou2009usability} presented the usability analysis of the various anonymous service applications including Tor. Various anonymity tools were installed on a machine and usability, ease of installation and use was analyzed.

\section{Discussion}
\label{sec: Discussion}

In this section, we present the discussion and our findings of Tor network. In the first part, we present the performance metrics used to evaluate the Tor network in different research works. In the second part, we present our findings of Tor research works referred in this study. In the last part, we show our findings for open research areas in the field of Tor network which may be used for future research works.

\subsection{Tor Performance Metrics}
\label{sec:Tor Performance Metrics}

Analyzing the performance metrics is a crucial task for future research, analysis, simulations and experiments in the Tor network.
Table \ref{table: Performance Metrics used in various researches} presents the performance metrics of Tor used in various studies. No clear patterns were observed, so, authors described the metrics used in individual studies. A brief overview of the table shows that throughput (bandwidth) and latency are the most frequently used metrics. However, every research formalized its own performance metric based upon the requirement of the experiment.

\begin{table*}
  \centering
  \small
  \caption{Performance metrics used in various research works.}\label{table: Performance Metrics used in various researches}
  \begin{tabular}{|p{1.8cm}|p{7cm}|p{8cm}|}
    \hline
    Domain & Performance Metrics & Research Works \\
    \hline \hline
    Quality of Service of Tor & Throughput; Bandwidth; Packet rate; Bit rate; Goodput & Mendonca \emph{et al.} \cite{mendonca2012flexible}, Zhang \emph{et al.} \cite{zhang2011application}, Pries \emph{et al.} \cite{pries2008new}, Jin and Wang \cite{jin2009effectiveness}, Marks \emph{et al.} \cite{marks2010unleashing}, Panchenko \emph{et al.} \cite{panchenko2008performance}, Chen and Pasquale \cite{chen2010toward}, Karaoglu \emph{et al.} \cite{karaoglu2012multi}, Li \emph{et al.} \cite{li2012relay}, Panchenko \emph{et al.} \cite{panchenko2012improving}, Snader and Borisov \cite{snader2011improving}, Houmansadr \emph{et al.}, Pries \emph{et al.} \cite{pries2008performance}, Tschorsch and Scheuermann \cite{tschorsch2011tor}, Andersson and Panchenko \cite{andersson2007practical}, Doswell \emph{et al.} \cite{doswell2013novel}, Jansen \emph{et al.} \cite{jansen2010recruiting}, Moghaddam \emph{et al.}, Weinberg \emph{et al.} \cite{weinberg2012stegotorus}, Jansen \emph{et al.} \cite{jansen2012methodically}, Ehlert \cite{ehlert2011i2p}, Barbera \emph{et al.} \cite{barbera2013cellflood}, Hopper \cite{hoppershort}, Nowlan \emph{et al.} \cite{nowlan2013reducing}, Ngan \emph{et al.} \cite{dingledine2010building}, Panchenko \emph{et al.} \cite{panchenko2008performance}, Wang \emph{et al.} \cite{wang2013empirical}, Jansen \emph{et al.} \cite{jansen2010recruiting}, Tang and Goldberg \cite{tang2010improved}, AlSabah \emph{et al.}, Wack \emph{et al.} \cite{wacek2013empirical}, Geddes \emph{et al.} \cite{geddes2013low}, AlSabah \emph{et al.} \cite{alsabah2011defenestrator}, Jansen and Hopper \emph{et al.} \cite{jansen2011shadow}, Jansen \emph{et al.} \cite{jansen2012throttling}   \\
    \cline{2-3}
    & Latency; Webpage loading time; Round trip time; Download Time; Router latency; Circuit setup duration; Boot strap duration; Time to first byte; Time to last byte; Ping reply delay; Per hop latency; SYN and SYN ACK difference; Delay per cell; Jitter; Inter packets delay distribution  & Mendonca \emph{et al.} \cite{mendonca2012flexible}, Overlier and Syverson \cite{overlier2006locating}, Loesing \emph{et al.} \cite{loesing2008performance}, Chan-Tin \emph{et al.} \cite{chan2013revisiting}, Herzberg \emph{et al.} \cite{herzberg2011camouflaged}, Murdoch and Danezis \cite{murdoch2005low}, Zhang \emph{et al.} \cite{zhang2008novel}, Akhoondi \emph{et al.} \cite{akhoondi2012lastor}, Panchenko \emph{et al.} \cite{panchenko2008performance}, Li \emph{et al.} \cite{li2012tmt}, Dhungel and Steiner \cite{dhungel2010waiting}, Andersson and Panchenko \cite{andersson2007practical}, Doswell \emph{et al.} \cite{doswell2013novel}, AlSabah \emph{et al.}, Moghaddam \emph{et al.},  Weinberg \emph{et al.} \cite{weinberg2012stegotorus},     Evans \emph{et al.} \cite{evans2009practical}, Wang \emph{et al.} \cite{wang2012congestion}, Jansen \emph{et al.} \cite{jansen2012methodically}, Ehlert \cite{ehlert2011i2p}, Hopper \cite{hoppershort}, Winter and Lindskog \cite{winter2012great}, Edmundson \emph{et al.} \cite{edmundson14security}, Ngan \emph{et al.} \cite{dingledine2010building}, Panchenko \emph{et al.} \cite{panchenko2008performance}, Lenhard \emph{et al.} \cite{lenhard2009performance}, Wack \emph{et al.} \cite{wacek2013empirical}, Geddes \emph{et al.} \cite{geddes2013low}, AlSabah \emph{et al.} \cite{alsabah2011defenestrator}, Jansen and Hopper \emph{et al.} \cite{jansen2011shadow}, Snader and Borisov \cite{snader2008tune}, Jansen \emph{et al.} \cite{jansen2012throttling}, Chen \emph{et al.} \cite{chen2009xpay}, Smits \emph{et al.} \cite{smits2011bridgespa}, Gopal and Heninger \cite{gopal2012torchestra}, Panchenko \emph{et al.} \cite{panchenko2012improving}, Panchenko \emph{et al.} \cite{panchenko2008performance}      \\
    \hline
    Performance of Tor's breaching attempts & True Positive; True Negative; False Positive; False Negative; Region of Convergence; Recognition rate; Mis-recognition rate; Accuracy; Recall; Precision; F-measure & Chakravarti \emph{et al.} \cite{chakravarty2008identifying}, Barker \emph{et al.} \cite{barker2011using}, Chan-Tin \emph{et al.} \cite{chan2013revisiting}, Akhoondi \emph{et al.} \cite{akhoondi2012lastor}, Danner \emph{et al.} \cite{danner2012effectiveness}, Gilad and Herzberg \cite{gilad2012spying}, Panchenko \emph{et al.} \cite{panchenko2011website}, Song \emph{et al.}, Wang and Goldberg \cite{wang2013improved}, Wang and Goldberg \cite{wang2013improved}, Elices \emph{et al.} \cite{elices2011fingerprinting}, Bai \emph{et al.} \cite{bai2008traffic}, AlSabah \emph{et al.}, Wagner \emph{et al.} \cite{wagner2012breaking}\\
    \cline{2-3}
    & Timing Attack correlation & Overlier and Syverson \cite{overlier2006locating}, Zhang \emph{et al.} \cite{zhang2011application}, Pries \emph{et al.} \cite{pries2008new}, Wang \emph{et al.} \cite{wang2009novel}, Houmansadr \emph{et al.},  Murdoch and Danezis \cite{murdoch2005low}, Song \emph{et al.}, Panchenko \emph{et al.} \cite{panchenko2008performance}  \\
    \cline{2-3}
    & Compromised relays; Compromised circuits; Compromised Streams; Time for first compromised stream; Failure rate; Compromise time; Compromised links; Compromised Tunnels; Detection rate; Compromised Clients; Compromised router bandwidth; Compromise probability; Congestion attack time & Overlier and Syverson \cite{overlier2006locating}, Sulaiman and Zhioua \cite{sulaiman2013attacking}, Chen and Pasquale \cite{chen2010toward}, Li \emph{et al.} \cite{li2012relay}, Li \emph{et al.} \cite{li2012tmt}, Bauer \emph{et al.} \cite{bauer2007low}, Johnson \emph{et al.} \cite{johnson2013users}, Evans \emph{et al.} \cite{evans2009practical}, Danner \emph{et al.} \cite{danner2012effectiveness}, Chakravarty \emph{et al.} \cite{chakravarty2011detecting}, Snader and Borisov \cite{snader2008tune}, Panchenko \emph{et al.} \cite{panchenko2011website}, Elahi \emph{et al.} \cite{elahi2012changing}, Abbott \emph{et al.} \cite{abbott2007browser}, Bauer \emph{et al.} \cite{bauer2009predicting}, Wang \emph{et al.} \cite{wang2012congestion}\\
    \hline
    Analysis of Tor & Packet Sizes; Probability difference plots; Energy plots; Recipient probabilities; Queued messages length; Anonymity vs performance; Router bandwidth; HTTP content distribution; Tor servers; Tor traffic; Generated Paths; Client resource usuage; Tor load per circuit; Node Connection pattern; IP TTL difference; Service, browser, file format usuage; Tor location usuage; Boot strap time; Exit traffic stats; Tor bridges statistics; hidden service descriptor request rate; Botnet decay rate; Tor overhead; Router statistics
     & Barker \emph{et al.} \cite{barker2011using}, Loesing \emph{et al.} \cite{loesing2008performance}, Benmeziane \emph{et al.}, Jin and Wang \cite{jin2009effectiveness}, Zhang \emph{et al.} \cite{zhang2008novel}, Liu and Wang \cite{liu2009random}, Liu and Wang \cite{liu2009anti}, Dhungel and Steiner \cite{dhungel2010waiting}, Chaabane \emph{et al.} \cite{chaabane2010digging}, Mulazzani \emph{et al.} \cite{mulazzani2010anonymity}, Moghaddam \emph{et al.}, Edman and Syverson \cite{edman2009awareness}, Barbera \emph{et al.} \cite{barbera2013cellflood}, Hopper \cite{hoppershort}, Winter and Lindskog \cite{winter2012great}, Huber \emph{et al.} \cite{huber2010tor}, Blond \emph{et al.} \cite{blond2011one}, Lenhard \emph{et al.} \cite{lenhard2009performance}, McCoy \emph{et al.} \cite{mccoy2008shining}, Wang \emph{et al.} \cite{wang2013rbridge}, Biryukov \emph{et al.} \cite{biryukov2013trawling}, Loesing \emph{et al.} \cite{loesing2010case}, Chen \emph{et al.} \cite{chen2009xpay}, Panchenko \emph{et al.} \cite{panchenko2011website}, Elahi \emph{et al.} \cite{elahi2012changing}, Marks \emph{et al.} \cite{marks2010unleashing}\\
     \cline{2-3}
      & Empirical Evaluations: Usability analysis, security model analysis, general discussion, proposed mechanism validation  & Clark \emph{et al.} \cite{clark2007usability}, Abou-Tair \emph{et al.} \cite{abou2009usability}, Goldberg \cite{goldberg2006security}, Kuhn \emph{et al.}, Barthe \emph{et al.} \cite{barthe2010robustness}, Gros \emph{et al.} \cite{gros2010protecting}\\

    \hline
  \end{tabular}
\end{table*}

\subsection{Survey Findings}

In this section, we summarize our findings for the onion router by comparing all studies with a deep focus over the key concepts and ideas used in different research works.
We divide our research evaluations in three subcategories considering (1) research areas, (2) research platforms, and (3) performance metrics.

\subsubsection{Research Areas}

The majority of Tor research (nearly 55\%) covering anonymity is focused over deanonymization of Tor network. Around 20\% studies are related to the path selection mechanism. Only 25\% research studies are on performance analysis and improvement mechanism of Tor network. According to Dingledine (the co-founder of Tor project), majority research works focus their attention on the breaching Tor.

\emph{Deanonymization:}

In the deanonymization track, 35\% of the studies design deanonymization attacks for Tor while 21\% deanonymize Tor using traffic analysis. 16\% focus on improvements to bypass deanonymization while 14\% study fingerprinting mechanisms to identify Tor traffic on the Internet. Only 9\% identify hidden services while 2\% focus on anonymity mechanisms without using Tor.

All deanonymization related Tor studies have exploited its inherent weaknesses. Compromised relays are the most exploited weaknesses followed by traffic interception and protocol messages. Very few studies focus on the compromised autonomous systems, browsers, servers, decoy traffic, and flag cheating.

\emph{Path Selection:}

In the path selection track, 87\% of the studies focus on the design of new path selection algorithms and 13\% research works analyze  currently developed algorithms.

Our analysis shows that anonymity and performance (bandwidth and latency) are the most important parameters used in the design and analysis of path selection algorithms. Relays have been incorporated in the design of path selection algorithms covering both location and capacity of relays. Other parameters include autonomous systems, hops, multi-path mechanism and load.

\emph{Performance Analysis and Architectural Improvements:}

In the performance analysis and improvement track, 32\% of the research works cover analysis and 27\% of studies focus on performance improvement mechanisms. 29\% of the studies provide general analysis of Tor covering usability and sociability issues. 10\% of the research works focus on modeling of the Tor network while 2\% address client mobility.

Analysis of various research studies show that performance (latency and bandwidth), relay selection, and anonymity are the most used parameters. Other studies also pay attention to queues, QoS, protocol messages and traffic shaping.

\subsubsection{Research Platforms}

An interesting feature revealed in analysis is the fact that 60\% of the research works were conducted by performing real-world experiments on the Tor network. Although special measures were taken to protect the identity of users but majority research works failed to analyze legal or ethical requirements of capturing user data and performing experiments by developing attacks in real network. Only 27\% of studies developed their own simulator and 13\% conducted analysis without experiments or simulations.

\emph{Experiments:}\\
Our survey shows that 86\% of the research works developed their own testbed for experiments. The majority of studies deployed 1-2 clients with 1-3 servers for experiments. Research works covering relays used 1-3 relays. However, some research works increased the number of relays by using virtual machines and PlanetLab. A limited number of studies used cloud-based setups.

\emph{Simulations:}\\
Interestingly, 75\% of the research works developed their own simulator without any common parameters used for Tor network. ns-3, OMNET, C and Java were used for the development of custom simulators. 13\% of the research works used ExperimenTor. Our research shows that ExperimenTor is the most common toolkit used by majority of the research. 8\% and 3\% of the studies used Shadow simulator and ModelNet, respectively.

\subsubsection{Performance Metrics}

We considered the performance metrics used in various works. Analysis shows that no hard and fast rule exists for use of performance metrics. Every study developed its own metrics to measure performance, anonymity and QoS. Moreover, no baseline techniques exist for the comparison of results.

\subsection{Open Research Areas}

Our survey shows that majority of the research works are concentrated in a few domains. However, a number of major challenges exist owing to the peer-to-peer nature of Tor.
A number of key areas have also been identified by the Tor team. We identified the following areas which require further research.

\begin{enumerate}
  \item \emph{Data Estimation:}
    Estimation of key network statistics is the most critical task in the Tor network because it is a peer-to-peer network. No one can see the entire traffic so it is not possible to estimate the size of Tor network. Some of the statistics requiring attention are as follows:
    \begin{itemize}
    \item Number of clients in the network: Peer-to-peer networks make it impossible to estimate the total traffic statistics because no user can see the complete traffic.
    \item Capabilities of relays: There is limited information available about the relays which are the most crucial parameters in path selection. Incorporation of relay capabilities into anonymity of Tor and performance model is a key research area as done in a number of studies.
    \item Performance of the network: Estimation of network performance at any given time is a crucial task. Owing to the P2P nature, only health of relays is known to the Tor administration. \emph{How is the network performing at any given instant?} is still a crucial task.
    \item Number of clients connecting via bridges: Tor authorities provide secret relay addresses to clients who can't access Tor due to blockage of relays in their location. However, very little is known about the quantity of clients connecting through bridges and their traffic statistics.
    \item Exit network traffic: Significant research is required about the exit network traffic. All clients pass their data through relays and very little is known about the statistics of traffic exiting exit relays.
    \end{itemize}
  \item \emph{Analysis:}
        Deep analysis of the current Tor network is required. Analysis may be based upon an extension of previous research into path length, anonymity, latency, etc. Analysis of the optimal performance parameters is required. 
  \item \emph{Measurement and Attack tools:}
        Development of novel attack methodologies to identify the shortcomings of the current Tor network. Tor has no automatic mechanism to identify anomalies and assess the health of the network. Attack tools should be developed which should prevent attacks occurring from compromised relays and servers. Comprised relays are vulnerable to botnet based attacks comprising of DDOS attacks, fingerprinting attacks etc. Despite large amount of research in botnet attacks, it is still open to research which would make Tor a more stable and secure network.
  \item \emph{Defenses against Attacks:}
        Develop novel defense methodologies to counter attacks on the Tor network. Although majority research works have focused on the development of novel attack methodologies, very little is known about viable counter-measures. Our survey shows that relays are mostly vulnerable because they can be deployed by any eavesdropper. Counter-measures against congestion attacks, latency measuring attacks, throughput measuring attacks, etc. can help in the improvement of Tor.
\end{enumerate}

\section{Conclusion}
\label{sec: Conclusion}

This paper deals with the survey, classification, quantification and comparative analysis of various research works covering Tor network.
To the author's best knowledge, no other survey/research has performed such a deep and thorough analysis of Tor studies.
Our study shows that Tor research areas can be broadly classified into (1) deanonymization, (2) path selection, (3) analysis and performance improvements.
More than half studies carried out address `deanonymization' with major subdivisions into deanonymization `attacks' and `traffic analysis' attacks.
In the `path selection' area, more than $85\%$ of the studies have focused on the development of new algorithms.
In the `analysis and performance improvement' area, the majority of studies are a mixed bag, followed by analysis, followed by performance improvement studies.
Our analysis of Tor platforms shows that $60\%$ of studies performed experiments while $27\%$ performed simulations.
Among experiments, $86\%$ of the studies deployed private testbeds.
Among simulations, $75\%$ developed their own simulators.
Analysis of parameters (used in various studies) shows that their is no little consistency across various studies.
However, a majority of the studies used variations of throughput and latency for performance analysis.

\section{Bibliography}
\bibliographystyle{unsrt}
\bibliography{tor}

\begin{thebibliography}{100}
\expandafter\ifx\csname url\endcsname\relax
  \def\url#1{\texttt{#1}}\fi
\expandafter\ifx\csname urlprefix\endcsname\relax\def\urlprefix{URL }\fi
\expandafter\ifx\csname href\endcsname\relax
  \def\href#1#2{#2} \def\path#1{#1}\fi

\bibitem{Brandom2013}
R.~Brandom, {FBI} agents tracked {H}arvard bomb threats despite {T}or, THE
  VERGE. Available at
  http://www.theverge.com/2013/12/18/5224130/fbi-agents-tracked-harvard-bomb-threats-across-tor.

\bibitem{Khrennikov2016}
I.~Khrennikov, Russians {F}ind {W}ays to {B}ypass {L}atest {W}eb {B}an,
  Bloomberg. Available at
  http://www.bloomberg.com/news/articles/2016-02-01/russians-turn-to-tor-and-anonymox-to-bypass-web-blocking.

\bibitem{Hern2016}
A.~Hern, {US} defence department funded {C}arnegie {M}ellon research to break
  {T}or, The guardian. Available at
  https://www.theguardian.com/technology/2016/feb/25/us-defence-department-funding-carnegie-mellon-research-break-tor.

\bibitem{Graham-Smith2016}
D.~Graham-Smith, Extreme online security measures to protect your digital
  privacy – a guide, The guardian. Available at
  https://www.theguardian.com/technology/2016/jul/03/online-security-measures-digital-privacy-guide.

\bibitem{Neal2016}
D.~Neal, Mozilla will have to wait to find out how the {FBI} cracked {T}or, The
  Inquirer. Available at
  http://www.theinquirer.net/inquirer/news/2458121/mozilla-wants-to-know-how-the-fbi-cracked-tor.

\bibitem{alsabah2016performance}
M.~AlSabah, I.~Goldberg, Performance and security improvements for tor: A
  survey, ACM Computing Surveys (CSUR) 49~(2) (2016) 32.

\bibitem{koch2016anonymous}
R.~Koch, M.~Golling, G.~D. Rodosek, How {A}nonymous {I}s the {T}or {N}etwork?
  {A} {L}ong-{T}erm {B}lack-{B}ox {I}nvestigation, Computer 49~(3) (2016)
  42--49.
\newblock \href {http://dx.doi.org/10.1109/MC.2016.73}
  {\path{doi:10.1109/MC.2016.73}}.

\bibitem{alsabah2015performance}
M.~AlSabah, I.~Goldberg, Performance and {S}ecurity {I}mprovements for {T}or:
  {A} {S}urvey., IACR Cryptology ePrint Archive 2015 (2015) 235.

\bibitem{conrad2014survey}
B.~Conrad, F.~Shirazi, A {S}urvey on {T}or and {I2P}, Proceedings of 9th
  International Conference on Internet Monitoring and Protection (ICIMP) (2014)
  22.

\bibitem{jagerman2014fifteen}
R.~Jagerman, W.~Sabee, L.~Versluis, M.~de~Vos, J.~Pouwelse, The fifteen year
  struggle of decentralizing privacy-enhancing technology, arXiv preprint
  arXiv:1404.4818.

\bibitem{ren2010survey}
J.~Ren, J.~Wu, Survey on anonymous communications in computer networks,
  Computer Communications 33~(4) (2010) 420--431.

\bibitem{johnson2007chaum}
P.~C. Johnson, A.~Kapadia, From {C}haum to {T}or and {B}eyond: {A} {S}urvey of
  {A}nonymous {R}outing {S}ystems, Dartmouth College (Technical Report),
  Available at
  http://www.cs.dartmouth.edu/~ccpalmer/classes/cs55/Content/Resources/JohnsonKapadiaSurvey.pdf.

\bibitem{goldschlag1999onion}
D.~Goldschlag, M.~Reed, P.~Syverson, Onion routing, Communications of the ACM
  42~(2) (1999) 39--41.

\bibitem{torMetricPortal2016}
TorMETRICS, Relays and bridges in the network, Available at
  https://metrics.torproject.org/.

\bibitem{reed1996proxies}
M.~G. Reed, P.~F. Syverson, D.~M. Goldschlag, Proxies for anonymous routing,
  in: Annual Computer Security Applications Conference, IEEE, 1996, pp.
  95--104.

\bibitem{goldschlag1996hiding}
D.~M. Goldschlag, M.~G. Reed, P.~F. Syverson, Hiding routing information, in:
  International Workshop on Information Hiding, Springer, 1996, pp. 137--150.

\bibitem{reed1998anonymous}
M.~G. Reed, P.~F. Syverson, D.~M. Goldschlag, Anonymous connections and onion
  routing, IEEE Journal on Selected areas in Communications 16~(4) (1998)
  482--494.

\bibitem{syverson2001towards}
P.~Syverson, G.~Tsudik, M.~Reed, C.~Landwehr, Towards an analysis of onion
  routing security, in: Designing Privacy Enhancing Technologies, Springer,
  2001, pp. 96--114.

\bibitem{bresson2001provably}
E.~Bresson, O.~Chevassut, D.~Pointcheval, J.-J. Quisquater, Provably
  authenticated group diffie-hellman key exchange, in: Proceedings of the 8th
  ACM conference on Computer and Communications Security, ACM, 2001, pp.
  255--264.

\bibitem{pries2008new}
R.~Pries, W.~Yu, X.~Fu, W.~Zhao, A new replay attack against anonymous
  communication networks, in: International Conference on Communications (ICC),
  IEEE, 2008, pp. 1578--1582.

\bibitem{barbera2013cellflood}
M.~V. Barbera, V.~P. Kemerlis, V.~Pappas, A.~D. Keromytis, Cellflood: Attacking
  {T}or onion routers on the cheap, in: European Symposium on Research in
  Computer Security - ESORICS, Springer, 2013, pp. 664--681.

\bibitem{wang2013improved}
T.~Wang, I.~Goldberg, Improved website fingerprinting on {T}or, in: Proceedings
  of the Workshop on Privacy in the Electronic Society (WPES), ACM, 2013, pp.
  201--212.

\bibitem{gordon2016official}
A.~Gordon, S.~Hernandez, The Official (ISC) 2 Guide to the SSCP CBK, John Wiley
  \& Sons, 2016.

\bibitem{softether2016}
S.~V. Project, University of {T}sukuba, https://www.softether.org.

\bibitem{janusvm2016}
J.~A. I.~P. Appliance, http://janusvm.peertech.org.

\bibitem{proxpn2016}
pro{XPN}, https://www.proxpn.com.

\bibitem{USAIP2016}
USAIP, https://usaip.eu.

\bibitem{VPNreactor2016}
VPNreactor, https://www.vpnreactor.com/.

\bibitem{xbbrowser2009}
X.~N. AG, S.~Topletz, x{B} {B}rowser, http://xerobank.com/.

\bibitem{hotspotshield2016}
A.~Inc, Hotspot {S}hield, http://www.hotspotshield.com/.

\bibitem{AdvTor2016}
Advanced {O}nion {R}outer, https://sourceforge.net/projects/advtor/.

\bibitem{securitykiss2016}
Security{KISS}, https://www.securitykiss.com/.

\bibitem{ultrsurf2016}
UltraReach, Ultra{S}urf, http://ultrasurf.us/.

\bibitem{cyberghost2016}
C.~S.R.L, Cyber{G}host {VPN}, http://www.cyberghostvpn.com/.

\bibitem{freegate2016}
D.~I. T.~I. (DIT), Freegate, http://dit-inc.us/freegate.html.

\bibitem{tails2016}
Tails, https://tails.boum.org/.

\bibitem{privatix2016}
Privatix, http://www.mandalka.name/privatix/.

\bibitem{ixquick2016}
S.~H. B.V., Ixquick, https://www.ixquick.com/.

\bibitem{duckduckgo2016}
I.~DuckDuckGo, Duck{D}uck{G}o, https://duckduckgo.com/.

\bibitem{anonymousemail2016}
Anonymous {E}mail, https://anonymousemail.me/.

\bibitem{safemail2016}
Safe-mail, http://www.safe-mail.net/.

\bibitem{hushmail2016}
Hushmail, https://www.hushmail.com/.

\bibitem{10minutemail2016}
10minutemail, https://10minutemail.com.

\bibitem{yopmail2016}
Y{OP}mail, http://www.yopmail.com.

\bibitem{arp2014torben}
D.~Arp, F.~Yamaguchi, K.~Rieck, Torben: Deanonymizing {T}or communication using
  web page markers, Tech. rep., IFI-TB-2014-01, University of G{\"o}ttingen
  (2014).

\bibitem{overlier2006locating}
L.~Overlier, P.~Syverson, Locating hidden servers, in: Symposium on Security
  and Privacy, IEEE, 2006, pp. 15--pp.

\bibitem{elices2011fingerprinting}
J.~A. Elices, F.~Perez-Gonzalez, C.~Troncoso, Fingerprinting {T}or's hidden
  service log files using a timing channel, in: International Workshop on
  Information Forensics and Security (WIFS), IEEE, 2011, pp. 1--6.

\bibitem{zhang2011application}
L.~Zhang, J.~Luo, M.~Yang, G.~He, Application-level attack against {T}or's
  hidden service, in: International Conference on Pervasive Computing and
  Applications (ICPCA), IEEE, 2011, pp. 509--516.

\bibitem{biryukov2013trawling}
A.~Biryukov, I.~Pustogarov, R.~Weinmann, Trawling for {T}or hidden services:
  {D}etection, measurement, deanonymization, in: Symposium on Security and
  Privacy, IEEE, 2013, pp. 80--94.

\bibitem{bai2008traffic}
X.~Bai, Y.~Zhang, X.~Niu, Traffic identification of {T}or and web-mix, in:
  International Conference on Intelligent Systems Design and Applications
  (ISDA), Vol.~1, IEEE, 2008, pp. 548--551.

\bibitem{barker2011using}
J.~Barker, P.~Hannay, P.~Szewczyk, Using {T}raffic {A}nalysis to {I}dentify the
  {S}econd {G}eneration {O}nion {R}outer, in: International Conference on
  Embedded and Ubiquitous Computing (EUC), IEEE, 2011, pp. 72--78.

\bibitem{alsabah2012enhancing}
M.~AlSabah, K.~Bauer, I.~Goldberg, Enhancing {T}or's performance using
  real-time traffic classification, in: Proceedings of the conference on
  Computer and Communications Security, ACM, 2012, pp. 73--84.

\bibitem{houmansadr2013parrot}
A.~Houmansadr, C.~Brubaker, V.~Shmatikov, The parrot is dead: Observing
  unobservable network communications, in: Symposium on Security and Privacy,
  IEEE, 2013, pp. 65--79.

\bibitem{chakravarty2008identifying}
S.~Chakravarty, A.~Stavrou, A.~D. Keromytis, Identifying proxy nodes in a {T}or
  anonymization circuit, in: International Conference on Signal Image
  Technology and Internet Based Systems (SITIS), IEEE, 2008, pp. 633--639.

\bibitem{winter2012great}
P.~Winter, S.~Lindskog, How the great firewall of {C}hina is blocking {T}or,
  Free and Open Communications on the Internet (FOCI).

\bibitem{sulaiman2013attacking}
M.~A. Sulaiman, S.~Zhioua, Attacking {T}or through {U}npopular {P}orts, in:
  International Conference on {D}istributed {C}omputing {S}ystems {W}orkshops
  (ICDCSW), IEEE, 2013, pp. 33--38.

\bibitem{chan2013revisiting}
E.~Chan-Tin, J.~Shin, J.~Yu, Revisiting {C}ircuit {C}logging {A}ttacks on
  {T}or, in: International Conference on Availability, Reliability and Security
  (ARES), IEEE, 2013, pp. 131--140.

\bibitem{wang2009novel}
X.~Wang, J.~Luo, M.~Yang, Z.~Ling, A novel flow multiplication attack against
  {T}or, in: International Conference on Computer Supported Cooperative Work in
  Design (CSCWD), IEEE, 2009, pp. 686--691.

\bibitem{wagner2012breaking}
C.~Wagner, G.~Wagener, R.~State, A.~Dulaunoy, T.~Engel, Breaking {T}or
  anonymity with game theory and data mining, Concurrency and Computation:
  Practice and Experience 24~(10) (2012) 1052--1065.

\bibitem{benmeziane2010tor}
S.~Benmeziane, N.~Badache, Tor network limits, Tech. rep., CERIST Digital
  Library, Availablle at http://dl.cerist.dz/handle/CERIST/317 (2010).

\bibitem{jansen2014sniper}
R.~Jansen, F.~Tschorsch, A.~Johnson, B.~Scheuermann, The {S}niper {A}ttack:
  {A}nonymously {D}eanonymizing and {D}isabling the {T}or {N}etwork, in:
  Network and Distributed Systems Security Symposium (NDSS), San Diego, CA,
  USA, 2014.

\bibitem{abbott2007browser}
T.~G. Abbott, K.~J. Lai, M.~R. Lieberman, E.~C. Price, Browser-based attacks on
  {T}or, in: Privacy Enhancing Technologies, Springer, 2007, pp. 184--199.

\bibitem{evans2009practical}
N.~S. Evans, R.~Dingledine, C.~Grothoff, A {P}ractical {C}ongestion {A}ttack on
  {T}or {U}sing {L}ong {P}aths, in: USENIX Security Symposium, 2009, pp.
  33--50.

\bibitem{bauer2007low}
K.~Bauer, D.~McCoy, D.~Grunwald, T.~Kohno, D.~Sicker, Low-resource routing
  attacks against {T}or, in: Proceedings of the workshop on Privacy in
  electronic society, ACM, 2007, pp. 11--20.

\bibitem{edman2009awareness}
M.~Edman, P.~Syverson, {AS}-awareness in {T}or path selection, in: Proceedings
  of the conference on Computer and Communications Security, ACM, 2009, pp.
  380--389.

\bibitem{blond2011one}
S.~L. Blond, P.~Manils, C.~Abdelberi, M.~A.~D. Kaafar, C.~Castelluccia,
  A.~Legout, W.~Dabbous, One bad apple spoils the bunch: exploiting {P2P}
  applications to trace and profile {T}or users, arXiv preprint
  arXiv:1103.1518.

\bibitem{geddes2013low}
J.~Geddes, R.~Jansen, N.~Hopper, How low can you go: Balancing performance with
  anonymity in {T}or, in: International Symposium on Privacy Enhancing
  Technologies (PETS), Springer, 2013, pp. 164--184.

\bibitem{chakravarty2011detecting}
S.~Chakravarty, G.~Portokalidis, M.~Polychronakis, A.~D. Keromytis, Detecting
  traffic snooping in {T}or using decoys, in: Recent Advances in Intrusion
  Detection (RAID), Springer, 2011, pp. 222--241.

\bibitem{johnson2013users}
A.~Johnson, C.~Wacek, R.~Jansen, M.~Sherr, P.~Syverson, Users get routed:
  Traffic correlation on {T}or by realistic adversaries, in: Proceedings of the
  conference on Computer \& Communications Security (CCS), ACM, 2013, pp.
  337--348.

\bibitem{murdoch2005low}
S.~J. Murdoch, G.~Danezis, Low-cost traffic analysis of {T}or, in: Symposium on
  Security and Privacy, IEEE, 2005, pp. 183--195.

\bibitem{chakravarty2014effectiveness}
S.~Chakravarty, M.~V. Barbera, G.~Portokalidis, M.~Polychronakis, A.~D.
  Keromytis, On the {E}ffectiveness of {T}raffic {A}nalysis {A}gainst
  {A}nonymity {N}etworks {U}sing {F}low {R}ecords, in: International Conference
  on Passive and Active Measurement (PAM), Springer, 2014, pp. 247--257.

\bibitem{zhang2008novel}
J.~Zhang, H.~Duan, J.~Wu, A {N}ovel {M}ethod to {P}revent {T}raffic {A}nalysis
  in {L}ow-{L}atency {A}nonymous {C}ommunication {S}ystems, in: Proceedings of
  the International Conference on Computer and Electrical Engineering, IEEE,
  2008, pp. 906--911.

\bibitem{song2013anonymize}
M.~Song, G.~Xiong, Z.~Li, J.~Peng, L.~Guo, A de-anonymize attack method based
  on traffic analysis, in: International ICST Conference on Communications and
  Networking in China (CHINACOM), IEEE, 2013, pp. 455--460.

\bibitem{panchenko2011website}
A.~Panchenko, L.~Niessen, A.~Zinnen, T.~Engel, Website fingerprinting in onion
  routing based anonymization networks, in: Proceedings of the annual workshop
  on Privacy in the Electronic Society (WPES), ACM, 2011, pp. 103--114.

\bibitem{jin2009effectiveness}
J.~Jin, X.~Wang, On the effectiveness of low latency anonymous network in the
  presence of timing attack, in: International Conference on Dependable Systems
  \& Networks (DSN), IEEE, 2009, pp. 429--438.

\bibitem{gilad2012spying}
Y.~Gilad, A.~Herzberg, Spying in the dark: {TCP} and {T}or traffic analysis,
  in: International Symposium on Privacy Enhancing Technologies, Springer,
  2012, pp. 100--119.

\bibitem{gros2010protecting}
S.~Gro{\v{s}}, M.~Salki{\'c}, I.~{\v{S}}ipka, Protecting {TOR} exit nodes from
  abuse, in: Proceedings of the International Convention on Information and
  Communication Technology, Electronics and Microelectronics (MIPRO), IEEE,
  2010, pp. 1246--1249.

\bibitem{winter2014spoiled}
P.~Winter, R.~K{\"o}wer, M.~Mulazzani, M.~Huber, S.~Schrittwieser, S.~Lindskog,
  E.~Weippl, Spoiled {O}nions: Exposing {M}alicious {T}or {E}xit {R}elays, in:
  International Privacy Enhancing Technologies Symposium (PETS), Springer,
  2014, pp. 304--331.

\bibitem{xin2009design}
L.~Xin, W.~Neng, Design {I}mprovement for {T}or {A}gainst {L}ow-{C}ost
  {T}raffic {A}ttack and {L}ow-{R}esource {R}outing {A}ttack, in: International
  Conference on Communications and Mobile Computing (CMC), Vol.~3, IEEE, 2009,
  pp. 549--554.

\bibitem{backes2012provably}
M.~Backes, I.~Goldberg, A.~Kate, E.~Mohammadi, Provably secure and practical
  onion routing, in: Computer Security Foundations Symposium (CSF), IEEE, 2012,
  pp. 369--385.

\bibitem{marks2010unleashing}
D.~Marks, F.~Tschorsch, B.~Scheuermann, Unleashing {T}or, {B}it{T}orrent \&
  co.: {H}ow to relieve {TCP} deficiencies in overlays, in: Conference on Local
  Computer Networks (LCN), IEEE, 2010, pp. 320--323.

\bibitem{nowlan2013reducing}
M.~F. Nowlan, D.~Wolinsky, B.~Ford, Reducing latency in {T}or circuits with
  unordered delivery, in: USENIX Workshop on Free and Open Communications on
  the Internet (FOCI), 2013.

\bibitem{danner2012effectiveness}
N.~Danner, S.~Defabbia-Kane, D.~Krizanc, M.~Liberatore, Effectiveness and
  detection of denial-of-service attacks in {T}or, ACM Transactions on
  Information and System Security (TISSEC) 15~(3) (2012) 11.

\bibitem{herzberg2011camouflaged}
A.~Herzberg, E.~Porat, N.~Soffer, E.~Waisbard, Camouflaged {P}rivate
  {C}ommunication, in: Privacy, security, risk and trust (PASSAT),
  international conference on social computing (socialcom), IEEE, 2011, pp.
  1159--1162.

\bibitem{mendonca2012flexible}
M.~Mendonca, S.~Seetharaman, K.~Obraczka, A flexible in-network {IP}
  anonymization service, in: International Conference on Communications (ICC),
  IEEE, 2012, pp. 6651--6656.

\bibitem{murdoch2006hot}
S.~J. Murdoch, Hot or not: Revealing hidden services by their clock skew, in:
  Proceedings of the conference on Computer and Communications Security (CCS),
  ACM, 2006, pp. 27--36.

\bibitem{chakravarty2008linkwidth}
S.~Chakravarty, A.~Stavrou, A.~D. Keromytis, Linkwidth: a method to measure
  link capacity and available bandwidth using single-end probes, Computer
  Science Department Technical Report (CUCS Tech Report) CUCS-002, Columbia
  University.

\bibitem{oh2017fingerprinting}
S.~E. Oh, S.~Li, N.~Hopper, Fingerprinting keywords in search queries over tor,
  Proceedings on Privacy Enhancing Technologies 2017~(4) (2017) 251--270.

\bibitem{daemen2013design}
J.~Daemen, V.~Rijmen, The design of Rijndael: AES-the advanced encryption
  standard, Springer Science \& Business Media, 2013.

\bibitem{standard2001announcing}
N.-F. Standard, Announcing the advanced encryption standard (aes), Federal
  Information Processing Standards Publication 197 (2001) 1--51.

\bibitem{marks_unleashingtor_thesis}
D.~Marks, F.~Tschorsch, B.~Scheuermann, D.~Marks, F.~Tschorsch, B.~Scheuermann,
  Unleashing tor, bittorrent and co.: How to relieve tcp deficiencies in
  overlays (extended version), Heinrich Heine University, Dusseldorf, Germany.

\bibitem{borisov2007denial}
N.~Borisov, G.~Danezis, P.~Mittal, P.~Tabriz, Denial of service or denial of
  security?, in: Proceedings of the 14th ACM conference on Computer and
  communications security, ACM, 2007, pp. 92--102.

\bibitem{dingledine2004tor}
R.~Dingledine, N.~Mathewson, P.~Syverson, Tor: The second-generation onion
  router, Tech. rep., DTIC Document (2004).

\bibitem{akhoondi2012lastor}
M.~Akhoondi, C.~Yu, H.~V. Madhyastha, {LASTor}: A low-latency {AS}-aware {T}or
  client, in: Symposium on Security and Privacy (SP), IEEE, 2012, pp. 476--490.

\bibitem{chen2010toward}
F.~Chen, J.~Pasquale, Toward improving path selection in {T}or, in: Global
  Telecommunications Conference (GLOBECOM), IEEE, 2010, pp. 1--6.

\bibitem{karaoglu2012multi}
H.~T. Karaoglu, M.~B. Akgun, M.~H. Gunes, M.~Yuksel, Multi {P}ath
  {C}onsiderations for {A}nonymized {R}outing: {C}hallenges and
  {O}pportunities, in: International Conference on New Technologies, Mobility
  and Security (NTMS), IEEE, 2012, pp. 1--5.

\bibitem{panchenko2012improving}
A.~Panchenko, F.~Lanze, T.~Engel, Improving performance and anonymity in the
  tor network, in: Performance Computing and Communications Conference (IPCCC),
  2012 IEEE 31st International, IEEE, 2012, pp. 1--10.

\bibitem{li2012tmt}
C.~Li, Y.~Xue, L.~He, L.~Wang, {TMT}: A new {T}unable {M}echanism of {T}or
  based on the path length, in: International Conference on Cloud Computing and
  Intelligent Systems (CCIS), Vol.~2, IEEE, 2012, pp. 661--665.

\bibitem{panchenko2008performance}
A.~Panchenko, L.~Pimenidis, J.~Renner, Performance analysis of anonymous
  communication channels provided by {T}or, in: International Conference on
  Availability, Reliability and Security (ARES), IEEE, 2008, pp. 221--228.

\bibitem{liu2009random}
X.~Liu, N.~Wang, Random{W}alk-{B}ased {T}or {C}ircuit {B}uilding {P}rotocol,
  in: International Conference on Computational Intelligence and Security
  (CIS), Vol.~2, IEEE, 2009, pp. 335--340.

\bibitem{liu2009improved}
X.~Liu, N.~Wang, An {I}mproved {T}or {C}ircuit-{B}uilding {P}rotocol, in:
  International Joint Conference on Artificial Intelligence (JCAI), IEEE, 2009,
  pp. 671--675.

\bibitem{snader2011improving}
R.~Snader, N.~Borisov, Improving security and performance in the {T}or network
  through tunable path selection, Transactions on Dependable and Secure
  Computing 8~(5) (2011) 728--741.

\bibitem{li2012relay}
C.~Li, Y.~Xue, Y.~Dong, D.~Wang, Relay recommendation system ({RRS}) and
  selective anonymity for {T}or, in: Global Communications Conference
  (GLOBECOM), IEEE, 2012, pp. 833--838.

\bibitem{tang2010improved}
C.~Tang, I.~Goldberg, An improved algorithm for {T}or circuit scheduling, in:
  Proceedings of the conference on Computer and Communications Security (CCS),
  ACM, 2010, pp. 329--339.

\bibitem{wang2012congestion}
T.~Wang, K.~Bauer, C.~Forero, I.~Goldberg, Congestion-aware path selection for
  {T}or, in: Financial Cryptography and Data Security, Springer, 2012, pp.
  98--113.

\bibitem{snader2008tune}
R.~Snader, N.~Borisov, A {T}une-up for {T}or: {I}mproving {S}ecurity and
  {P}erformance in the {T}or {N}etwork, in: Network and Distributed System
  Symposium (NDSS), Vol.~8, 2008, p. 127.

\bibitem{elahi2012changing}
T.~Elahi, K.~Bauer, M.~AlSabah, R.~Dingledine, I.~Goldberg, Changing of the
  guards: A framework for understanding and improving entry guard selection in
  {T}or, in: Proceedings of the workshop on Privacy in the electronic society,
  ACM, 2012, pp. 43--54.

\bibitem{bauer2009predicting}
K.~Bauer, D.~Grunwald, D.~Sicker, Predicting {T}or path compromise by exit
  port, in: International Performance Computing and Communications Conference
  (IPCCC), IEEE, 2009, pp. 384--387.

\bibitem{wacek2013empirical}
C.~Wacek, H.~Tan, K.~S. Bauer, M.~Sherr, An {E}mpirical {E}valuation of {R}elay
  {S}election in {T}or, in: Network and Distributed System Symposium (NDSS),
  2013.

\bibitem{johnson2017avoiding}
A.~Johnson, R.~Jansen, A.~D. Jaggard, J.~Feigenbaum, P.~Syverson, Avoiding the
  man on the wire: Improving tor's security with trust-aware path selection, In
  the Proceedings of the Network and Distributed Security Symposium - NDSS,
  2017.

\bibitem{dingledine2007deploying}
R.~Dingledine, N.~Mathewson, P.~Syverson, Deploying low-latency anonymity:
  Design challenges and social factors, IEEE Security \& Privacy 5~(5) (2007)
  83--87.

\bibitem{abou2009usability}
D.~e. D.~I. Abou-Tair, L.~Pimenidis, J.~Schomburg, B.~Westermann, Usability
  inspection of anonymity networks, in: Proceedings of the World Congress on
  Privacy, Security, Trust and the Management of e-Business, IEEE, 2009, pp.
  100--109.

\bibitem{clark2007usability}
J.~Clark, P.~C. Van~Oorschot, C.~Adams, Usability of anonymous web browsing: an
  examination of {T}or interfaces and deployability, in: Proceedings of the
  symposium on Usable privacy and security, ACM, 2007, pp. 41--51.

\bibitem{edmundson14security}
A.~Edmundson, S.~AKornfeld, J.~A. Kroll, E.~W. Felten, Security {A}udit of
  {S}afeplug ``{T}or in a {B}ox'', in: Workshop on Free and Open Communications
  on the Internet (FOCI), USENIX Association.

\bibitem{barthe2010robustness}
G.~Barthe, A.~Hevia, Z.~Luo, T.~Rezk, B.~Warinschi, Robustness {G}uarantees for
  {A}nonymity, in: Computer Security Foundations Symposium (CSF), IEEE, 2010,
  pp. 91--106.

\bibitem{mulazzani2010anonymity}
M.~Mulazzani, M.~Huber, E.~R. Weippl, Anonymity and monitoring: how to monitor
  the infrastructure of an anonymity system, IEEE Transactions on Systems, Man,
  and Cybernetics, Part C: Applications and Reviews 40~(5) (2010) 539--546.

\bibitem{huber2010tor}
M.~Huber, M.~Mulazzani, E.~Weippl, Tor {HTTP} usage and information leakage,
  in: Communications and Multimedia Security, Springer, 2010, pp. 245--255.

\bibitem{mccoy2008shining}
D.~McCoy, K.~Bauer, D.~Grunwald, T.~Kohno, D.~Sicker, Shining light in dark
  places: Understanding the {T}or network, in: Privacy Enhancing Technologies,
  Springer, 2008, pp. 63--76.

\bibitem{loesing2010case}
K.~Loesing, S.~J. Murdoch, R.~Dingledine, A case study on measuring statistical
  data in the {T}or anonymity network, in: Financial Cryptography and Data
  Security, Springer, 2010, pp. 203--215.

\bibitem{chen2009xpay}
Y.~Chen, R.~Sion, B.~Carbunar, Xpay: Practical anonymous payments for {T}or
  routing and other networked services, in: Proceedings of the workshop on
  Privacy in the electronic society, ACM, 2009, pp. 41--50.

\bibitem{jansen2012methodically}
R.~Jansen, K.~S. Bauer, N.~Hopper, R.~Dingledine, Methodically modeling the tor
  network, in: Workshop on Cyber Security Experimentation and Test (CSET),
  USENIX, 2012.

\bibitem{jansen2011shadow}
R.~Jansen, N.~Hooper, Shadow: Running {T}or in a box for accurate and efficient
  experimentation, Tech. rep., Minnesota University, Department of Computer
  Science and Engineering (No. TR-11-020) (2011).

\bibitem{bauer2011experimentor}
K.~S. Bauer, M.~Sherr, D.~Grunwald, Experimentor: A {T}estbed for {S}afe and
  {R}ealistic {T}or {E}xperimentation, in: Workshop on Cyber Security
  Experimentation and Test (CSET), USENIX, 2011.

\bibitem{dhungel2010waiting}
P.~Dhungel, M.~Steiner, I.~Rimac, V.~Hilt, K.~W. Ross, Waiting for anonymity:
  Understanding delays in the {T}or overlay, in: International Conference on
  Peer-to-Peer Computing (P2P), IEEE, 2010, pp. 1--4.

\bibitem{loesing2008performance}
K.~Loesing, W.~Sandmann, C.~Wilms, G.~Wirtz, Performance measurements and
  statistics of {T}or hidden services, in: International Symposium on
  Applications and the Internet (SAINT), IEEE, 2008, pp. 1--7.

\bibitem{ehlert2011i2p}
M.~Ehlert, {I2P} {U}sability vs. {T}or {U}sability {A} {B}andwidth and
  {L}atency {C}omparison, in: Seminar Report, Humboldt University of Berlin,
  2011.

\bibitem{pries2008performance}
R.~Pries, W.~Yu, S.~Graham, X.~Fu, On performance bottleneck of anonymous
  communication networks, in: International Symposium on Parallel and
  Distributed Processing (IPDPS), IEEE, 2008, pp. 1--11.

\bibitem{liu2009anti}
X.~Liu, N.~Wang, Anti-misbehavior {S}ystem for {T}or {N}etwork, in:
  International Joint Conference on INC, IMS and IDC (NCM), IEEE, 2009, pp.
  257--261.

\bibitem{wang2013empirical}
X.~Wang, J.~Shi, B.~Fang, L.~Guo, An empirical analysis of family in the {T}or
  network, in: International Conference on Communications (ICC), IEEE, 2013,
  pp. 1995--2000.

\bibitem{tschorsch2011tor}
F.~Tschorsch, B.~Scheuermann, Tor is unfair -- {A}nd what to do about it, in:
  Conference on Local Computer Networks (LCN), IEEE, 2011, pp. 432--440.

\bibitem{chaabane2010digging}
A.~Chaabane, P.~Manils, M.~A. Kaafar, Digging into anonymous traffic: A deep
  analysis of the {T}or anonymizing network, in: International Conference on
  Network and System Security (NSS), IEEE, 2010, pp. 167--174.

\bibitem{hoppershort}
N.~Hopper, Challenges in protecting {T}or hidden services from botnet abuse,
  in: International Conference on Financial Cryptography and Data Security,
  Springer, 2014, pp. 316--325.

\bibitem{lenhard2009performance}
J.~Lenhard, K.~Loesing, G.~Wirtz, Performance measurements of {T}or hidden
  services in low-bandwidth access networks, in: Applied Cryptography and
  Network Security, Springer, 2009, pp. 324--341.

\bibitem{goldberg2006security}
I.~Goldberg, On the security of the {T}or authentication protocol, in: Privacy
  Enhancing Technologies, Springer, 2006, pp. 316--331.

\bibitem{jansen2010recruiting}
R.~Jansen, N.~Hopper, Y.~Kim, Recruiting new tor relays with braids, in:
  Proceedings of the 17th ACM conference on Computer and communications
  security, ACM, 2010, pp. 319--328.

\bibitem{dingledine2010building}
R.~Dingledine, D.~S. Wallach, et~al., Building incentives into {T}or, in:
  Financial Cryptography and Data Security, Springer, 2010, pp. 238--256.

\bibitem{wang2013rbridge}
Q.~Wang, Z.~Lin, N.~Borisov, N.~Hopper, rbridge: User {R}eputation based {T}or
  {B}ridge {D}istribution with {P}rivacy {P}reservation, in: Network and
  Distributed System Security Symposium (NDSS), 2013.

\bibitem{smits2011bridgespa}
R.~Smits, D.~Jain, S.~Pidcock, I.~Goldberg, U.~Hengartner, Bridgespa: Improving
  {T}or bridges with single packet authorization, in: Proceedings of the
  workshop on Privacy in the electronic society, ACM, 2011, pp. 93--102.

\bibitem{mohajeri2012skypemorph}
H.~Mohajeri~Moghaddam, B.~Li, M.~Derakhshani, I.~Goldberg, Skypemorph: Protocol
  obfuscation for {T}or bridges, in: Proceedings of the conference on Computer
  and Communications Security (CCS), ACM, 2012, pp. 97--108.

\bibitem{weinberg2012stegotorus}
Z.~Weinberg, J.~Wang, V.~Yegneswaran, L.~Briesemeister, S.~Cheung, F.~Wang,
  D.~Boneh, Stego{T}orus: a camouflage proxy for the {T}or anonymity system,
  in: Proceedings of the conference on Computer and Communications Security
  (CCS), ACM, 2012, pp. 109--120.

\bibitem{gopal2012torchestra}
D.~Gopal, N.~Heninger, Torchestra: Reducing interactive traffic delays over
  {T}or, in: Proceedings of the workshop on Privacy in the electronic society,
  ACM, 2012, pp. 31--42.

\bibitem{alsabah2011defenestrator}
M.~AlSabah, K.~Bauer, I.~Goldberg, D.~Grunwald, D.~McCoy, S.~Savage, G.~M.
  Voelker, Defenestra{T}or: Throwing out windows in {T}or, in: Privacy
  Enhancing Technologies, Springer, 2011, pp. 134--154.

\bibitem{jansen2012throttling}
R.~Jansen, P.~F. Syverson, N.~Hopper, Throttling {T}or {B}andwidth {P}arasites,
  in: USENIX Security Symposium, 2012, pp. 349--363.

\bibitem{mani2017historvarepsilon}
A.~Mani, M.~Sherr, Histor$\varepsilon$: Differentially private and robust
  statistics collection for tor, in: Network and Distributed System Security
  Symposium (NDSS), 2017.

\bibitem{doswell2013novel}
S.~Doswell, N.~Aslam, D.~Kendall, G.~Sexton, The novel use of {B}ridge {R}elays
  to provide persistent {T}or connections for mobile devices, in: International
  Symposium on Personal Indoor and Mobile Radio Communications (PIMRC), IEEE,
  2013, pp. 3371--3375.

\bibitem{andersson2007practical}
C.~Andersson, A.~Panchenko, Practical anonymous communication on the mobile
  internet using {T}or, in: International Conference on Security and Privacy in
  Communications Networks and the Workshops (SecureComm), IEEE, 2007, pp.
  39--48.

\end{thebibliography}

\end{document}